\documentclass{CSML}

\def\dOi{11(4:9)2015}
\lmcsheading%
{\dOi}
{1--41}
{}
{}
{Nov.~\phantom05, 2014}
{Dec.~11, 2015}
{}

\ACMCCS{[{\bf Theory of computation}]: Logic---Equational logic and rewriting}

\usepackage{latexsym,xspace,amssymb,amsmath,graphicx,url} 
\usepackage[all]{xy}
\usepackage{verbatim}
\usepackage{prooftree}
\usepackage[british]{babel}
\usepackage{stmaryrd}
\usepackage{amsfonts}
\usepackage{mathtools}
\usepackage{graphics} 
\usepackage{hyperref}

\newcommand\tauNabla[2]{\llbracket #1\rrbracket^{\nabla}_{\Lambda_{#2}}}
\newcommand\tauDelta[2]{\llbracket #1\rrbracket^{\Delta}_{\Lambda_{#2}}}
\newcommand{\Nterm}{\Delta\vdash t}
\newcommand\piOf[2]{\pi_{#1}\cdot #2}

\def\pia{\pi(a)}
\def\piX{\pi\cdot X}

\let\emptyset\varnothing

\def\suppi{support(\pi)}

\newcommand\metaVar[1]{#1^{n}_{i}}

\def\hash{{\#}}
\def\apart{\hash}

\def\Atoms{\ensuremath{A}\xspace}
\def\Vars{\ensuremath{V}\xspace}
\def\Id{\mathtt{Id}}

\newcommand\Tran[2]{(#1\ #2)}

\newcommand\NEW[0]{\reflectbox{\ensuremath{\mathsf{N}}}}
\newcommand\nw[1]{#1^{\scalebox{.4}{\NEW}}}

\newcommand\tuple[1]{{{(}#1{)}}}

\newcommand\aleq{\mathrel{{\approx}_{\scriptstyle {\alpha}}}}

\newcommand\act[0]{{\cdot}}

\newcommand\To{\Rightarrow} 

\newcommand\abs[1]{[#1]}

\newcommand\sm{{\mapsto}}

\newcommand\ds{ds}

\newcommand\rulefont[1]{\ensuremath{\mathrm{\scriptstyle \bf (#1)}}}

\newcommand\vr{\mathtt{var}}

\def\Id{\mathtt{Id}}

\newcommand\cent[0]{\vdash}

\def\eq{=}

\newcommand\ra{\rightarrow}
\newcommand{\lCRS}{\mbox{\sf lam}}
\newcommand{\App}{\mbox{\sf app}}
\newcommand{\Diff}{\mbox{\sf diff}}
\newcommand{\Sin}{\mbox{\sf sin}}
\newcommand{\Cos}{\mbox{\sf cos}}
\newcommand{\Mult}{\mbox{\sf mult}}

\newcommand{\CLeft}{\mathit{Left}}
\newcommand{\CRight}{\mathit{Right}}

\theoremstyle{plain}\newtheorem{property}[thm]{Property}

\newcommand{\nf}{nf_\sigma}

\newcommand     \comp[3]        {#1^{#2}_{\Phi_{#3}}}    
\newcommand     \compil[2]      {\lbrack\!\lbrack #1 \rbrack\!\rbrack_{\Phi_{#2}}}
\newcommand\new[0]{\reflectbox{\ensuremath{\mathsf{N}}}} 
 
\begin{document}
\title[From Nominal to Higher-Order Rewriting and Back Again]
      {From Nominal to Higher-Order Rewriting and Back Again}

\author[J.~Dom\'{i}nguez]{Jes\'{u}s Dom\'{i}nguez}
\address{Department of Informatics, King's College London, Strand WC2R 2LS, UK}
\email{\{jesus.dominguez\_alvarez, maribel.fernandez\}@kcl.ac.uk} 
\author[M.~Fern\'{a}ndez]{Maribel Fern\'{a}ndez}
\address{\vspace{-18 pt}}
%
%

\begin{abstract}
We present a translation function from nominal rewriting systems
(NRSs) to combinatory reduction systems (CRSs), transforming closed
nominal rules and ground nominal terms to CRSs rules and terms,
respectively, while preserving the rewriting relation.  We also
provide a reduction-preserving translation in the other direction,
from CRSs to NRSs, improving over a previously defined translation.
These tools, together with existing translations between CRSs and
other higher-order rewriting formalisms, open up the path for a
transfer of results between higher-order and nominal rewriting. In
particular, techniques and properties of the rewriting relation, such
as termination, can be exported from one formalism to the other.
\end{abstract}

\maketitle
\section{Introduction}\label{introduction}
Programs and logical systems often include binding operators.
Term \mbox{rewriting} systems~\cite{Baader1988,Dershowitz2003}, in their standard form,
do not provide support for reasoning on binding
structures. This motivated the study of combinations of first-order rewriting
systems with the $\lambda$-calculus~\cite{Barendregt1984}, which 
offers a notion of variable binding and substitution.  
Combinatory reduction systems (CRSs)~\cite{Klop1980,Klop1993} 
are well-known examples of \emph{higher-order rewriting} formalisms, where
a meta-language based on the untyped $\lambda$-calculus was incorporated 
to a first-order rewriting framework.
Other approaches followed, such as HRSs~\cite{Nipkow1991} and
 ERSs~\cite{Khasidashvili1990,Glauert2005} for example.

 Techniques to prove confluence and termination of higher-order
 rewriting systems were studied in~\cite{Mayr1998,Klop1993,Hamana2010}
 amongst others. However, the syntax and type restrictions imposed on
 rules in these systems have prevented the design of completion
 procedures for higher-order rewriting systems~\cite{Nipkow1998}.

More recently, the nominal approach~\cite{Gabbay2002,Pitts2003} has
been used to design rewriting systems with support for
binding~\cite{Fernandez2007}.  Nominal rewriting systems do not rely
on the $\lambda$-calculus, instead, two kinds of variables are used:
\textit{atoms}, which can be abstracted but behave similarly to
constants thus allowing explicit manipulation, and meta-level
variables or just \textit{variables}, which are first-order in that
they cannot be abstracted and substitution does not avoid capture of
unabstracted atoms.  On nominal
terms~\cite{Urban2003,Urban2004,Fernandez2007} $\alpha$-equivalence is
axiomatised using bijective mappings on atoms, known as
\textit{permutations}, and a \textit{freshness relation} between atoms
and terms. Nominal syntax enjoys many useful properties, for instance,
unification modulo $\alpha$-equivalence is decidable and
unitary~\cite{Urban2003,Urban2004} and nominal matching is
linear~\cite{Calves2009}. Nominal rewriting can be implemented
efficiently if rules are \textit{closed} (roughly speaking, closed
rules do not contain free atoms, and preserve the free/abstracted
status of atoms during reduction --- a natural restriction, which is
also imposed on CRSs, HRSs and ERSs by definition).

The availability of efficient algorithms to solve unification problems
on nominal terms motivated the study of the relationship between
higher-order and nominal syntax in a series of
papers~\cite{Cheney2005,Fernandez2004,Levy2012}.  In this paper, we
focus on the relationship between  higher-order and nominal
\emph{rewriting}, specifically between  CRSs and NRSs.  The
translations provided in \cite{Cheney2005,Levy2012} preserve the
unifiability relation, whereas ours preserves the rewriting relation, which is
key to the translation of properties such as confluence and
termination.  We define a translation function from closed NRS rules
and ground nominal terms to CRS rules and terms, preserving the
rewriting relation.  Then, we give a translation function from CRSs to
NRSs, improving over a previous translation described
in~\cite{Fernandez2004}.  Since we now have reduction-preserving
translations in both directions, properties and techniques developed
for one formalism can be exported to the other (e.g., termination
techniques based on the construction of a well-founded reduction
ordering). A Haskell implementation of the translation functions,
along with a tool to prove termination using the nominal recursive
path ordering~\cite{Fernandez2012}, are available
from~\cite{Dominguez2014d,Dominguez2014c}.

\paragraph*{Related work.}
CRSs, HRSs and ERSs are well-known examples of higher-order rewriting formalisms.
 A comparison of various higher-order rewriting formalisms, with many
 interesting examples, is provided in~\cite{Raamsdonk1999}; see
 also~\cite{Jouannaud2005} for a concise presentation of higher-order
 rewrite systems.  In \cite{Oostrom1994,Glauert2005}, CRSs are compared
 with HRSs and ERSs respectively, and in~\cite{Bertolissi2006} CRSs
 are expressed in terms of the
 $\rho$-calculus~\cite{Cirstea2001,Cirstea2001a}.  
In \cite{Kop2011}, a termination-preserving translation between Algebraic Functional
   Systems and other higher-order formalisms is presented.  Although
 in this paper we focus on the relationship between NRSs and CRSs,
 thanks to the existing translations between CRSs and other
 higher-order rewriting formalisms, this is sufficient to obtain a
 bridge between nominal and higher-order rewriting.

Our work is closely related to the work reported
in~\cite{Cheney2005,Levy2012}: Cheney~\cite{Cheney2005} represented
higher-order unification as nominal unification, and Levy and
Villaret~\cite{Levy2012} transformed nominal unification into
higher-order unification, providing a translation that preserves
unifiers. Our translation differs from~\cite{Cheney2005,Levy2012}
in that our requirement is to have a mapping of NRS  ground terms and rules to
CRS terms and rules in such a way that reductions are preserved.

This paper is an updated and extended version
of~\cite{Dominguez2014b}.  We have included here, in addition to the
translation from NRSs to CRSs given in~\cite{Dominguez2014b}, all the
proofs previously omitted due to space constraints as well as a
translation from CRSs to NRSs, improving on a previous result given
in~\cite{Fernandez2004}.  We provide detailed explanations, and illustrate the
translations with examples.
 
\paragraph*{Overview of the paper.} 
The rest of the paper is organised as follows.  In section
\ref{sec:mwr} we recall both formalisms, CRSs and NRSs, as defined
in \cite{Klop1980} and \cite{Fernandez2007} respectively.  In section
\ref{sec:formalCRS} we describe in detail the translation of nominal
terms to CRS meta-terms, while in section \ref{sec:rules} we extend
it to take into account rules and substitution.  In Section
\ref{sec:redRel} we prove that nominal rewrite steps can be simulated
in CRSs via the translation function.  Section \ref{sec:toNRS}
presents a translation from CRSs to NRSs.  In section
\ref{sec:examples} we show examples of application of the
translations.  Section \ref{sec:conclusions} concludes and discusses
future work.

\section{Preliminaries}
\label{sec:mwr}
We start by briefly recalling  the main concepts 
of nominal rewrite systems and combinatory  reduction 
systems --- two rewriting formalisms
that extend the syntax of first-order terms and 
the notion of rewriting, to facilitate the
specification of systems with binding operators. 
We refer the reader
to~\cite{Klop1993,Fernandez2007}
for more details and examples.

\subsection{Nominal Rewriting}\label{sec:NomRew}

A \emph{nominal signature} $\Sigma$ is a set of term-formers, or 
\emph{function symbols}, $f,g,\ldots$, each with a fixed
arity.  Fix a countably infinite set ${\mathcal X}$
 of \emph{variables} ranged over by $X,Y,Z, \ldots$, and a countably infinite
set ${\mathcal A}$ of \emph{atoms} ranged over by $a,b,c,\ldots$, and assume that
$\Sigma$, ${\mathcal X}$ and ${\mathcal A}$ are pairwise disjoint.
We follow the \emph{permutative convention}~\cite[Convention~2.3]{Gabbay2008} 
 where $a,b,c,\ldots$
range over \emph{distinct} atoms unless stated otherwise.

\textit{Permutations} $\pi$ are bijections on ${\mathcal A}$ such that
the set of atoms for which $a \neq \pi(a)$ is finite; this is called
the \emph{support} of $\pi$, written as $\suppi$.  
A \emph{swapping} is a pair of atoms, written $(a\ b)$.
Permutations are represented by lists of
swappings, $\Id$ denotes the \textit{identity permutation}.  We write
$\pi^{-1}$ for the inverse of $\pi$ and $\pi \circ \pi'$ for the
composition of $\pi'$ and $\pi$.  For example, if
$\pi=\Tran{a}{b}\Tran{b}{c}$, then $\pi(a)\eq b$,
 $\pi^{-1}=\Tran{b}{c}\Tran{a}{b}$ and $\pi^{-1}(a)\eq c$.

\begin{defi}[Syntax]
\emph{Nominal terms}, or just \emph{terms} if there is no ambiguity,
are generated by the grammar
\[s,t  ::=   a \mid \pi\act X \mid   \abs{a}s \mid f s \mid (s_1,\ldots,s_n)\]
and called, respectively, atoms, moderated variables or simply
variables, abstractions, function applications (which must respect the arity of
the function symbol) and tuples; if the arity of $f$ is $0$ we may omit the 
parentheses in the application.  We abbreviate $\Id\act X$ as $X$
if there is no ambiguity.  An abstraction $\abs{a}t$ is intended to
represent $t$ with $a$ bound; we say that the scope of $\abs{a}$ is $t$.
Call occurrences of $a$
\textit{abstracted} if they are in the scope of an abstraction, and
\textit{unabstracted} (or free) otherwise.  
\end{defi}
For example, $f(X,\Tran{a}{b}\act X)$ is a
nominal term, and so is $f([a]X,[b]b)$.
The latter term has $X$ in the scope of $\abs{a}$ and
$b$ in the scope of $\abs{b}$.
For more examples, 
we refer the reader to~\cite{Urban2003,Urban2004,Fernandez2007}.

\begin{defi}\label{def:Atms&VarsOcc}
The functions $\Vars(t)$ and $\Atoms(t)$ are used to compute the sets
of variables and atoms  in a nominal term $t$, respectively.
They are inductively defined as follows:
\[
\begin{array}{c}
\Vars(a)\eq\emptyset \qquad \Vars(\abs{a}t)\eq\Vars(t)\qquad \Vars(\piX)\eq\{X\}\\[1ex]
\Vars(fs)\eq\Vars(s)\qquad\Vars(\tuple{s_1,\ldots,s_n})\eq\Vars(s_1)\cup\ldots\cup\Vars(s_n)\\[1ex]
\\
\Atoms(a)\eq\{a\} \qquad \Atoms(\abs{a}t)\eq\Atoms(t)\cup\{a\} \qquad \Atoms(\piX)\eq\suppi\\[1ex]
\Atoms(fs)\eq\Atoms(s)\qquad\Atoms(\tuple{s_1,\ldots,s_n})\eq\Atoms(s_1)\cup\ldots\cup\Atoms(s_n)
\end{array}
\]
\textit{Ground terms} have no variables: $\Vars(t)= \emptyset$.
\end{defi}
Notice that $\Vars(t)$ is a syntactic notion, whereas $\Atoms(t)$ takes into account 
the semantics of permutations (represented as lists of swappings). For example,
 $\Atoms (f(a,\, [b]\Tran{c}{d}\Tran{e}{f}\Tran{f}{e}\act X))$
$=\{a,b,c,d\}$.

\begin{defi}[Positions and subterms of nominal terms]
\label{def:Pos(s)}
Let $s$ be a nominal term. The set $\mathcal{P}os(s)$ of \emph{positions} in $s$ 
is a set  of strings of positive integers, inductively defined below. We also define below
the subterms of $s$:  $s|_p$ denotes the subterm of $s$ at position $p$.
\begin{itemize}
\item if $s\eq a$ or $s\eq\pi\act X$, then $\mathcal{P}os(s)=\{\epsilon\}$ and $s|_\epsilon = s$, where $\epsilon$ denotes the empty string;
\item if $s\eq\abs{a}t$, then $\mathcal{P}os(s)=\{\epsilon\}\cup\{1\cdot p\mid p\in\mathcal{P}os(t)\}$, $s|_\epsilon = s$ and $s|_{1\cdot p} = t|_p$;
\item if $s\eq f t$, then $\mathcal{P}os(s)=\{\epsilon\}\cup\{1\cdot p\mid p\in\mathcal{P}os(t)\}$, $s|_\epsilon = s$ and $s|_{1\cdot p} = t|_p$;
\item if $s\eq\tuple{t_1,\ldots,t_n}$, then $\mathcal{P}os(s)=\{\epsilon\}\cup{\overset{n}{\underset{i\eq 1}{\bigcup}}}\{i\cdot p\mid p\in\mathcal{P}os(t_i)\}$, $s|_\epsilon = s$ and $s|_{i\cdot p} = t_i|_p$.
\end{itemize}
The position $\epsilon$ is called the \emph{root position} of the term $s$, and the symbol at this position is called the \emph{root symbol} of $s$.
\end{defi}

We now extend the action of permutations to terms. Recall we use the permutative convention, so atoms $a,b,c$ are considered distinct among them.

\begin{defi}[Permutation action] 
\label{def:permAct}
The \emph{action of a permutation} $\pi$ \emph{on a term} $t$, written
$\pi\act t$, is defined by induction: $\Id \act t = t$ and
$\Tran{a}{b}\pi\act t = \Tran{a}{b}\act (\pi\act t)$, where a swapping
acts inductively on the structure of terms as follows:
\[
\begin{array}{c}
\Tran{a}{b}\act a = b \quad \Tran{a}{b}\act b = a \quad \Tran{a}{b}\act c =c
\\[1ex]
\Tran{a}{b}\act (\pi \act X) = (\Tran{a}{b} \circ \pi) \act X  \qquad
\Tran{a}{b}\act \abs{c}t=\abs{ c}\Tran{a}{b}\act t \\[1ex]
\Tran{a}{b}\act \abs{a}t=\abs{b}\Tran{a}{b}\act t \qquad
\Tran{a}{b}\act \abs{b}t=\abs{a}\Tran{a}{b}\act t \\[1ex]
\Tran{a}{b}\act f t= f \Tran{a}{b}\act t\qquad
\Tran{a}{b}\act (t_1,\ldots,t_n) = (\Tran{a}{b}\act t_1,\ldots,\Tran{a}{b}\act t_n).
\end{array}
\]
\end{defi}\bigskip

\noindent Substitutions map variables to terms, and act on terms 
without avoiding capture of atoms, according to the following definition.

\begin{defi}[Substitution]
\textit{Substitutions} are generated by the grammar: 
\[\sigma::=\Id\mid [X\sm s]\sigma\]
We use the same notation for the identity substitution and permutation, and also for
composition,  since there will be no ambiguity.  

Write $t\sigma$ for the application of $\sigma$ on $t$, defined as follows: 
$$t\Id = t \quad t[X\sm s]\sigma = (t[X\sm s])\sigma \quad where$$ 
\[\begin{array}{c}
a[X\sm s] =  a \qquad (\pi\act X)[X\sm s]= \pi\act s \qquad (\pi\act Y)[X\sm s]= \pi\act Y ~~(X \neq Y)\\[1ex]
(\abs{a}t)[X\sm s] = \abs{a}(t[X\sm s])
\qquad
(f t)[X\sm s] = f t[X\sm s]\\[1ex]
(t_1,\ldots,t_n)[X\sm s] = (t_1[X\sm s],\ldots,t_n[X\sm s]).
\end{array}
\]

The \emph{domain} of a substitution $\sigma$, $dom( \sigma)$, is the set
of variables such that $X\sigma\not\eq X$.
The restriction of a substitution $\sigma$ to a set of variables $V$,
written $\sigma|_{V}$, is defined as $\sigma|_{V}\eq [X\ra X\sigma\mid X\in V]$.
\end{defi}

The semantics of nominal terms is defined using nominal sets~\cite{PittsA:nomsetbook}. A 
$Perm(\mathcal{A})$-set is a set $T$ equipped with a permutation action, such that 
$\Id\act x = x$ and $\pi\act(\pi'\act x) = (\pi\circ\pi')\act x$ for each object $x \in T$.
A set $S$ of atoms  \emph{supports} $x \in T$ if for all atoms $a,b\not\in S$, 
$\Tran{a}{b}\act x = x$. A \emph{nominal set} is a $Perm(\mathcal{A})$-set where
 each element has finite support. Nominal terms 
form a nominal set, using $\alpha$-equivalence as equality~\cite{PittsA:nomsetbook}.  
To define the support of a term, we introduce the notion of
\emph{freshness}. The support set of a term $t$, abbreviated
$supp(t)$, is the complement of the set of fresh atoms in $t$.  When a
term $t$ is ground, $supp(t)$ coincides with the syntactic notion of
unabstracted atoms in $t$.

\begin{defi}[Freshness]
\label{def.constraints}
A \emph{freshness}  \emph{constraint} is a pair $a\# t$
 of an atom and a term. 
A \emph{freshness context} (ranged over by $\Delta,\nabla,\Gamma$), is a set 
of constraints of the form $a\# X$.  \emph{Freshness judgements},
written $\Delta \cent a\#t$,
are derived using the rules below.
\[
 \begin{prooftree} 
\justifies \Delta \cent a\apart b
 \using \rulefont{\#ab}
\end{prooftree}
\quad
 \begin{prooftree} 
\pi^{\text{-}1}\act a\apart X \in \Delta
\justifies  \Delta \cent a\apart \pi\act X 
 \using\rulefont{\#X}
\end{prooftree}
 \quad
 \begin{prooftree} 
 \Delta \cent a\apart s
 \justifies \Delta \cent a\apart f s
 \using\rulefont{\#f}
\end{prooftree}\]\\
\[\begin{prooftree} 
\Delta \cent a\apart s_1\: \cdots \:  \Delta \cent a\apart s_n 
 \justifies \Delta \cent a\apart (s_1, \ldots, s_n)
 \using\rulefont{\#tupl}
 \end{prooftree}
\quad
\begin{prooftree} 
\justifies  \Delta \cent  a\apart \abs{a}s
 \using\rulefont{\#[a]}
 \end{prooftree}
 \quad
 \begin{prooftree}  
\Delta \cent a\apart s
\justifies  \Delta \cent a\apart \abs{b}s
 \using\rulefont{\#[b]}
 \end{prooftree}
 \]
\end{defi}\medskip

\noindent For example, $a \# X \cent b \# \Tran{a}{b}\act X$ can be derived using rule 
$\rulefont{\#X}$, since $\Tran{a}{b}\act a = b$.

In nominal languages, one is interested in terms $t$ that have finite support,
because for them there exists always a fresh atom $a$ such that $a\apart t$
(recall the set $\mathcal{A}$ of atoms  is infinite).

\begin{defi}[$\alpha$-equivalence]
An \emph{$\alpha$-equality}  \emph{constraint} is a pair $s\aleq t$
of terms. 
\emph{Equivalence judgements},
written $\Delta\cent s\aleq t$,
are derived using the rules below,
where $\ds(\pi,\pi') =\{a\in \mathcal{A} \mid \pi \act a\neq\pi' \act a\}$ (difference set).
\[
\begin{prooftree} 
\justifies \Delta \cent a\aleq a 
\using\rulefont{{\aleq}a} 
\end{prooftree}
\qquad
\begin{prooftree} 
\forall a \in \ds(\pi,\pi'): a\# X \in \Delta
\justifies \Delta \cent \pi{\act} X\aleq \pi'{\act} X 
\using\rulefont{{\aleq}X}
\end{prooftree}
\qquad
\begin{prooftree} 
\Delta \cent s\aleq t
\justifies
\Delta \cent f s\aleq f t
\using\rulefont{{\aleq}f}
\end{prooftree}
\]
\[
\begin{prooftree} 
\Delta \cent s_1\aleq t_1\:\cdots\:\Delta \cent s_n\aleq t_n
\justifies
\Delta \cent (s_1,\ldots,s_n)\aleq (t_1,\ldots,t_n)
\using\rulefont{{\aleq}tupl}
\end{prooftree}
\]
\[
\begin{prooftree} 
\Delta \cent s\aleq t
\justifies 
\Delta \cent \abs{a}s\aleq \abs{a}t 
\using\rulefont{{\aleq}[a]}
\end{prooftree}
\qquad
\begin{prooftree} 
\Delta \cent \Tran{b}{a}{\act} s\aleq t  \quad  
\Delta \cent b \# s
\justifies 
\Delta \cent \abs{a}s\aleq \abs{b}t 
\using\rulefont{{\aleq}[b]}
\end{prooftree}
\]

Let $P_i$ be a freshness or $\alpha$-equality constraint (for $1 \leq i \leq n$). 
We write $\Delta \cent P_1, \ldots, P_n$ when proofs of $\Delta \cent P_i$ 
exist (for $1 \leq i\leq n$), using the derivation rules above.
\end{defi}

The relation $\aleq$ is indeed an equivalence relation~\cite{Urban2003,Urban2004}.
\begin{exa}
We can derive $a\apart X \cent \abs{a}\Tran{a}{b}\act X\aleq\abs{b}X$ as follows.
\[\begin{prooftree}
\begin{prooftree}
ds(\Tran{b}{a}\Tran{a}{b},\Id)=\emptyset
\justifies
a\apart X \cent \Tran{b}{a}\Tran{a}{b}\act X\aleq X
\using\rulefont{\aleq X}
\end{prooftree}
\quad
\begin{prooftree}
\justifies
a\apart X \cent b\apart\Tran{a}{b}\act X
\using\rulefont{\apart X}
\end{prooftree}
\justifies
a \# X \cent \abs{a}\Tran{a}{b}\act X\aleq\abs{b}X
\using\rulefont{\aleq\abs{b}}
\end{prooftree}
\]
\end{exa}

\begin{property}[\cite{Fernandez2007}, Lemma 23]
\label{prop:support}
For any $a \in \mathcal{A}$, if $\Delta \cent a \# s$ and $\Delta \cent s \aleq t$ then 
$\Delta \cent a \# t$.  Hence, if $\Delta \cent s \aleq t$ then $s$ and $t$ have the same support set.
\end{property}

\begin{defi}[Nominal rewrite system]
A \emph{nominal rewrite rule} $R= \nabla\cent l\to r$ is a tuple of 
 a freshness 
context $\nabla$ and terms $l$ and $r$ such that $\Vars(r)\cup\Vars(\nabla) \subseteq \Vars(l)$.  

A \emph{nominal rewrite system} (NRS) 
is an equivariant set $\mathcal{R}$ of nominal rewrite rules, 
that is, a set of nominal rules that is closed under permutations. 
We shall generally equate a  set of rewrite rules with its
equivariant closure.
\end{defi}

\begin{defi}
We extend the notions given in 
Definition~\ref{def:Atms&VarsOcc} for both variables, $\Vars(t)$, and 
atoms, $\Atoms(t)$,  to include rules, contexts, substitutions, etc.
Particularly,
for contexts
$\Atoms(\Delta)\eq\{a\mid a\apart X\in\Delta\mbox{ for some } X\}$
and for substitutions,
$\Atoms(\sigma)\eq\{\Atoms(X\sigma)\mid X\in dom(\sigma)\}$.
\end{defi}

\begin{exa}

\label{ex:fol}
The following rules are used to compute prenex normal forms in first-order
logic. The signature has term-formers $\mbox{\sf forall,~ exists,~ not,~ and}$.
Intuitively, equivariance means that the choice of atoms in rules is not important
(see~\cite{Fernandez2007} for more details), therefore we could change below $a$ to $b$, 
for instance.
\[\begin{array}{lcll}
a\# P &\cent & \mbox{\sf and}(P, \mbox{\sf forall}([a]Q)) &\ra \mbox{\sf forall}( [a] \mbox{\sf and}(P, Q))\\
a\# P &\cent & \mbox{\sf and}(\mbox{\sf forall}( [a]Q) , P) &\ra \mbox{\sf forall}( [a] \mbox{\sf and}(Q, P))
\\
a\# P &\cent & \mbox{\sf or}(P , \mbox{\sf forall}( [a]Q)) &\ra \mbox{\sf forall}( [a] \mbox{\sf or}(P, Q))\\
a\# P &\cent & \mbox{\sf or}(\mbox{\sf forall}( [a]Q) , P) &\ra \mbox{\sf forall}( [a] \mbox{\sf or}(Q, P))\\

a\# P &\cent & \mbox{\sf and}(P, \mbox{\sf exists}( [a]Q)) &\ra \mbox{\sf exists}( [a] \mbox{\sf and}(P, Q))\\
a\# P &\cent & \mbox{\sf and}(\mbox{\sf exists}([a]Q),  P) &\ra \mbox{\sf exists}([a] \mbox{\sf and}(Q, P))
\\
a\# P &\cent & \mbox{\sf or}(P,  \mbox{\sf exists}( [a]Q) &\ra \mbox{\sf exists} ([a] \mbox{\sf or}(P, Q))\\
a\# P &\cent & \mbox{\sf or}(\mbox{\sf exists}( [a]Q), P) &\ra \mbox{\sf exists} [a] \mbox{\sf or}(Q, P)
\\
&\cent & \mbox{\sf not}(\mbox{\sf exists}( [a]Q)) &\ra \mbox{\sf forall}( [a] \mbox{\sf not}(Q))\\
&\cent & \mbox{\sf not}(\mbox{\sf forall}( [a]Q)) &\ra \mbox{\sf exists}( [a] \mbox{\sf not}( Q)).
\end{array}
\]
\end{exa}\medskip

\noindent Nominal rewriting~\cite{Fernandez2007} operates 
on \emph{terms-in-contexts},  written $\Delta\cent s$ or just $s$ if
 $\Delta=\emptyset$.  
Below,  $C[~]$ varies over terms with exactly one
occurrence of a distinguished variable
$\Id\act \text{-}$, or
just $ \text{-}$.  We write $C[s]$ for $C[\text{-}\sm
  s]$, and  $\Delta \cent \nabla\theta$ for
 $\{\Delta \vdash a \# X\theta \mid a \# X \in \nabla\}$.

\begin{defi}[Nominal rewriting]
\label{rewrite-step}
A term \emph{$s$ rewrites with $R=\nabla\cent l\to r$ to $t$ in
  $\Delta$}, written $\Delta\cent s\ra_{R} t$ (as usual, we assume $V(R)\cap
(V(\Delta)\cup V(s))=\emptyset$), 
if $s= C[s']$ 
and there exists $\theta$ such that
$\Delta \cent \nabla\theta$,   $\Delta \cent l\theta \aleq s'$, and 
$\Delta\cent C[r\theta]\aleq t$. 
Since $\Delta$ does not change during rewriting, a rewriting derivation
is written $\Delta \cent   s_1\ra_{R} s_2 \ra_R \ldots \ra_R s_n$, abbreviated as 
$\Delta \cent s_1 \to^* s_n$. 
\end{defi}

When rules are closed, nominal rewriting can be efficiently implemented
using nominal matching (then, there is no need to consider equivariance). We
define closed rewriting below, after defining closed terms.

Closed terms are, roughly speaking, terms without unabstracted atoms,
 such that variables behave uniformly with respect to their support.
We give a definition below.

\begin{defi}[Closedness]
\label{def:closed}
A term-in-context $\Delta\vdash t$ is closed if it satisfies the following conditions:
\begin{enumerate}
\item if  $t|_p = a$ then $t|_p$ is in the scope of an abstraction for $a$;
\item if $\pi\cdot X$ occurs in the scope of an abstraction of $\pi\cdot a$ then 
any occurrence of $\pi'\cdot X$ occurs in the scope of an abstraction of $\pi'\cdot a$ or
 $a\#X \in \Delta$;
\item for any pair $\pi_1\act X, \pi_2\act X$ occurring in $t$, and 
$a \in ds(\pi_1,\pi_2)$, if $a$ is not abstracted in one of the occurrences 
then $a\#X \in \Delta$.
\end{enumerate}
A rewrite rule $\nabla\cent l\to r$ is closed if $\nabla\cent (l,r)$ is a closed term.
\end{defi}

The first condition in the definition specifies that no atom occurs
unabstracted in a closed term. The second condition states that if an atom $a$
in an instance of a variable $\pi\act X$ is captured (i.e. $\pi\act X$
is under an abstraction for $\pi\act a$) then it is captured in all
occurrences of $X$, otherwise it is fresh for $X$. The third condition says
that if two occurrences of $X$ have different suspended permutations,
then any atom in the difference set that could occur in an instance of
$X$ is captured.

For example, $[a]f(X,a)$ is closed, but $f(X,a)$ and $f(X,[a]X)$ are
not, however $a \# X \cent f(X,[a]X)$ is closed. All the rewrite rules 
in Example~\ref{ex:fol} are closed.
 
Closedness can be easily checked using the nominal matching
algorithm~\cite{Calves2009} as follows.  First, given a term in
context $\nabla \cent t$, or more generally, a pair $P = \nabla\cent
(l,r)$ (this could be a rule $R= \nabla\cent l\to r$), let us write
$\nw{P}= \nw{\nabla}\cent (\nw{l}, \nw{r})$ to denote a
\emph{freshened variant} of $P$, i.e., a version where the atoms and
variables have been replaced by `fresh' ones.  We shall always
explicitly say what $\nw{P}$ is freshened for when this is not
obvious.  For example, a freshened version of $(a\apart X\cent f(X)\to
X)$ with respect to itself and to $a'\apart X\cent a'$ is $(a''\apart
X'\cent f(X')\to X')$.  We will write $A(P')\apart V(P)$ to mean that all
atoms in $P'$ are fresh for each of the variables occurring
in $P$.  Let $ \nw{\nabla}\cent \nw{t}$ be a freshened version of
$\nabla \cent t$. Then $\nabla \cent t$ is \emph{closed} if there
exists a substitution $\sigma$ such that $\nabla,A(\nw{\nabla}\cent
\nw{t})\apart V(\nabla \cent t)\cent \nw{\nabla}\sigma$ and
$\nabla,A(\nw{\nabla}\cent \nw{t})\apart V(\nabla \cent t)\cent
\nw{t}\sigma \aleq t$.  A similar check can be done for nominal
rewrite rules.

\begin{defi}[Closed rewriting]
\label{def:closedR}
Let $\nw{R}$ be a freshened version of the rule $R$ with respect to
$\Delta$, $s$, $t$ (i.e., a version where the atoms and
variables in $R$ have been replaced by fresh ones;
as shown in~\cite{Fernandez2007}, it does not
matter which particular freshened $\nw{R}$ we choose). 
We write
$\Delta\cent s\ra^c_R t$ if $\Delta,\Delta'\cent
s\ra_{\nw{R}}t$, where $\Delta' = A(\nw{R})\apart V(\Delta,s)$,
and call this a \emph{closed rewriting step}. 
The subindex $R$ may be omitted if it is clear from the context.
\end{defi}
 
Closed NRSs inherit properties of first-order
rewriting systems such as the Critical Pair Lemma~\cite{Fernandez2007}.

\begin{exa}\label{ex:NomStep}
We show a  closed rewriting step for the term 
$\cent \mbox{\sf and} (X, \mbox{\sf forall}
 ([b]\mbox{\sf  f}(b)))$ using the first
rule in Example~\ref{ex:fol}:
$$\cent \mbox{\sf and} (X, \mbox{\sf forall}
 ([b]\mbox{\sf  f}(b)))\to^c \mbox{\sf forall}([a'] 
 \mbox{\sf and}(X, \mbox{\sf f}(a')))$$ 
To generate it, we first obtain a freshened variant of the rule
 with respect to itself and the given term: 
 $a'\# P' \cent  \mbox{\sf and}(P', \mbox{\sf forall}([a']Q')) 
\ra \mbox{\sf forall}( [a'] \mbox{\sf and}(P', Q'))$. Notice that 
there is a rewrite step 
$$a'\# X \cent \mbox{\sf and} (X, \mbox{\sf forall}
 ([b]\mbox{\sf  f}(b)))\to \mbox{\sf forall}([a'] 
 \mbox{\sf and}(X, \mbox{\sf f}(a'))).$$
using the matching substitution 
 $\theta= [P'\mapsto X][Q'\mapsto f(a')]$, since $a'\# P'\theta$ holds.
\end{exa}

\subsection{Combinatory Reduction Systems}
\label{sec:CRS}
 
A combinatory reduction system (CRS)~\cite{Klop1980,Klop1993} is a pair consisting of an
alphabet $\mathbb{A}$ and a set of rewrite rules. 

 The CRS \emph{alphabet} $\mathbb{A}$ consists of:
 \begin{enumerate}
 \item  a countably infinite set $\mathcal{V}$ of variables ranged over by $a,b,c, \ldots$; 
 \item a countably infinite set $\mathcal{MV}$ of meta-variables
 with fixed arities, written as $Z^n_i$ where $n$ is the arity of
 $Z^n_i$ (when there is no ambiguity, $n$ is omitted);
  \item  an abstraction operator $[\cdot]\cdot$;
 \item function symbols $f,g,\ldots$ with fixed arities; and
 \item improper symbols `(\textquoteright, `)\textquoteright and `,\textquoteright.
\end{enumerate} 
\begin{defi}[Syntax] 
\label{def:CRS-syntax}
CRS \emph{meta-terms} are generated by the grammar
\[
s,t ::= a\mid Z^n_i t\mid [a]t\mid f t \mid \tuple{t_1,\ldots,t_n}~~~ (n \geq 0)
\]
Only variables can be abstracted;  
in a \emph{function application} $f t$ (resp.~\emph{meta-application} 
$Z_i^n t$), $t$ is a $n$-tuple respecting the arity
of the function symbol $f$  (resp.~meta-variable $Z_i^n$); 
when the  arity is 0, we omit the brackets in applications and
meta-applications (so $Z^0_i$ is a meta-term). 
\end{defi}

\begin{defi}
 We write $MV(t)$ and $Var(t)$ for the set of meta-variables
 and variables occurring in a meta-term $t$, respectively (the same notation 
is used for rules, etc.).
 They are inductively defined as follows:
 \[
\begin{array}{ll}
MV(a)\eq\ \emptyset &  
MV(\metaVar{Z} t)\eq MV(t)\cup\{Z^n_i\} \qquad
MV(f t )\eq MV(t)\\
[1ex]
MV(\abs{a}t)\eq MV(t) &
MV(\tuple{t_1,\ldots,t_n})\eq MV(t_1)\cup\cdots\cup MV(t_n)\\

\\
Var(a)\eq\ a &  
Var(\metaVar{Z} t )\eq Var(t)\qquad\quad\qquad
Var(f t )\eq Var(t)\\
[1ex]
Var(\abs{a}t)\eq Var(t)\cup\{a\} &
Var(\tuple{t_1,\ldots,t_n})\eq Var(t_1)\cup\cdots\cup Var(t_n)\\
\end{array}
\]
\end{defi}\medskip

\noindent In CRSs a distinction is made between \emph{meta-terms} and
\emph{terms}. Meta-terms are the expressions built from the symbols in
the alphabet, in the usual way (see Definition~\ref{def:CRS-syntax}).
Variables that occur in the scope of the abstraction operator are
\emph{bound}, and \emph{free} otherwise. Meta-terms are defined modulo
renaming of bound variables, that is, a meta-term represents an
$\alpha$-equivalence class.  \emph{Terms} are meta-terms that do not
contain meta-variables, and are also defined modulo
$\alpha$-equivalence. A (meta-)term is closed if every variable
occurrence is bound.

\begin{defi}
\label{def:Pos(s)CRS}
Let $s$ be a CRS meta-term. 
The set $\mathcal{P}os(s)$ of \emph{positions} in $s$ is a set  of strings 
of positive integers, which is
inductively defined as follows:
\begin{itemize}
\item if $s\eq a$, then $\mathcal{P}os(s)=\{\epsilon\}$,
where $\epsilon$ denotes the empty string;
\item if $s\eq\metaVar{Z} t$ then $\mathcal{P}os(s)=\{\epsilon\}\cup\{1\cdot p\mid p\in\mathcal{P}os(t)\}$;
\item if $s\eq\abs{a}t$, then $\mathcal{P}os(s)=\{\epsilon\}\cup\{1\cdot p\mid p\in\mathcal{P}os(t)\}$;
\item if $s\eq f t$, then $\mathcal{P}os(s)=\{\epsilon\}\cup\{1\cdot p\mid p\in\mathcal{P}os(t)\}$;
\item if $s\eq \tuple{t_1,\ldots,t_n}$, then $\mathcal{P}os(s)=\{\epsilon\}\cup{\overset{n}{\underset{i\eq 1}{\bigcup}}}\{ i\cdot p\mid p\in\mathcal{P}os(t_i)\}$.
\end{itemize}
The position $\epsilon$ is called the \emph{root position} of the term $s$, and the symbol at this position is called the \emph{root symbol} of $s$.
\end{defi}

\begin{defi}[CRS rewrite rules]
A \emph{rewrite rule} is a pair of meta-terms, written $l \Rightarrow r$, where
$l, r$ are closed, $l$ has the form $f(s_1,\ldots,s_n)$ where $n\geq 0$
(when $n=0$ we omit the parentheses), $MV(r)\subseteq MV(l)$, and $MV(l)$ 
occur only in the form
$Z^n_i(a_1,\ldots,a_n)$, where $a_1,\ldots,a_n$ are pairwise distinct
bound variables.
\end{defi}
 
\begin{exa}
\label{beta-CRS}
The $\beta$-reduction rule for the $\lambda$-calculus
is written: 
$$\App(\lCRS(\abs{a}Z(a)),Z') \Rightarrow Z(Z')$$
where 
$Z$ is a unary meta-variable and $Z'$ is 0-ary.
\end{exa}

The reduction relation is defined on terms.  To
extract from rules the actual rewrite relation, each meta-variable is
replaced by a special kind of $\lambda$-term, and in the obtained term
all $\beta$-redexes and the residuals of these $\beta$-redexes are
reduced (i.e., a complete development is performed).
Formally, the rewrite
relation is defined using \emph{substitutes} and \emph{valuations}. 
\begin{defi}[Substitute]
An $n$-ary substitute is an expression of the form $\underline{\lambda} (a_1 \ldots
a_n).s$, where $s$ is a term and $a_1,\ldots,a_n$ are different
variables bound in $\underline{\lambda} (a_1 \ldots
a_n).s$.
We use a meta-lambda $\underline{\lambda}$ to emphasise that this is part of
the meta-language.

An $n$-ary substitute $\underline{\lambda} (a_1 \ldots
a_n).s$ may be applied to a $n$-tuple
$(t_1,\ldots,t_n)$ of terms, resulting in the following simultaneous
substitution:
\[
(\underline{\lambda} (a_1 \ldots
a_n).s)(t_1,\ldots,t_n)\eq s[a_1 \mapsto t_1,\ldots,a_n \mapsto t_n]
\]
where we denote by $s[a\sm t]$ the capture-avoiding substitution of
 variable $a$ by term $t$ in the CRS term $s$.
\end{defi}

\begin{defi}[Valuation]
A valuation $\sigma$  assigns an $n$-ary substitute to each
$n$-ary meta-variable:
\[
\sigma(Z^n_i)\eq\underline{\lambda}(a_1,\ldots,a_n).s.
\]
It is extended to a mapping from meta-terms to
terms as follows:
\begin{enumerate}
\item First,  replace all meta-variables in the term  for 
their images in $\sigma$ as shown below.
\[
\begin{array}{lr}
\sigma(a)\eq a\mbox{ for } a\in\mathcal{V}\qquad\qquad\qquad
\sigma(\abs{a}t)\eq\abs{a}\sigma(t)&
\sigma(f t)\eq f \sigma(t)\\
[1ex]
\sigma(\tuple{t_1,\ldots,t_n})\eq\tuple{\sigma(t_1),\ldots,\sigma(t_n)}&\sigma(\metaVar{Z} t)\eq\sigma(Z^n_i)\sigma(t).
\end{array}
\]
\item Next, perform the developments of the $\beta$-redexes created. 
\end{enumerate}
\end{defi}

Valuations must satisfy some \emph{safety conditions}. Before stating the
conditions, we recall a standard naming convention used in CRSs, originally 
stated by Barendregt for the $\lambda$-calculus~\cite{Barendregt1984}.

\begin{rem}[Barendregt's variable convention]
\label{rem:Barendregt}
 CRSs adopt the following naming conventions: 
\begin{itemize}
\item
all bound variables are chosen to be different among them, that is, each binder
uses a distinct variable name;
\item
bound variables are also chosen to be different from free variables.
\end{itemize}
\end{rem}\medskip

\noindent In CRSs, rewriting is performed under the following conditions.
\begin{defi}[Safety conditions]
\label{def:safety}
   The CRS rule $l\Rightarrow r$ \emph{is safe for} the valuation 
$\sigma$ if free variables occurring
in substitute $\sigma(Z)$ are different from the bound variables in both $l,r$ for all
$Z\in dom(\sigma)$ and also, we say $\sigma$ is safe with respect to itself when there
are no two substitutes $\sigma(Z),\sigma(Z')$ where a free variable in $\sigma(Z)$
occurs bound in $\sigma(Z')$ or vice versa.
\end{defi}

In the rest of the paper we adopt, without loss of generality,
Barendregt's convention for CRSs and assume that all valuations are safe
with respect to themselves and the reduction rules to which they are
applied.

A \emph{context} is a term with an occurrence of a special symbol $[\:]$ called hole.
A rewrite step is now defined in the usual way.
\begin{defi}[Rewrite step]
Let $\sigma$ be a valuation and $C[\:]$ a context.
If $l \Rightarrow r$ is a rewrite rule, then $C[\sigma(l)] \Rightarrow
C[\sigma(r)]$ is a rewrite (or reduction) step.
\end{defi}

\begin{exa}
\label{ex:redCRS}
The following is a rewrite step using the
 $\beta$-rule given in  Example~\ref{beta-CRS}:
\begin{center}
$\App(\lCRS(\abs{a}f(a,a)),t) \Rightarrow_{\beta} f(t,t).$
\end{center}
To generate the reduction, a
 valuation $\sigma$ that maps $Z$ to $\underline{\lambda} (b).f(b,b)$ and $Z'$ to
 the term $t$ is applied to the rule. 
Then,
 $\sigma(\App(\lCRS(\abs{a}Z(a)),Z'))$ is the term
 $\App(\lCRS(\abs{a}f(a,a)),t)$ obtained by first replacing $Z$ and
 $Z'$ as indicated by $\sigma$ and then reducing the $\beta$-redex $(\underline{\lambda}
 (b).f(b,b))(a)$.  Also, $\sigma(Z(Z'))$ is the term $f(t,t)$ obtained by first
 replacing $Z$ and $Z'$, resulting in $(\underline{\lambda}( b).f(b,b))(t)$, and 
 then $\beta$-reducing to $f(b,b)\{b\sm t\}$.
\end{exa}

\subsection{Symmetric groups}\label{sec:Symmetry}
The following definitions and properties, well-known in group theory~\cite{sagan2001}, will be useful later, in Section~\ref{sec:toNRS}, when
translating CRSs into NRSs.

\begin{defi}
The \emph{symmetric group} $\mathcal{S}_n$ is the group of bijections (permutations) of $\{a_1,\ldots,a_n\}$ to itself.
A standard notation for the permutation that maps $a_i$ to $\pi(a_i)$ is
the \emph{two-line notation} or  \emph{array form}
\[
\left(\begin{array}{ccccc}
a_1&a_2&a_3&\ldots&a_n\\
\pi(a_1)&\pi(a_2)&\pi(a_3)&\ldots&\pi(a_n)
\end{array}\right)
\]
Under composition of mappings, the permutations of $\{a_1,\ldots, a_n\}$ are a group.

A permutation $\pi\in\mathcal{S}_n$ is a \emph{$k$-cycle} if there are distinct elements
$a_1,a_2,\ldots,a_k$ such that $\pi(a_1)\eq a_2,\pi(a_2)\eq a_3,\ldots,\pi(a_k)\eq a_1$
and $\pi$ fixes every other element.
A $2$-cycle permutation is known as a \emph{transposition}, or swapping.
There is a standard notation for \emph{$k$-cycle forms}:
\[\tuple{a_1,a_2,a_3,\ldots,a_k}\]
A pair of  cycles $\tuple{a_1,\ldots,a_n}$ and $\tuple{a'_1,\ldots,a'_n}$ are \emph{disjoint}
when the sets $\{a_1,\ldots,a_n\}$ and $\{a'_1,\ldots,a'_n\}$ are disjoint.
\end{defi}

\begin{lem}\label{lem:disjointCycles}
Every permutation is uniquely expressible as a product of disjoint cycles.

Disjoint cycles commute.
\end{lem}

\begin{thm}[Product of transpositions]\label{theo:prodTrans}
Every permutation in $\mathcal{S}_n$, $n > 1$, can be expressed as a product 
of $2$-cycles.
\end{thm}

Theorem~\ref{theo:prodTrans} justifies our choice of representation for permutations as lists of swappings in Section~\ref{sec:NomRew}.

\begin{property}[$k$-cycle as a product of transpositions]\label{def:toTrans}
A $k$-cycle 
$\tuple{a_1 , a_2 , \ldots, a_{k-1} , a_k}$ in $\mathcal{S}_n$  can be decomposed into transpositions:
$$(a_1 , a_2 , \ldots , a_{k-1} , a_k )\eq \Tran{a_1}{a_k}\Tran{a_1}{a_{k-1}}\ldots\Tran{a_1}{a_2}$$
following the grammar of permutations given in Section~\ref{sec:NomRew}.
\end{property}
Using this property, when given a permutation defined as a bijection, we
can find its corresponding list of swappings by first writing it as 
a product of disjoint cycles, and then 
decomposing each cycle into $2$-cycles as shown in the example below.

\begin{exa}\label{exa:cycleDecomp}
The following  bijective function:
\[
\begin{array}{lll}
f(a)=c & f(b)=d & f(c)=e\\
f(d)=f & f(e)=g & f(f)=h\\
f(g)=a & f(h)=b & f(i)=i\\
\end{array}
\]
has the following array form associated with it.
\[
f\eq\left(\begin{array}{c}
a\quad b\quad c\quad d\quad e\quad f\quad g\quad h\quad i\\
c\quad d\quad e\quad f\quad g\quad h\quad a\quad b\quad i 
\end{array}\right)
\]
In this case, the first
row represents the elements in the domain, in lexicographic order,
and the second row their respective image.

To convert array form notation into cycle notation we follow these steps:
\begin{itemize}
\item  Start with the smallest letter in the set, in this case $a$, since
$f(a)\eq c$ we begin the cycle by writing
\[(a,c,\ldots)\ldots\]
Notice we could start with any letter since there are a number of equivalent representations
of $f$ in cycle form.
Non-unique representation does not alter the action of the permutations in $f$.
\item Next, $c$ maps to $e$, so we continue building the cycle
\[(a,c,e,\ldots)\ldots\]
\item Continuing in this way we construct $(a,c,e,g,\ldots)\ldots$ and since
$g$ maps back to $a$, then we close off the cycle
\[(a,c,e,g)\ldots\]
\item Next, we pick the smallest letter that does not appear in any previously constructed
cycle, this is letter $b$ in this case, and repeat the previous steps to construct a new cycle:
\[(a,c,e,g)(b,d,f,h)\ldots\]
\item Finally the last letter $i$ is picked and the cycle is constructed.
In this case $i$ maps to itself:
\[(a,c,e,g)(b,d,f,h)(i)\]
\item Decomposition into $2$-cycle form by application of 
Property~\ref{def:toTrans} produces:
\[\Tran{b}{h}\Tran{b}{f}\Tran{b}{d}\Tran{a}{g}\Tran{a}{e}\Tran{a}{c}\]
where the $1$-cycle $(i)$ is discarded since it produces $\Tran{i}{i}$.
\end{itemize}
\end{exa}

\noindent This simple method of converting bijective mappings in array form into
a 2-cycle representation of permutations, while preserving the action
of the mappings, is instrumental for a correct translation of
meta-applications in CRSs into permutations suspended on variables in
NRSs.  We postpone further discussion along with the formal definition
of the conversion procedure for Section~\ref{sec:toNRS}, where CRS
rules and terms are translated to NRSs.

\section{Translating  from Nominal to CRS Syntax}
\label{sec:formalCRS}
In this section, we give an overview of the main issues surrounding 
the translation between NRS and CRS syntax, along with our approach to solve them.
Further examples and formal proofs are given after defining the translation function.


\subsection{Overview of the Problem}\label{sec:overview}
In order to design a function that transforms NRSs to CRSs, we must
take into account the following distinctions between formalisms:
\begin{itemize}
\item CRS rules are closed by definition, but this is not the case for
  nominal rules.  Thus, the translation is restricted to closed NRS rules.
\item CRSs make a distinction between meta-terms and terms and
  rewriting is defined only on terms. Such a distinction does not
  occur in NRSs.  To solve such issue, nominal rewriting is restricted
  to operate only on ground terms.
\item NRSs contain a (possibly empty) set of freshness assumptions to
  avoid accidental name capture. Such a mechanism does not exist in
  CRSs, where (meta-)terms are defined modulo $\alpha$.  Therefore,
  freshness assumptions must be considered when constructing both
  meta-applications and CRS substitutes.
\item Nominal variables have arity zero whereas CRS meta-variables may
  have non-zero arity.  Hence, a unique arity for all occurrences of a
  meta-variable must be correctly enforced when applying the
  translation function.
\item A moderated variable $\pi\cdot X$ contains a suspended
  permutation $\pi$ which is applied immediately after instantiation
  of $X$. There are no permutations in CRSs.  Intuitively, we must
  observe the potential effects of applying the permutation to any
  instantiation of $X$ and translate accordingly to CRSs to simulate
  the action of $\pi$.
\item Nominal substitution allows capture, whereas substitution in CRSs
   is non-capturing. Therefore, the translation algorithm
  must recognise abstractions and build CRS substitutes that 
  replicate the behaviour of the substitutions in NRSs.
\end{itemize}

\noindent First, we discuss the simulation of capturing substitution in CRSs.
For this, we use an auxiliary function $\Lambda$ which traverses a
nominal term $t$ and outputs, for each nominal variable $X$ in $t$, a
set of distinct atoms that occur abstracted above \emph{any} of the
occurrences of $X$ in $t$.  For instance, if $t\eq
g\tuple{\abs{a}\abs{b}X,\abs{a}\abs{b}X}$ then
$\Lambda_t(X)\eq\{a,b\}$.  Atoms in $\Lambda_t(X)$ would be captured
if $X$ is instantiated by a term that contains these atoms free, e.g.:
$\sigma(X)=f(a,b)$.  However, in CRSs, a distinct representative of
the term class would be chosen if any of the $a,b$ variables were to
appear free in a substitute.  Since variable capture must be allowed,
$\Lambda_t (X)$ is used to create the variable binding list for a
substitute of a meta-variable $X$, in the case of the example,
$\sigma\eq[X\sm\underline{\lambda}(a.b).f(a,b)]$.  Moreover, as CRS
rules must be closed, $\Lambda_t (X)$ also aids in constructing the
list of bound variables (thus occurring in $\Lambda_t (X)$) associated
with each occurrence of a meta-variable $X$ in a rule.
Therefore, the example above would translate to $\hat{t}\eq
g\tuple{\abs{a}\abs{b}X\tuple{a,b},\abs{a}\abs{b}X\tuple{a,b}}$.
Indeed, our translation algorithm outputs closed CRS meta-terms when
applied to closed nominal terms.

Additionally, when translating a term-in-context $\Delta\cent t$, we
take into account the freshness constraints in $\Delta$ such that, if
$a\in\Lambda_t(X)$ and $a\# X\in\Delta$, then $a$ is not
considered when simulating variable capture in CRSs, since $a$ cannot
occur free in any substitution $\sigma$ for $X$.  However, there are
some special cases.  The application of a permutation $\pi$ to
$\sigma$ after instantiation may alter the final outcome of the
instantiation.  CRSs rely on the list of variable arguments and
binders along with the mechanism of $\beta$-reduction to simulate the
process of swapping atoms.  This means that variables that were not
initially part of the argument list are now introduced back into the
meta-application in order to deal with the necessary renamings.

In our translation, the list of variable arguments occurring in each
meta-application is ordered with respect to a total
ordering.\footnote{We have chosen a lexicographic ordering but any
  other total ordering also works.}  By doing so, there is no relation
between the position of each variable in the argument list of the
meta-application and the position of the abstractions in the CRS
meta-term.  On the other hand, there is a bijection between the
variables in the binding list added to the substitute for $X$ and the
variables in the argument list, as expected.  Furthermore, when
translating nominal substitutions, we will show that, for all
occurrences of a meta-application, each substitute built by the
translation function is equivalent modulo $\alpha$.  This property
allows us to choose just one of the occurrences (we choose the leftmost
one) to work with and then apply it back to all the occurrences in the
translated CRS term.  We give examples below.

Permutations are the main cause of variations among occurrences of the
same variable in a term when instantiated, leading to possible
modifications of the binding structure.  Consider for example, \[t\eq
~\cent f([a][b]X, ~[a][b]\Tran{a}{b}\act X)\] and the
substitution $\sigma\eq[X\mapsto g(a,b)]$, which produces the term
\[t\sigma\eq f([a][b]g(a,b), ~[a][b]g(b,a))\]
where atoms $a,b$ have been swapped on the second occurrence of $X$.
How should we take into account these permutations in the CRS syntax?.

Intuitively, we apply each $\pi$ directly to the set of atoms
$\Lambda_t(X)=\{a,b\}$, for each occurrence of $\piX$ in $t$,
resulting in two argument lists: $(a, b)$ for the first occurrence of
$X$ and $(b, a)$ for the second one.  However, this approach is not
effective when we encounter occurrences of $\piX$ where swappings in
$\pi$ contain atoms which do not occur abstracted above $X$, therefore
not contained in $\Lambda$.  Take for instance the nominal term \[s\eq
[a]\Tran{a}{b}\act X\] where $\Lambda_s (X)=\{a\}$.  A direct
application of $\pi$ to $\Lambda_s(X)$ results in the CRS
meta-term \[\hat{s}\eq [a]X(b).\] We immediately notice two problems
with this translation: the possibility of atom $b$ occurring in an
instantiation of $X$ in $s$ has not been accounted for in the CRS
translation $\hat{s}$.  We expect $b$ to be renamed to $a$ by the
swapping $\Tran{a}{b}$ and captured by the abstraction in $s$, yet $a$ does
not appear in the variable argument list after application of $\pi$ to
$\Lambda_s (X)$ in $\hat{s}$.  Furthermore, if our goal is to translate NRS rules
into CRS rules, and CRS rules are closed by definition, the
application of $\pi$ to $\Lambda_s(X)$ produces a list of atoms no
longer bound above $X$, and thus not suitable for CRSs.

Alternatively, we order lexicographically the set
$\pi^{-1}\act\Lambda_s(X)$, obtaining an
intermediate list: $xs$.  It contains those atoms that could be
captured if occurring free in a substitution for $X$ (notice in the
example $xs\eq\{b\}$ where $b$ is captured because of $\Tran{a}{b}$).
In other words, $xs$ is the \emph{initial list} of binders that allow
our translation to capture variables.  Next, $\pi$ is applied to $xs$
to return an ordered and filtered version of $\Lambda_s(X)$ as a list,
which we call $\overline{xs}$ (in the example, $\overline{xs}\eq a$).
The list $\overline{xs}$ contains variables occurring in $\Lambda$ and
thus bound so that it can finally be displayed as the variable
argument list for the meta-application of an occurrence of $X$.
Therefore, at this point we have a bijection from the atoms in $xs$ to
those in $\overline{xs}$ such that $xs_i\mapsto\overline{xs}_i$ is the
mechanism that maps a captured variable $a\in xs$ at position $i$ to the
variable  $\pi(a)\in\overline{xs}$ at position $i$ when a
substitution is provided.  In addition, $\pi$ must also be applied to
the nominal substitution $\sigma(X)$ prior translation, to rename
atoms in $\suppi$ not in scope of $\Lambda_t(X)$.  As a result, $\pi$
is also applied to the binding list $xs, \pi\act xs$, to preserve the
original binding structure when added to $\pi\act\sigma(X)$, i.e.,
$\underline{\lambda}(\pi\act xs).(\pi\act\sigma(X))$.  Note that  $\pi\act
xs$ is $\overline{xs}$.

A more detailed explanation of the  algorithm is provided after its definition.
Now, we look at the examples again, applying the new approach:
\[(t\eq ~\cent  f([a][b]X, ~[a][b]\Tran{a}{b}\act X), \sigma\eq[X\mapsto g(a,b)])\]
is translated as
\[(\hat{t}\eq  f([a][b]X(a,b), ~[a][b]X(b,a)), \hat{\sigma}\eq[X\mapsto \underline{\lambda}(a.b).\, g(a,b)])\]
since the leftmost one is chosen, where
$$\hat{\sigma}(\hat{t})=  f([a][b]g(a,b), ~[a][b]g(b,a))$$
Also, \[s\eq [a]\Tran{a}{b}\act X\]
translates to the CRS meta-term $$\hat{s}\eq  [a]X (a)$$
which is now closed.
And a nominal substitution $\sigma\eq[X\sm g(a,b))]$ to instantiate nominal term $s$ would translate into
the valuation $\hat{\sigma}\eq[X\sm \underline{\lambda}(a).g(b,a)]$.

We have discussed the main issues in the translation
of NRSs to CRSs, together with strategies to solve such
issues. In the  rest of the section we formalise this
approach and provide examples.


\subsection{Translating Nominal Terms}\label{sec:terms}
For each nominal signature $\Sigma$, and sets $\mathcal{A}$ and
$\mathcal{X}$ of atoms and variables, we consider a CRS alphabet
containing $\Sigma$, variables $\mathcal{A}$ and meta-variables
$\mathcal{X}$.

First we define an auxiliary function, $\Lambda$, to compute,
for each  nominal term $t$, and each variable $X$ in $t$,
the set of atoms that may be captured
when $X$ is instantiated.

Intuitively, $\Lambda_t(X)=\{a_1,\ldots, a_n\}$ 
if $X\in \Vars(t)$ has $k$ occurrences in $t$, $A_i$  is the set of atoms abstracted above the $i$th occurrence of $X$, and $\{a_1,\ldots, a_n\} = A_1 \cup \ldots \cup A_k$. In other words, $\Lambda_t(X)$ is the set of all the atoms abstracted above occurrences of $X$ in $t$.

\begin{defi}[Mapping $\Lambda_t$]\label{def:Lambda_t}
For each nominal term $t$, the mapping $\Lambda_t:\Vars (t)\ra\mathcal{P}(\mathcal{A})$ is defined by $\Lambda_t(X)=\Lambda'(\emptyset,t)(X)$, where $\Lambda'(\act, \act)$ is an
auxiliary function defined inductively over the structure of $t$ as follows:

\begin{tabular}{lcl}\\
$\Lambda'(A, a)(X)$&=& $\emptyset,$\\
$\Lambda'(A, \pi\act X)(X)$&=&$ A,$\\
$\Lambda'(A, \pi\act Y)(X)$&=& $\emptyset,$\\
$\Lambda'(A, [a]s)(X)$&=& $\Lambda'(A\cup\{a\}, s)(X),$\\
$\Lambda'(A, f s)(X)$&=&$ \Lambda'(A, s)(X),$\\
$\Lambda'(A, (s_1, \ldots, s_n))(X)$&=&$ \Lambda'(A, s_1)(X)\cup\ldots\cup\Lambda'(A,s_n)(X)$.
\end{tabular}
\end{defi}\medskip

\noindent For example, if $t\eq \tuple{\abs{a}X,\abs{b}X,\abs{c}Y}$ then 
 $\Lambda_t(X)\eq\{a,b\}$ and $\Lambda_t(Y)\eq\{c\}$.

Next, we define the translation of nominal terms into CRS (meta-)terms.

\begin{defi}[Term translation]\label{def:groundterm}
Let $\Delta\vdash t$ be a nominal term-in-context 
and $\Lambda_t$  as  in Definition~\ref{def:Lambda_t}.
Then $\mathcal{T}(\Delta,t) = \llbracket t \rrbracket^{\Delta}_{\Lambda_t}$, where $\llbracket \act \rrbracket^{\Delta}_{\Lambda_t}$ is an auxiliary function defined by induction
over the structure of nominal terms as follows:

\begin{tabular}{llcl}\\
\textbf{(atom)} & $\tauDelta {a}{t}$ & = & $a$,\\ 
\textbf{(var)}& $\tauDelta {\piX}{t}$ & = & $X (\overline{xs})$ where\\
&&& $\lefteqn{\overline{xs} \triangleq \pi\cdot xs \mbox{ (we omit }(\overline{xs})\mbox{ if empty)}}$\\
 &&& $\lefteqn{xs \triangleq \mathtt{toAscList}
 ([\pi^{-1}\act\Lambda_t(X)]-\{a\, |\, a \# X\in \Delta\})}$,\\ 
\textbf{(abs)}& $\tauDelta {[a]s}{t}$&=&$[a]\tauDelta {s}{t}$,\\ 
\textbf{(fun)}&$\tauDelta {f s}{t}$&=&$f \tauDelta {s}{t} $,\\ 
\textbf{(tuple)}&$\tauDelta {(s_1, \ldots, s_n)}{t}$&=&$(\tauDelta {s_1}{t}, \ldots, \tauDelta {s_n}{t})$.
\end{tabular}
where \texttt{toAscList} is a function that builds a sorted list\footnote{List of atoms in ascending lexical order.} from a set of atoms. When there is no ambiguity, we refer to the translation of a term $t$ as $\hat{t}$.
\end{defi}

The interesting case in the translation is that of a variable. 
Intuitively, the list $xs$ contains
atoms $a_i$ that will be captured if occurring free in any instance
of $X$, because $\pi\act X$ occurs in the scope of an abstraction
for $\pi\act a_i$. Note that if $a \not\in \overline{xs}$ then if
$\pi^{-1}\act a$ is in a substitution for $X$, then $a$ will be in
$supp(t)$ (it will not be captured). In other words, $a \not\in
\overline{xs}$ implies that if $\pi^{-1}\act a \in supp(X)$ then $a
\in supp(t)$.

The examples below illustrate the translation of nominal terms
to CRS terms.
Prior to that, we highlight a property of the translation and introduce some terminology.

\begin{lem}[Preservation of variables as meta-variables in the translation]
\label{lem:preserVarsNRS2CRS}
Suppose $\Delta\cent t$ is a term-in-context and  $\hat{t}\eq\tauDelta{t}{t}$ its translation
by Definition~\ref{def:groundterm}.
Then,
$\piX$ occurs in $t$ at position $p$ if and only if
$X\tuple{\overline{xs}}$ occurs in $\hat{t}$ at position $p$ for some $\overline{xs}$.
\end{lem}
\proof
By induction on the structure of $t$ and the fact that there is a one-to-one correspondence between the elements of $t$ and $\hat{t}$ as  shown 
in the syntax-directed translation function in Definition~\ref{def:groundterm}, namely 
a variable $\piX$ occurs at position $p$ in $t$ if and only if there exists a meta-variable $X$ at $p$ in its translation $\hat{t}$
along with a list of variable arguments $\overline{xs}$.
\qed

\begin{rem}[Translation of a moderated variable]\label{rem:translationX}
Let  $\mathcal{T}(\Delta,t)\eq\hat{t}$, where 
$\Delta\cent t$ is a term-in-context and $\mathcal{T}$ is the translation 
function given in Definition~\ref{def:groundterm}. 
If $\piX$ occurs in $t$ at position $p$ then 
$X\tuple{\overline{xs}}\eq\tauDelta{\piX}{t}$ occurs in $\hat{t}$ at position $p$, by Lemma~\ref{lem:preserVarsNRS2CRS}.
We say that $X\tuple{\overline{xs}}$ \emph{is the translation of} $\piX$ \emph{in} $\hat{t}$
and call $\overline{xs}$ \emph{the arguments of} $X$.
\end{rem}

\begin{exa}
According to Definition~\ref{def:groundterm}, the (closed) nominal
term $\cent [a][b] X$ is translated as the CRS meta-term
$[a][b]X(a,b)$, where we include both variables in the
meta-application as they may appear free in a substitution for $X$.

Freshness constraints are taken into account in the translation of
variables.  Consider $a\hash X\cent [a][b] X$, which is translated to
the CRS meta-term $[a][b]X(b)$.  Since any substitution for $X$ must
satisfy $\Delta$ (that is, $\cent a\#X\sigma$), $a$ is not included in
the arguments of $X$.

However, a freshness constraint does not always produce this effect, 
it depends on the  permutations  in the term.
We adjust our example to show this.
Consider
$$a\hash X\cent [a][b]\Tran{a}{b}\act X.$$
In this case we should take into account  the
mapping $b\mapsto a$ but not $a\mapsto b$ since $a\hash X\in\Delta$.
Our translation outputs the CRS meta-term $$[a][b]X(a),$$ which suggests that
$a$ may occur free in an instance of $X$ contradicting the nominal
constraint $a\# X\in\Delta$.
However, since any nominal substitution
$\sigma$ that instantiates $X$ must also satisfy $\Delta$, the atom
$a$ does not occur unabstracted in $X\sigma$ or in its CRS translation.
Hence the mapping $a\mapsto b$ is discarded.
\end{exa}

\begin{exa}\label{example (ac)}
Let $t$ be the (closed) nominal term $\cent [a][b]\Tran{a}{c}\act X$,
 where $\Delta=\emptyset$. Then $\Lambda_t (X)=\{a,b\}$, $xs=[b,c]$, 
and $\overline{xs}=[b,a]$.
The swapping $\Tran{a}{c}$ maps the abstracted atom $a$ 
to the unabstracted atom $c$.
This particular kind of mapping cannot be explicitly represented at term level thus it
will be applied to the substitute that instantiates $X$, if any.
This  is shown in more detail in Section \ref{sec:rules}
when describing the translation of substitutions.
The translation function for terms (Definition~\ref{def:groundterm})
 produces the meta-term $$[a][b]X(b,a)$$
which effectively takes into account the rest of the mappings in the permutation, 
that is,  $b\ \sm\ b, c\ \sm\ a$, generating a closed CRS meta-term.
\end{exa}

The following example illustrates the case of a term with multiple occurrences of 
a variable with different suspended permutations.

\begin{exa}\label{exa:diffPerms}
The translation of the (closed) nominal term-in-context
\[
f\apart X\cent g\tuple{\abs{a}\abs{b}\abs{c}\abs{d}\Tran{c}{f}\Tran{a}{b}\act X,\abs{a}\abs{b}\abs{c}\abs{d}\Tran{c}{f}\Tran{d}{b}\Tran{a}{d}\act X}\] 
is the CRS meta-term
\[
g\tuple{\abs{a}\abs{b}\abs{c}\abs{d} X\tuple{b,a,d},\abs{a}\abs{b}\abs{c}\abs{d}X\tuple{b,d,a}}
\]
where
\begin{itemize}
\item $\Lambda_t(X)\eq\{a,b,c,d\}$,
\item $xs_1\eq xs_2\eq [a,b,d]$   (the sub-indices  are used to refer to the two
 occurrences of $X$),
\item $\pi_1\act xs_1\eq\overline{xs}_1\eq [b,a,d]$ and $\pi_2\act xs_2\eq\overline{xs}_2\eq [b,d,a]$ 
\end{itemize}
Notice that if atom $c$ occurs unabstracted in an instance $X\sigma$
 of $X$, there exists a mapping $c\,\sm f$ in
both permutations that renames $c$ to $f$ when
translating $X\sigma$.  This renaming cannot be dealt with
at term level because it would involve including variable
$f$ in the list of arguments and the result would not be closed.
However, the function that translates  nominal substitutions (see
Definition~\ref{def:withSubs}) takes care of this
renaming.
\end{exa}

The following lemma states the uniqueness of the intermediate list of atoms $xs$ 
computed by the translation function for each of the occurrences of a variable $X$ 
in a \emph{closed} nominal term.  This is important when proving preservation of
arities among meta-variable occurrences and closedness of the
translated CRS meta-term.

\begin{lem}[\textbf{Equivalence}]
\label{lem:xsEquivalence}
Let $\Delta\vdash t$ be a closed term-in-context and
$\mathcal{T}(\Delta,t)= \hat{t}$ its CRS translation.  If $\pi_1\cdot
X$ and $\pi_2\cdot X$ are two occurrences of the same variable $X$ in
$t$, and $X(\overline{xs_1})$, $X(\overline{xs_2})$ are their respective
translations in $\hat{t}$, then $\pi_1^{-1}\cdot \overline{xs_1}
\eq \pi_2^{-1}\cdot \overline{xs_2}$.

\end{lem}
\proof
This is a consequence of the definition of $\overline{xs}$ and the fact that the term is closed.
More precisely, by Definition \ref{def:groundterm}, the translation of $\pi\cdot X$ is  $X(\overline{xs})$ 
where $\overline{xs} \triangleq \pi\cdot xs$ and
$xs \triangleq \mathtt{toAscList}([\pi^{-1}\act\Lambda_t(X)]-\{a\, |\, a \# X\in \Delta\})$.\\
It is sufficient to prove that if an atom $a \in xs_1$  at a position $i$
then $a \in xs_2$ at a position $j$ such that $i\eq j$, and vice versa.

Now, for any $a$ such that  $a\in xs_1$, it is also the case that
$ \pi_1 (a)\in\Lambda_t(X)$ (since $\pi^{-1}\circ\pi\act a\eq a$) and $a\# X\not\in\Delta$ by definition of $xs$, then either\\ 
(1) $\pi_2(a)\not\in \Lambda_t(X)$, or\\
(2) $\pi_2(a)\in \Lambda_t(X)$, thus $a\in xs_2$,\\
No other cases are possible.

In case (1), any substitution of $X$ containing  atom $a$ unabstracted is in
the scope of
an abstraction in $\pi_1\cdot X$ since $\pi_1 (a)\in\Lambda_t (X)$, but unabstracted under $\pi_2\cdot X$, since $\pi_2 (a)\not\in\Lambda_t (X)$.
Since the term is closed it must be the case that $a\# X\in\Delta$ by Definition~\ref{def:closed},
contradicting the fact that $a\in xs_1$.
Hence it is the case that $\pi_2(a)\in \Lambda_t(X)$ too, as stated in (2).
Thus, we have established that
for each $a$ in $xs_1$ at some position $i$,   $a\in xs_2$
at some position $j$.
Similarly, we can prove $a\in xs_{2}$ implies $a\in xs_{1}$.
Since $\mathtt{toAscList}$ is applied to both $xs_{1}, xs_{2}$,
then $i\eq j$ leading to $xs_{1}\eq xs_{2}$.

Therefore,  for any pair $X (\overline{xs}_1), X (\overline{xs}_2)$ in $\hat{t}$,
$\pi^{-1}_{1}\cdot \overline{xs}_1\eq\pi^{-1}_{2}\cdot \overline{xs}_2$.
\qed

Next we prove that the translation function  produces 
CRS (meta-)terms (in particular, meta-applications respect variable arities).

\begin{property}[\textbf{Arity}]\label{pro:arity}
Let $\Delta\vdash t$ be a closed term-in-context and 
$\mathcal{T}(\Delta,t)= \hat{t}$ its CRS translation. 
For each occurrence of $X$ in $t$, there is a corresponding occurrence of 
$X$ in $\hat{t}$; moreover, there exists $n$ (the arity of $X$) such that 
all the occurrences of $X$ in  $\hat{t}$ are in meta-applications of arity $n$. 
In other words, in the translated term all the occurrences of $X$ respect the arity of $X$.
\end{property}

\proof The translation is syntax directed.  For every $\pi\cdot X$ in
$t$, $\tauDelta {\piX}{t} = X (\overline{xs})$ where, by Property
\ref{lem:xsEquivalence}, $\pi^{-1}\cdot \overline{xs}$ is a unique
list of variables for all occurrences of $X (\overline{xs})$ in
$\hat{t}$ .  This leads to $\overline{xs}$ having the same length $n$
for all $X$ in $\hat{t}$, where $n$ is the arity of $n$.  
\qed

\begin{property}[\textbf{Preservation of closedness}]\label{property:closedCRSterm}\hfill
\begin{enumerate}[label=\({\alph*}]
\item[(a)] If $\Delta \vdash t$ is a closed nominal term then its 
CRS translation $\hat{t}$ (according to Definition \ref{def:groundterm}) is 
a closed CRS meta-term. 
\item[(b)]Moreover, if the nominal term $t$ is ground, then its translation  is a CRS term.
\end{enumerate}
\end{property}
\proof
First we prove that the translation $\hat{t}$ is a CRS meta-term.
This is due to:
\begin{enumerate}
\item Our translation respecting the structure of $t$, which maps atoms to variables, moderated variables to meta-applications consisting of a meta-variable and its corresponding list of arguments in the form $X^n (a_1,\ldots, a_n)$ (where the arity $n\geq0$ can be read directly from the meta-term and thus omitted), nominal abstraction to CRS abstraction, nominal functions to CRS functions and nominal tuples to CRS tuples.
\item As a direct consequence of Properties \ref{lem:xsEquivalence} and \ref{pro:arity}, every meta-application $X^n (\overline{xs})$ respects the  arity $n = |\overline{xs}|$ for all occurrences of $X$ in $\hat{t}$.
\end{enumerate}
Hence, if $t$ is ground, $\hat{t}$ is a CRS term.

It remains to prove that $\hat{t}$ is closed.  By definition of
closedness of a nominal term, see Definition~\ref{def:closed}, every
occurrence of an atom in $t$ is in the scope
of an abstraction, since our translation respects the structure of
$t$, see $(1)$.  Finally, we prove that variable arguments in a
meta-application are bound.  For this we show that for all occurrences
of $X (\overline{xs})$ in $\hat{t}$, $\overline{xs}$ is a list of
bound variables:
by definition of $\mathcal{T}(\Delta, t)$ we know
that $\overline{xs}=\pi\act xs$ for some $\pi$ such that $\pi\cdot X$
occurs in $t$.  Also, $xs \triangleq
\mathtt{toAscList}([\pi^{-1}\act\Lambda_t(X)]-\{a \mid a \# X\in
\Delta\})$, that is, for each $a\in xs$ it is a requirement that
$\pia\in \Lambda_t(X)$ (since $\pi^{-1}\circ\pi\act a\eq a$).  If
$\pi(a)\in \Lambda_t(X)$, by Definition \ref{def:Lambda_t}, $\pi(a)$
must occur abstracted above $X$. Moreover, as a consequence of
Property \ref{lem:xsEquivalence}, this is the case for all occurrences
of $X$, otherwise $a\# X\in\Delta$ leading to
$\pi(a)\not\in\overline{xs}$.

Therefore all the variables in $\overline{xs}$ are bound,  for all occurrences of $X (\overline{xs})$ in $\hat{t}$.
Hence we conclude that  $\hat{t}$ is a closed meta-term.
\qed

The following auxiliary lemmas are used to prove that
$\alpha$-equivalent closed nominal terms have the same CRS
translation.  Intuitively this is the case since CRS terms are, by
definition, considered modulo $\alpha$.

\begin{lem}
\label{lem:supp-alpha-equiv}
If $\Delta\cent s\aleq t$ and $\Delta\cent s$, $\Delta\cent t$ are closed, then 
for any variable $X$ such that $\pi_s\act X$ occurs in $s$ and $\pi_t\act X$ occurs 
in $t$,  $\pi_s^{-1}\act \overline{xs}_s\eq \pi_t^{-1}\act \overline{xs}_t$, where $X\tuple{\overline{xs}_s}$ and $X\tuple{\overline{xs}_t}$ are the translations
of $\pi_s\act X$ in $s$ with arguments $\overline{xs}_s$  and $\pi_t\act X$ in $t$
with arguments $\overline{xs}_t$, respectively.
\end{lem}
\proof
Direct consequence of Lemma~\ref{lem:xsEquivalence} and
the fact that $\alpha$-equivalent terms have the same support.
\qed

\begin{lem}
\label{lem:transl-perm-term}
Let $\Delta\cent s$ be a nominal term-in-context and
$\Lambda_s$ as defined in Definition~\ref{def:Lambda_t}.
Then 
$\llbracket \Tran{a}{b}\act s'\rrbracket^{\Delta}_{\Tran{a}{b}\act\Lambda_{s}}$ $=\tauDelta{s'}{s}\{a\sm b,b\sm a\}$
for all subterms $ s' $ of $ s $.
\end{lem}
\proof
By induction on the structure of $s'$.

\begin{itemize}
\item[$\bullet$] The case $(s'\eq c)$.
 There are three cases to consider: $c\eq a$, $c\eq b$, and  $c\not\eq a, c\not\eq b$.\\
\begin{tiny}
$\bullet$
\end{tiny} Suppose $c\eq a$.
Then, by Definition~\ref{def:groundterm},
$\llbracket b\rrbracket^{\Delta}_{\Tran{a}{b}\act\Lambda_{s}}=b$
$=\tauDelta{a}{s}\{a\sm b,b\sm a\}$.\\
\begin{tiny}
$\bullet$
\end{tiny} Suppose $c\eq b$.
Then, by Definition~\ref{def:groundterm},
$\llbracket a\rrbracket^{\Delta}_{\Tran{a}{b}\act\Lambda_{s}}=a$
$=\tauDelta{b}{s}\{a\sm b,b\sm a\}$.\\
\begin{tiny}
$\bullet$
\end{tiny} Suppose $c\not\eq a$ and $c\not\eq b$.
By Definition~\ref{def:groundterm},
$\llbracket c\rrbracket^{\Delta}_{\Tran{a}{b}\act\Lambda_{s}}=c$
$=\tauDelta{c}{s}\{a\sm b,b\sm a\}$.\\

\item[$\bullet$] The case $(s'\eq \piX)$.
\ By Definition~\ref{def:groundterm}, we have
$\llbracket \Tran{a}{b}\act(\pi\act X)\rrbracket^{\Delta}_{\Tran{a}{b}\act\Lambda_{s}}=\llbracket (\Tran{a}{b}\circ\pi)\act X\rrbracket^{\Delta}_{\Tran{a}{b}\act\Lambda_{s}}
=X(\overline{xs})$ and also $\tauDelta{\piX}{s}\{a\sm b,b\sm a\}=X(\overline{xs'}\{a\sm b, b\sm a\})$.

For this case we must take into account each atom $c\in\overline{xs}$
at a position $p$, then prove it also occurs at $p$ in the argument list
$\overline{xs'}\{a\sm b,b\sm a\}$.

We consider first the case where $\overline{xs}=\emptyset$ then
the case where $\overline{xs}\not =\emptyset$.

~$(1)$ For the case where $\overline{xs}=\emptyset$, it is
also the case that $\Tran{a}{b}\act\Lambda_s(X)=\emptyset$.
Hence 
$\overline{xs'}\{a\sm b,b\sm a\}=\emptyset$ by Definition~\ref{def:groundterm} and the result follows.

~$(2)$ For the case where $\overline{xs}\not =\emptyset$, we
distinguish cases depending on whether $c=a$, $c=b$ and finally for
any other atom $c$ such that $a\not=c\not=b$.  Notice that for any
atom $c$, if $c\in\overline{xs}$ at $p$ then, observing
Definition~\ref{def:groundterm},
$\overline{xs}\eq(\Tran{a}{b}\circ\pi)\act xs$ such that, by property
of permutations, $(\pi^{-1}\circ\Tran{a}{b})\act\overline{xs}\eq xs$
thus $(\pi^{-1}\circ\Tran{a}{b})\act c\in xs$ also at $p$.  Hence, $
c\in\Tran{a}{b}\act\Lambda_s(X)$ and $(\pi^{-1}\circ\Tran{a}{b})\act
c\# X\not\in\Delta$.  This remark is implicitly applied in the three
cases below.

\begin{tiny}
$\bullet$
\end{tiny} Suppose $c=a$ such that $a\in\overline{xs}$ at position $p$.
Then $a\in\Tran{a}{b}\act\Lambda_s(X)$, thus $b\in\Lambda_s(X)$.
Following Definition~\ref{def:groundterm} and the remark above,
if $a\in\overline{xs}$ at position $p$ then
$\pi^{-1}\act b\in xs$ at $p$, for both translations, 
as stated in Lemma~\ref{lem:xsEquivalence}, resulting in $\pi\circ \pi^{-1}\act b\in\overline{xs'}$ thus
$b\in\overline{xs'}$ by properties of permutations.
Hence $a\in\overline{xs'}\{a\sm b,b\sm a\}$   at $p$.

\begin{tiny}
$\bullet$
\end{tiny} Suppose $c=b$ such that $b\in\overline{xs}$ at position $p$.
Then $b\in\Tran{a}{b}\act\Lambda_s(X)$, thus $a\in\Lambda_s(X)$.
Following Definition~\ref{def:groundterm}
and the remark above,
if $b\in\overline{xs}$ at position $p$ then
$\pi^{-1}\act a\in xs$ at $p$, for both translations, 
as stated in Lemma~\ref{lem:xsEquivalence}, resulting in $\pi\circ \pi^{-1}\act a\in\overline{xs'}$ thus
$a\in\overline{xs'}$ by properties of permutations.
Hence $b\in\overline{xs'}\{a\sm b,b\sm a\}$   at $p$.

\begin{tiny}
$\bullet$
\end{tiny} Suppose now $c\not=a,c\not=b$ such that $c\in\overline{xs}$
at position $p$.\! Then $c\in\Tran{a}{b}\act\Lambda_s(X)$, thus $c\in\Lambda_s(X)$.
Following Definition~\ref{def:groundterm} and the remark above,
$\pi^{-1}\act c\in xs$ at $p$, for both translations,
as stated in Lemma~\ref{lem:xsEquivalence}, resulting in $\pi\circ \pi^{-1}\act c\in\overline{xs'}$ thus 
$c\in\overline{xs'}$ by properties of permutations.
Hence $c\in\overline{xs'}\{a\sm b,b\sm a\}$   at $p$.


\item[$\bullet$] The case $(s'\eq [a]t)$.
\ By Definition~\ref{def:groundterm},
$\llbracket [b]\Tran{a}{b}\act t\rrbracket^{\Delta}_{\Tran{a}{b}\act\Lambda_{s}}
=[b]\llbracket \Tran{a}{b}\act t\rrbracket^{\Delta}_{\Tran{a}{b}\act\Lambda_{s}}$
where $b\in\Tran{a}{b}\act\Lambda_s(X)$ for any $X\in\mathcal{X}$ occurring in $t$.
Take any variable $c$ not occurring free~in the CRS meta-term $\llbracket \Tran{a}{b}\act t\rrbracket^{\Delta}_{\Tran{a}{b}\act\Lambda_{s}}$.
Then one could choose another CRS representative $[c]\llbracket \Tran{a}{b}\act t\rrbracket^{\Delta}_{\Tran{a}{b}\act\Lambda_{s}}\{b\sm c\}$ in its  $\alpha$-equivalence class.
The induction hypothesis impliess
$[c](\llbracket \Tran{a}{b}\act
t\rrbracket^{\Delta}_{\Tran{a}{b}\act\Lambda_{s}})\{b\sm c\}$
$=[c](\tauDelta{t}{s}\{a\sm b,b\sm a\})\{b\sm c\}$ where
$(\tauDelta{t}{s}\{a\sm b,b\sm a\})\{b\sm c\}$
$=\tauDelta{t}{s}\{a\sm c\}\{b\sm a\}$.
Since variable $c$ does not occur free in $\llbracket \Tran{a}{b}\act t\rrbracket^{\Delta}_{\Tran{a}{b}\act\Lambda_{s}}$, we can assume
$[c](\tauDelta{t}{s}\{a\sm b,b\sm a\})\{b\sm c\}=[c]\tauDelta{t}{s}\{a\sm c, c\sm a\}\{b\sm a\}$ without loss of generality.

Furthermore,
$\tauDelta{[a]t}{s}\{a\sm b,b\sm a\}=[a](\tauDelta{t}{s}\{b\sm a\})$
where $a\in\Lambda_s(X)$ for any $X\in\mathcal{X}$ occurring in $t$.
Since there exists a variable $c$ not free in $\tauDelta{t}{s}\{b\sm a\}$ as explained above,
we choose an $\alpha$-equivalent $CRS$ meta-term
$[c](\tauDelta{t}{s}\{a\sm c,c\sm a\})\{b\sm a\}$ without loss of generality.
And the result follows.

The case ($s=[b]t$) is similarly solved and thus omitted here.

\item[$\bullet$] The case $(s'\eq [c]t)$.
\ By Definition~\ref{def:groundterm},
$\llbracket [c]\Tran{a}{b}\act t\rrbracket^{\Delta}_{\Tran{a}{b}\act\Lambda_{s}}
=[c]\llbracket \Tran{a}{b}\act t\rrbracket^{\Delta}_{\Tran{a}{b}\act\Lambda_{s}}$
where $c\in\Tran{a}{b}\act\Lambda_s(X)$ for any $X\in\mathcal{X}$ occurring in $t$.

Furthermore,
$\tauDelta{[c]t}{s}\{a\sm b,b\sm a\}=[c]\tauDelta{t}{s}\{a\sm b,b\sm a\}$
where $c\in\Lambda_s(X)$ for any $X\in\mathcal{X}$ occurring in $t$.

By induction hypothesis, 
$\llbracket \Tran{a}{b}\act t\rrbracket^{\Delta}_{\Tran{a}{b}\act\Lambda_{s}}$ $=\tauDelta{t}{s}\{a \sm b,b\sm a\}$ and the result follows.

\item[$\bullet$] The case $(s'\eq ft)$.
\ By Definition~\ref{def:groundterm},
$\llbracket f(\Tran{a}{b}\act t)\rrbracket^{\Delta}_{\Tran{a}{b}\act\Lambda_{s}}
=f(\llbracket \Tran{a}{b}\act t\rrbracket^{\Delta}_{\Tran{a}{b}\act\Lambda_{s}})$.

Furthermore,
$\tauDelta{ft}{s}\{a\sm b,b\sm a\}=f\tauDelta{t}{s}\{a\sm b,b\sm a\}$.

By induction hypothesis, 
$\llbracket \Tran{a}{b}\act t\rrbracket^{\Delta}_{\Tran{a}{b}\act\Lambda_{s}}$ $=\tauDelta{t}{s}\{a\sm b,b\sm a\}$ and the result follows.

\item[$\bullet$] The case $(s'\eq\tuple{s_1,\ldots,s_n})$.
\ Note that 
\[\llbracket \Tran{a}{b}\act\tuple{s_1,\ldots,s_n}\rrbracket^{\Delta}_{\Tran{a}{b}\act\Lambda_{s}}=
\llbracket (\Tran{a}{b}\act s_1,\ldots,\Tran{a}{b}\act
s_n)\rrbracket^{\Delta}_{\Tran{a}{b}\act\Lambda_{s}}\,,\]
 and by Definition~\ref{def:groundterm},
\[\llbracket (\Tran{a}{b}\act s_1,\ldots,\Tran{a}{b}\act s_n)\rrbracket^{\Delta}_{\Tran{a}{b}\act\Lambda_{s}}
=\tuple{\llbracket \Tran{a}{b}\act s_1\rrbracket^{\Delta}_{\Tran{a}{b}\act\Lambda_{s}},\ldots,
\llbracket \Tran{a}{b}\act s_n\rrbracket^{\Delta}_{\Tran{a}{b}\act\Lambda_{s}}}\,.\]
Furthermore,
\[\tauDelta{\tuple{s_1,\ldots,s_n}}{s}\{a\sm b,b\sm a\}=
\tuple{\tauDelta{s_1}{s}\{a\sm b,
b\sm a\},\ldots,\tauDelta{s_n}{s}\{a\sm b,b\sm a\}}.\]
By induction hypothesis, 
$\llbracket \Tran{a}{b}\act s_i\rrbracket^{\Delta}_{\Tran{a}{b}\act\Lambda_{s}}$ $=\tauDelta{s_i}{s}\{a\sm b,b\sm a\}$ where $0\leq i\leq n$ and the result follows.\qed
\end{itemize}

\begin{thm}[\textbf{Uniformity w.r.t. $\alpha$}]
\label{theo:alphaCRS}
Let $\Delta\cent t,\ \Delta\cent s$ be a pair of closed nominal
terms-in-context such that $\Delta\cent s\aleq t$ and let
$\hat{s}\eq\mathcal{T}(\Delta,s)$, $\hat{t}\eq\mathcal{T}(\Delta,t)$
be their respective CRS translations according to Definition
\ref{def:groundterm}.  Then $\hat{s}\eq\hat{t}$.
\end{thm}
\proof
We prove a more general property: 

Let  $\Delta\cent s'\aleq t'$ be a
subderivation of $\Delta\cent s\aleq t$. Hence,   
$s' = \pi\act s|_p$ and $t'=\rho\act t|_p$,  for some permutations $\pi$, $\rho$.
Then, $\tauDelta{s'}{\pi\act s}=\tauDelta{t'}{\rho\act t}$.

 From this property we deduce in particular
 $\hat{s}\eq\hat{t}$ when $p = \epsilon$.

To prove this property, we proceed by induction on the derivation and 
distinguish cases according to the last rule used.

\begin{itemize}
\item[$\bullet$]  $\rulefont{\aleq a}$.  In this case,  $s'\eq a \eq t'$.  By  Definition \ref{def:groundterm}, $\tauDelta{s'}{\pi\act s} = a = \tauDelta{t'}{\rho\act t}$.

\item[$\bullet$]  $\rulefont{\aleq X}$.
In this case, $s' = \pi_{s'}\act X$, $t' = \rho_{t'}\act X$, and
$\Delta \cent ds(\pi_{s'},\rho_{t'})\apart X$. The result follows 
by~Lemma~\ref{lem:supp-alpha-equiv}.

\item[$\bullet$]  $\rulefont{\aleq [a]}$. In this case, $s' = [a]s''$, $t' = [a]t''$, and $\Delta\cent s''\aleq t''$. The result follows by induction hypothesis.

\item[$\bullet$]  $\rulefont{\aleq [b]}$.
In this case, $s' = [a]s''$, $t' = [b]t''$, $\Delta\cent \Tran{b}{a}\act s''\aleq t''$
and  $\Delta\cent b\apart s''$. 

By induction hypothesis, $\llbracket \Tran{b}{a}\act s''\rrbracket^{\Delta}_{\Tran{b}{a}\circ\pi\act\Lambda_{s}} = \llbracket t''\rrbracket^{\Delta}_{\rho\act\Lambda_{t}}$.

By Lemma~\ref{lem:transl-perm-term}, $\llbracket \Tran{b}{a}\act s''\rrbracket^{\Delta}_{\Tran{b}{a}\circ\pi\act\Lambda_{s}} = \llbracket s''\rrbracket^{\Delta}_{\pi\act\Lambda_{s}}\{a\sm b, b\sm a\}$.
Since $\Delta\cent b\# s'$ then $\llbracket s''\rrbracket^{\Delta}_{\pi\act\Lambda_{s}}\{a\sm b, b\sm a\}\eq \llbracket s''\rrbracket^{\Delta}_{\pi\act\Lambda_{s}}\{a\sm b\}$.
Therefore, $\llbracket s''\rrbracket^{\Delta}_{\pi\act\Lambda_{s}}\{a\sm b\}=  \llbracket t''\rrbracket^{\Delta}_{\rho\act\Lambda_{t}}$.

Then, $\llbracket s'\rrbracket^{\Delta}_{\pi\act\Lambda_{s}}= [a]\llbracket s''\rrbracket^{\Delta}_{\pi\act\Lambda_{s}} = [b] \llbracket s''\rrbracket^{\Delta}_{\pi\act\Lambda_{s}}\{a\sm b\}$ since $b$ does not occur free in the translation of $s''$.
And  $ \llbracket t'\rrbracket^{\Delta}_{\rho\act\Lambda_{t}}= [b]\llbracket t''\rrbracket^{\Delta}_{\rho\act\Lambda_{t}}$. 
The result follows.

\item[$\bullet$]   $\rulefont{\aleq f}$.
In this case, $s' = fs''$, $t' = ft''$ and  $\Delta\cent s''\aleq t''$. The result 
follows directly by induction hypothesis.

\item[$\bullet$]   $\rulefont{\aleq tupl}$.
In this case, $s'= \tuple{s_1',\ldots, s_n'}$, $t' = \tuple{t_1',\ldots,t_n'}$
and $\Delta\cent s'_1\aleq t'_1,\ldots,\Delta\cent s'_n\aleq t'_n$. 
The result follows directly by induction hypothesis.\qed
\end{itemize}


\section{Transforming NRS Rules}\label{sec:rules}
NRS rules are more general than CRS rules in that unabstracted atoms may
occur in rules. In this section, we impose some conditions on NRS
rules to obtain a class of rules that can be translated to CRS
rules.
\begin{defi}[Standard nominal rule]\label{def:standard}
A nominal rule is called \emph{standard} when 
it is closed and the left-hand side has the form $f s$.
\end{defi}

\begin{defi}[Rule translation function]\label{def:rules}
Let $R\eq \nabla \vdash l\rightarrow r $ be a standard nominal rule.
The translation of $R$ is $ \mathcal{T}^{\mathcal{R}}(\nabla,l, r) =
\mathcal{T}(\nabla, l) \Rightarrow \mathcal{T}(\nabla, r)$, where
$\mathcal{T}(\Delta, t)$ is given in Definition~\ref{def:groundterm}.
\end{defi}

\begin{lem}[Well-defined rule translation]\label{lem:isRule} 
Let $R\eq \nabla \vdash l\rightarrow r $ be a standard nominal rule.
If $R'\eq \hat{l}\Rightarrow\hat{r}$ is its translation according to
Definition~\ref{def:rules}, then $R'$ is a CRS rule.

\end{lem}
\proof
First, note that if a nominal rule 
$\nabla\vdash l\to r$ is closed (i.e., $\nabla\cent (l,r)$ is closed),
then $\nabla\vdash l$
and $\nabla\vdash r$ are both closed terms. Hence:
\begin{itemize}
\item By Property \ref{property:closedCRSterm} both $\hat{l}$ and $\hat{r}$ are closed 
CRS meta-terms.
\item By definition of a nominal rule, the variables in $r$ are also in $l$.
It is easy to see, by induction on  Definition~\ref{def:groundterm}, that 
$\hat{r}$ contains only those meta-variables occurring in $\hat{l}$, and meta-variables occur only in meta-applications where the arguments are lists of bound variables respecting the arity
of the meta-variable (see Property~\ref{pro:arity}).
Moreover, $\tauNabla{l}{l} = \tauNabla{l}{(l,r)}$ and $\tauNabla{r}{r} = \tauNabla{r}{(l,r)}$.
This is because $\Vars(r)\subseteq\Vars (l)$, the rule is closed 
and $\nabla$ is shared by all functions.
\item By definition of a standard rule (see Definition~\ref{def:standard}), $l$ has the form $f s$.
\end{itemize}
Hence $R'$ is a CRS rule (see Section~\ref{sec:CRS}).
\qed

\begin{exa}\label{ie:prenex}
The (closed) nominal rules to compute prenex normal forms (see
Ex\-ample \ref{ex:fol})
can be translated  to CRS rules by application of our algorithm.  We show
the CRS translation computed by our Haskell implementation~(see~\cite{Dominguez2014d}):
\[\begin{array}{rll}
&\mbox{\sf and}(P, \mbox{\sf forall}([a]Q(a))) &\Rightarrow\mbox{\sf forall}( [a]\mbox{\sf and}(P, Q(a)))\\
&\mbox{\sf and}(\mbox{\sf forall}( [a]Q(a)) , P) &\Rightarrow\mbox{\sf forall}( [a]\mbox{\sf and}(Q(a), P))
\\
 &\mbox{\sf or}(P ,\mbox{\sf forall}( [a]Q(a))) &\Rightarrow\mbox{\sf forall}( [a]\mbox{\sf or}(P, Q(a)))\\
& \mbox{\sf or}(\mbox{\sf forall}( [a]Q(a)) , P) &\Rightarrow \mbox{\sf forall}( [a]\mbox{\sf or}(Q(a), P))\\
 &\mbox{\sf and}(P, \mbox{\sf exists}( [a]Q(a))) &\Rightarrow \mbox{\sf exists}( [a] \mbox{\sf and}(P, Q(a)))\\
 &\mbox{\sf and}(\mbox{\sf exists}([a]Q(a)),  P) &\Rightarrow \mbox{\sf exists}([a] \mbox{\sf and}(Q(a), P))
\\
 &\mbox{\sf or}(P,  \mbox{\sf exists}( [a]Q(a)) &\Rightarrow \mbox{\sf exists} ([a] \mbox{\sf or}(P, Q(a)))\\
 &\mbox{\sf or}(\mbox{\sf exists}( [a]Q(a)), P) &\Rightarrow \mbox{\sf exists} [a] \mbox{\sf or}(Q(a), P)
\\
 &\mbox{\sf not}(\mbox{\sf exists}( [a]Q(a))) &\Rightarrow \mbox{\sf forall}( [a] \mbox{\sf not}(Q(a)))\\
 &\mbox{\sf not}(\mbox{\sf forall}( [a]Q(a))) &\Rightarrow \mbox{\sf exists}( [a] \mbox{\sf not}( Q(a))).
\end{array}
\]\medskip

\noindent Note that the nominal variable $P$ becomes the CRS meta-variable $P$ of  arity 0. 
Hence, by definition (see \cite{Klop1993}), if a substitute of P contains the free variable $a$, 
then the bound variable $a$ in the meta-term will be renamed to avoid name clashes.
On the other hand, the nominal variable $Q$ becomes the CRS meta-variable $Q$ of
arity 1, which has the bound variable  $a$ as argument.
\end{exa}
\begin{exa}\label{ie:beta}

The next set of nominal rules are inspired by the simulation of 
$\beta$-reduction and $\eta$-reduction as defined in \cite{Fernandez2010}.
\[\begin{array}{rrlll}
(\beta_{\App}) & &\App(\lCRS([a] \App(X, X')), Y) &\ra\\
&&\hspace{3cm}\lefteqn{\App( \App (\lCRS( [a]X'), Y), \App(\lCRS([a]X), Y))),}\\
(\beta_{\mbox{\sf var}}) && \App(\lCRS([a]a), X) & \ra& X,\\
(\beta_{ \Lambda}) & a\# Y \cent&  \App(\lCRS([a]Y), X) & \ra& Y,\\
(\beta_{\lCRS}) & b\# Y\cent& \App(\lCRS([a] \lCRS ([b]X)), Y) & \ra& \lCRS([b] \App(\lCRS([a]X), Y)),\\
(\eta) & a\# X\cent& \lCRS ([a] \App(X,a)) &\ra& X.
\end{array}
\]
The CRS translation is:
\[\begin{array}{rlll}
(\beta_{\App})&  \App(\lCRS([a] \App(X(a), X'(a))), Y) &\Rightarrow\\
&\hspace{4cm} \lefteqn{ \App( \App (\lCRS( [a]X'(a)), Y), \App(\lCRS([a]X(a)), Y))),}\\
(\beta_{\mbox{\sf var}})&  \App(\lCRS([a]a), X) & \Rightarrow& X,\\
(\beta_{\sf \Lambda})&  \App(\lCRS([a]Y), X) & \Rightarrow& Y,\\
(\beta_{\lCRS})  & \App(\lCRS([a] \lCRS ([b]X(a, b))), Y) & \Rightarrow& \lCRS([b] \App(\lCRS([a]X(a, b)), Y)),\\
(\eta) & \lCRS ([a] \App(X,a)) &\Rightarrow& X.
\end{array}
\]
In rule ($\beta_{\lCRS}$), notice how both occurrences of the meta-variable $X$ share the same ordered list of bound variables, regardless of the fact that in the left-hand side, $[a]$ is above $[b]$ in the syntax tree while in the right-hand side it is the opposite.
This ensures that substitutions work well, as explained in more detail in the next section.
\end{exa}

\section{Simulating Nominal Rewrite Steps}\label{sec:redRel}
We consider next the relationship between the rewriting relation on
nominal terms generated by a set of nominal rules $\mathcal{R}$ and
the rewriting relation on CRS terms generated by its translation. Our goal is to show
that the rewriting relation is preserved when nominal terms and rules are translated to 
CRSs.

Translation of a rewrite relation  is not as straight-forward as one could expect.
The rewriting relation generated by a set of CRS rules is defined on terms, not on 
meta-terms. Recall that 
CRS substitutes are terms, containing no meta-variables, preceded by the 
binder $\underline{\lambda}$ and a list of pairwise distinct variables (the length
of the list corresponds with the arity of the meta-variable it substitutes).
In order to preserve the rewriting relation, we need to  consider only ground nominal substitutions.
Moreover, substitutions are not translated on their own,
but together with the term-in-context to be instantiated
(since permutations are also applied to the  
substitution in order to preserve the meaning of the term).
For this reason,  we will define a translation function for pairs
of a term-in-context $\Delta\vdash t$ and a substitution $\sigma$.

Moreover, there are swappings occurring in a permutation
that can only be dealt with by applying directly the permutation
to the nominal substitution before
translation.
These swappings correspond to mappings from atoms to unabstracted
atoms occurring in the term.
Dealing with these swappings at term level would contradict
the property of closedness of a CRS rule.
Take for instance the example 
\[ (~\cent  \Tran{a}{b}\act X,\; [X \sm f(a,b)]~ ).\]
The term $\cent \Tran{a}{b}\act X$ is trivially closed 
(no unabstracted atoms occur in the term
and there is only one variable). The CRS translation given in Definition
~\ref{def:groundterm} for nominal terms and
Definition \ref{def:withSubs}  for substitution, given below,  produce the pair \[ (~X, [X\sm
  f(b,a)]~)\] where the permutation $\Tran{a}{b}$ has been directly applied to the
  instantiation of $X$, $f(b,a)$ to construct the CRS
substitute and not in the nominal term, since neither $a$ nor $b$ are 
abstracted above $X$.
Further examples are considered after 
Definition~\ref{def:withSubs}, where we present the nominal substitution translation function.

We must ensure a nominal substitution is correctly translated
with respect to the nominal term it instantiates.
For this, we apply the function \texttt{toAscList} equally in
both Definition ~\ref{def:groundterm} and  Definition ~\ref{def:withSubs},
over the set of mappings $\Lambda$ (see Definition~\ref{def:Lambda_t}), 
which produces a fixed and ordered list of  atoms $[a_1, \ldots, a_n]$
for each nominal variable in the term. These lists are added to the 
substitutes for meta-variables, which have the form
 $\underline{\lambda}(a_1, \ldots, a_n).t$.

\begin{defi}[Substitution translation]\label{def:withSubs}
Let $\Delta\vdash t$ be a closed nominal term-in-context, $\Lambda_t$ 
as  in Definition~\ref{def:Lambda_t}, 
and $\sigma$ a nominal substitution satisfying $\Delta$, such that $\sigma =[X_i\mapsto t_i],\, 1\leq i\leq n$ where
$dom(\sigma)\subseteq\Vars (t)$ and $t\sigma$ is ground.

Then  $\mathcal{T}^{\mathcal{s}}(\Delta,t, \sigma) 
= [X_i\mapsto \underline{\lambda} (\overline{xs}_{i}).s_i]$ is defined  as follows:
\begin{itemize}
\item $\overline{xs}_{i}\triangleq \pi_i\cdot xs_i$ and,\\
\item $xs_i\triangleq \mathtt{toAscList}([\pi^{-1}_i\act\Lambda_t(X_i)]-\{a\, |\, a \# X_i\in \Delta\})$,\\
\item $s_i \triangleq \mathcal{T}(\Delta, \pi_i\cdot t_i)$ where $\pi_i$ is the permutation
suspended in the leftmost occurrence of $X_i$ in $t$.
\end{itemize}
Lemma \ref{lem:subs} justifies the use of the leftmost occurrence of
$\pi\cdot X$ in $t$.
Intuitively, each substitute generated by application of the translation function to distinct occurrences of a moderated variable is indeed $\alpha$-equivalent.
Hence the leftmost occurrence is used as a representative. 
\end{defi}

We denote by 
$(\hat{t},\hat{\sigma})$  the result of $(\mathcal{T}(\Delta,t),\mathcal{T}^{\mathcal{s}}(\Delta,t, \sigma))$. 

\begin{exa}
Consider the following pair of a nominal term-in-context and substitution
$$(~t=~\cent f( [a][b]\,  X, [b][a]\,  X),\; \sigma=[X\sm g(a,b)] ~)$$
where $$t\sigma = ~ f([a][b]g(a,b), [b][a]g(a,b)).$$
Then, applying $(\mathcal{T},\mathcal{T}^{\mathcal{s}})$ we obtain the pair 
$$ (~ \hat{t}=\;f([a][b]X(a, b), [b][a]X(a,b)),\; \hat{\sigma}=[X\sm \underline{\lambda}(a.b). g(a,b)]~).$$
The  CRS term $\hat{\sigma}(\hat{t})$ is computed as follows:
\[f([a][b](\underline{\lambda} (a,b). g(a,b))(a, b), [b][a](\underline{\lambda} (a,b). g(a,b))(a,b))\ra_{\underline{\beta}}\] \[f([a][b]g(a,b), [b][a]g(a,b))\]
which corresponds to the nominal term $t\sigma$.
\end{exa}

\begin{exa}
We revisit the nominal term $$\cent[a][b]\Tran{a}{c}\act X$$ given in example \ref{example (ac)}, for a more detailed view on  its translation to CRSs.\\
Assume we are given the pair $$(~\cent[a][b]\Tran{a}{c}\act X,\, \sigma=[X\sm f(a,b,c)]~).$$
Its CRS translation is 
$$(~[a][b]X(b,a),\, \hat{\sigma}=[X\sm \underline{\lambda}(b,a).f(c,b,a)]~)$$ with $xs=[b,c],\; \overline{xs}=[b,a]$
and $\Tran{a}{c}\act \sigma(X) = f(c,b,a)$.
Notice we must apply $\Tran{a}{c}$ to $xs$ in order to permit the capture of
the variable $a$ occurring in $\Tran{a}{c}\act\sigma(X)$.
Thus, the obtained meta-term mimics the behaviour of the nominal term and substitution. 

Later we formally prove the application of permutations to the nominal
substitution $\sigma$ does not affect the uniqueness of $\hat{\sigma}$
modulo $\alpha$, when more than one occurrence of the same
meta-variable appears in the term.  Intuitively, if there are more
occurrences of $X$, Property~\ref{property:closedCRSterm} implies the
initial list of bindings we call $xs$ is syntactically equivalent for
all occurrences of $X$ thus it binds the same atoms.  Considering
$\pi$ is applied to both $xs$ and  $X\sigma$,
renamings do not affect the structure of the binding.  Moreover, for
any other renaming of atoms $a\ \sm\ \pi(a)$ occurring during
application of $\pi\act\sigma(X)$, an identical mapping must exist in
all other occurrences of $\pi\act X$ else it contradicts the property
of closedness (see Definition~\ref{def:closed}) therefore $a\#
X\in\Delta$.

For instance, consider the following pair of a closed term-in-context and a substitution.
$$(a,c\# X\cent g([a][b][c]\Tran{a}{d}\Tran{e}{f}\act X, [a][b][c]\Tran{c}{d}\Tran{e}{f}\act X),  [X\sm f(b,d,e,f)]).$$
The term translation function produces a CRS meta-term 
$$ g([a][b][c]X(b,a),\, [a][b][c]X(b,c))$$
and the substitution translation produces the 
corresponding substitute $$[X\sm \underline{\lambda}(b,a).f(b,a,f,e)].$$
Note that, if translation of the substitution is done with respect to each 
variable in the term, we obtain:\\
 $[X\sm \underline{\lambda}(b,a).f(b,a,f,e)]$ and 
$[X\sm \underline{\lambda}(b,c).f(b,c,f,e)]$ for each occurrence of $X$. 
These substitutes are indeed $\alpha$-equivalent and
our algorithm outputs the leftmost $[X\sm \underline{\lambda}(b,a).f(b,a,f,e)]$.

Moreover, applying the lexical ordering directly to $\overline{xs}$ instead of $xs$
would produce substitutes which are no longer $\alpha$-equivalent, providing
incorrect instantiations.
\end{exa}

We now formalise the property of equivalence modulo $\alpha$ of
substitution occurrences after translation.  The intuition is they are
equivalent because of terms being closed (see
Definition~\ref{def:closed}) and sharing an initially equivalent
variable binding list, $xs$ (see Property~\ref{lem:xsEquivalence}).

\begin{lem}[$\alpha$-equivalence of substitutes]\label{lem:subs}
Let $\Delta\vdash t$ be a closed nominal term-in-context, 
$\Lambda_t$ as defined in  Definition~\ref{def:Lambda_t}, 
and $\sigma$ a  nominal substitution satisfying $\Delta$ such that 
$dom(\sigma) \subseteq V(t)$ and $t\sigma$ is ground.
Let $\pi_i\cdot X,\, \pi_j\cdot X$ be two occurrences of the same variable in $t$, and
let $[X\mapsto \underline{\lambda} (\overline{xs}_{i}).s_i]$ and $[X\mapsto \underline{\lambda} (\overline{xs}_{j}).s_j]$ be translations according to Definition~\ref{def:withSubs} but using $\pi_i$ and $\pi_j$ respectively. 
Then 
$[X\mapsto \underline{\lambda} (\overline{xs}_{i}).s_i]\aleq[X\mapsto \underline{\lambda} (\overline{xs}_{j}).s_j]$.
\end{lem}

\proof
By definition, $\mathcal{T}^{\mathcal{s}}(\Delta,t,[X_i\mapsto t_i])=[X_i\mapsto \underline{\lambda}\;(\overline{xs}_{i}).\; s_i]$ with $\overline{xs}_i\triangleq\pi_i\act xs_i$, 
$\, xs_i\triangleq \mathtt{toAscList}([\pi^{-1}_i\act\Lambda_t(X_i)]-\{a\, |\, a \# X_i\in \Delta\})$ and  $s_i\triangleq \mathcal{T}(\Delta, \pi_i\act t_i)$ where $\pi_i\cdot X_i$ is the leftmost occurrence
of $X_i$.

Hence, each atom $a\in support (\pi_i)$ with $\pi_i(a)\not\in\Lambda_t(X)$ must satisfy
$\pi_i(a)\eq\pi_j(a)$, so that  when $\pi_i,\pi_j$ are applied to each occurrence
of $t_i$ during translation,
they remain equivalent, therefore $s_i\eq s_j$.
Otherwise $a\in ds(\pi_i, \pi_j)$
such that $a\# X\in \Delta$
by Definition~\ref{def:closed}.
Since $\sigma$ must also satisfy $\Delta$,
it is the case that $a\# t_i$, 
and application of either $\pi_i$ or $\pi_j$ to $t_i$ produces no changes.

Now we look at the binding list  added to the substitute (i.e., $\overline{xs}_i,\overline{xs}_j$).
Property \ref{lem:xsEquivalence} states that $xs$ is shared by all occurrences,
and the term is closed, 
then it is the case that each variable in $xs$ binds the same variable in $t_i$ for
both occurrences.
It immediately follows that by application of $\pi_i,\pi_j$ to each
occurrence of $xs$ and $t_i$, the renaming of bound variables does not
affect the binding structure, hence they are $\alpha$-equivalent.
Finally, for any other atom $a\in\Lambda_t(X)$ but $a\not\in\overline{xs}$
it can only be that $\pi^{-1}(a)\# X\in\Delta$ hence $\pi^{-1}(a)\not\in xs$.
It also does not occur
free in $t_i$ since it must satisfy $\Delta$, as previously stated.
Therefore it does not alter the outcome of the translation.

This shows that the choice of the leftmost element in the translation does not affect correctness.
\qed

We are now ready to prove that substitutions are correctly translated.

\begin{lem}[Instantiation]\label{lem:instantiation} 
Let $\Delta\vdash t$ be a closed nominal term-in-context, 
$\Lambda_t$ as defined in Definition \ref{def:Lambda_t}, 
and $\sigma$ a substitution satisfying $\Delta$ such that $dom(\sigma)
\subseteq V(t)$ and $t\sigma$ is ground.

Suppose $(\tauDelta {t'}{t},\mathcal{T^{\mathcal{s}}}(\Delta, t, \sigma))=
(\hat{t}', \hat{\sigma})$,   for  any subterm $t'$ of $t$ (e.g. $t'\eq t$),
is a recursive call in the translation process.

Then $\llbracket t'\sigma\rrbracket^{\Delta}_{\Lambda_t}=
\hat{\sigma}(\hat{t'})$.
\end{lem}

\proof
By induction on the structure of $t'$.
\begin{description}
\item[(atom)]  If $t'\eq a$, the property holds trivially.
\item[(var)] If $t'\eq\piX$, then we distinguish cases with respect to $\Lambda_t(X)$.

\begin{itemize}
\item[(1)] If $\Lambda_t(X)=\emptyset$,
then it immediately follows that $(\llbracket \piX\rrbracket^{\Delta}_{\emptyset},~
\mathcal{T^{\mathcal{s}}}(\Delta, t, \sigma))=
(X, \hat{\sigma})$ where  $\hat{\sigma}(X)\eq s\triangleq \mathcal{T}(\Delta, \pi\act\sigma(X))$ by definition of substitution translation.
Therefore $\hat{\sigma}(X)=s$.
This is equivalent to $\llbracket \piX\sigma\rrbracket^{\Delta}_{\emptyset}=s$.

\item[(2)] If $\Lambda_t(X)\not=\emptyset$, then
 $(\tauDelta{\piX}{t},~\mathcal{T^{\mathcal{s}}}(\Delta, t, \sigma)) =
(X (\overline{xs}),\hat{\sigma}(X)\eq \underline{\lambda}(\overline{xs}).s)$,
where  $s=\mathcal{T}(\Delta, \pi\act\sigma(X))$.
Therefore $(X(\overline{xs}))\hat{\sigma}(X)= (\underline{\lambda}(\overline{xs}).s)(\overline{xs}) =s$.
This is also equivalent to $\tauDelta{(\piX)\sigma(X)}{t}=s$.
\end{itemize}

\item[(abs)] If $t'\eq [a]s$, then
$\tauDelta{([a]s)\sigma}{t}=\tauDelta{[a]s\sigma}{t}=[a]\tauDelta{s\sigma}{t}=[a]\hat{\sigma}(\hat{s})=\hat{\sigma}([a]\hat{s})$ where $\hat{\sigma}(\hat{s})=\tauDelta{s\sigma}{t}$ by the induction hypothesis.

\item[(fun)] If $t'\eq f s$, then
$\tauDelta{(f s)\sigma}{t}=\tauDelta {f (s\sigma)}{t}=f\tauDelta {(s\sigma)}{t}=f(\hat{\sigma}(\hat{s}))=\hat{\sigma}(f\hat{s})$ where $\hat{\sigma}(\hat{s})=\tauDelta{s\sigma}{t}$ by the induction hypothesis.

\item[(tuple)] If $t'\eq(s_1, \ldots, s_n)$, then
$\tauDelta{(s_1, \ldots, s_n)\sigma}{t}=\tauDelta{(s_1\sigma|_{\Vars
    (s_1)}, \ldots, s_n\sigma|_{\Vars (s_n)})}{t}=$\break $(\hat{\sigma}_1(\hat{s}_1), \ldots, \hat{\sigma}_n(\hat{s_n}))$ where each $\hat{\sigma_i}(\hat{s}_i) = \tauDelta{s_i\sigma|_{\Vars (s_i)}}{t}$ by the induction hypothesis.
And $\hat{\sigma}(\hat{s_i})|_{\Vars (s_i)}=\hat{\sigma}_i(\hat{s}_i)$, as a consequence of Lemma \ref{lem:subs}.\qed
\end{description}

\noindent Nominal variable translation depends both on  freshness context and
 abstractions occurring above the variable.
The translation function uses them to build
both the  list of arguments of a meta-application and the list of
binders added to a substitute, whereas the syntax-directed nature of the
translating function transforms the rest of the elements directly.  By
keeping track of the abstractions above a variable via  function
$\Lambda$, translating any subterm $t'$ of $t$ with $\Lambda_t (X)$ 
results in the same term as translating $t'$ within $t$, as stated
in the following lemma.

Note that $C[~]$ is a term, as explained in the paragraph above
Definition~\ref{rewrite-step}, and is translated to $\hat{C}[~]$ using
Definition~\ref{def:groundterm}.

\begin{lem}[Context]\label{lem:context}
Let $\Delta\cent t$ be a nominal term-in-context such that $t\eq C[s]$
(i.e., $t \eq C[-\mapsto s]$), where $s$ is a ground nominal term.
Assume $\tauDelta {C} {C}= \hat{C}$ and $\llbracket s
\rrbracket^{\Delta}_{\emptyset} = \hat{s}$.  Then $\tauDelta {C[s]}{t}
= \hat{C}[\hat{s}]$.
\end{lem}
\proof This is a particular case of Lemma \ref{lem:instantiation},
where $\sigma\eq [-\mapsto s]$. If $\mathcal{T^{\mathcal{s}}} (\Delta,
t, \sigma) = \hat{\sigma}$ then $\tauDelta {C[s]}{t}=\hat{C}[\hat{s}]$
since $s$ is ground, hence $\Lambda_C\eq\Lambda_{t}$.
\qed


We can now derive the main result of the paper: the preservation of the 
rewrite relation under the translation. 
\begin{thm}[Preservation of reduction]
\label{theo:redRel}
Let $R\eq\nabla\vdash l\rightarrow r$ be a standard nominal rule.  Let
$t$ be a ground nominal term and $\hat{t}=\mathcal{T}(\emptyset,t)$.
If $t\rightarrow_R u$ then $\hat{t}\Rightarrow_{R'}\hat{u}$ using
$R'\eq\mathcal{T}^\mathcal{R}(\nabla,l,r)$, and
$\hat{u}=\mathcal{T}(\emptyset,u)$.
\end{thm}

\proof
If  $t\rightarrow_R u$ then there exists $C, \sigma$ such that $t\aleq C[l\sigma]$ with $\sigma$ a ground nominal substitution satisfying $\nabla$ such that $dom(\sigma)\subseteq\Vars (l)$.

Also $R' \eq \mathcal{T}^\mathcal{R}(\nabla,  l, r)\eq\tauNabla{l}{l}\Rightarrow \tauNabla{r}{r}=  \hat{l}\Rightarrow\hat{r}$ by Definition \ref{def:rules}, where Lemma \ref{lem:isRule} asserts that the translation is a CRS rule.

If we have, by application of Definition \ref{def:withSubs}, $\mathcal{T^{\mathcal{s}}}(\nabla, l,\sigma)=\hat{\sigma}_l$  then, by Lemma  \ref{lem:instantiation}  $\tauNabla{l\sigma}{l\sigma}=\hat{\sigma}_l(\hat{l})$.
Hence we have $\hat{t}=\hat{C}[\hat{\sigma}_l(\hat{l})]$ by Lemmas \ref{lem:context} and Property~\ref{theo:alphaCRS}.

Similarly, since $u\aleq C[r\sigma]$ we have $\mathcal{T^{\mathcal{s}}}(\nabla, r, \sigma)=\hat{\sigma_r}$, leading to $\hat{u}= \hat{C}[\hat{\sigma}_r(\hat{r})]$ by application of Definition \ref{def:withSubs}, followed by  Lemmas \ref{lem:instantiation},  \ref{lem:context} and Property~\ref{theo:alphaCRS}.
Notice that $dom(\hat{\sigma}_r)\subseteq dom(\hat{\sigma}_l)$ and $\hat{\sigma}_r (\hat{X})\approx_{\alpha}\hat{\sigma}_l (\hat{X})$ by Lemma \ref{lem:subs}.

Hence we conclude by stating that if $l\sigma\rightarrow_R r\sigma$ then
$\hat{t}\Rightarrow_{R'}\hat{u}$ as expected.
\qed

\begin{cor}[Termination]\label{cor:termination}
Termination of the translated CRS implies termination of the NRS.

\end{cor}
\section{Translating from CRSs to NRSs: An Improved Approach}\label{sec:toNRS}

Our goal is to obtain a tool capable of translating rules back and
forth between CRSs and NRSs.  Based on the CRS to NRS translation
defined in~\cite{Fernandez2004}, we now provide an improved algorithm
to translate CRS rules to closed nominal rules.

Since CRSs follow Barendregt's naming convention (each
abstraction uses a different bound variable  to avoid name
clashes), we will work with closed rewriting (see
Definition~\ref{def:closedR}), where in each rewriting step a
freshened copy of the rule is chosen,  making
the translation easier.

\subsection{Translating  Meta-terms in CRS Rewrite Rules}
We begin by defining a pair of  auxiliary functions.

Function $\Phi$ provides the leftmost meta-application for each
meta-variable occurring in the left-hand side of a CRS rule $l$.
 More precisely,
$\Phi_l(Z_i^n)\eq[a_1,\ldots,a_n]$ if $Z_i^n(a_1,\ldots,a_n)$ is the
leftmost occurrence of $Z_i^n$ in $l$.  
We use it in the translation to ensure the preservation of
both closedness (see Lemmas~\ref{lem:closedLeft} \&
~\ref{lem:closedRight}) and the rewriting relation (see
Theorem~\ref{theo:CRSRewRel}).

The second auxiliary function, $\Psi$, is used to convert each
meta-application $\metaVar{Z}\tuple{a_1,\ldots,a_n}$ occurring
in the left-hand side $l$ of a CRS rule into a list $\pi_i$ of swappings,
such that $\pi_i\act Z_i$, when instantiated,
simulates the $\beta$-reduction of a valuation $\sigma$ applied to
each occurrence of $Z_i^n$.  To accomplish this, $\Psi$ is
parameterised by $\Phi_l(Z_i^n)$, and applied locally to each argument
list $\tuple{b_1,\ldots,b_n}$ in a meta-application of $Z_i^n$ which
is not the leftmost one, in order to preserve $\alpha$-equality along
the NRS translation.

Both auxiliary functions will be used when defining the translation
algorithm for left- and right-hand sides of CRS rules
(Definitions~\ref{def:lCRS} \&~\ref{def:rCRS}).  We start by providing
a formal definition of $\Phi$.

\begin{defi}\label{def:Phi}
Given a closed CRS meta-term $t$, the partial mapping $\Phi_t$ from
meta-variables to lists of variables is defined such that 
\[\Phi_t(Z^n_i) = [a_1,\ldots,a_n]\]
 if the leftmost occurrence of the meta-variable $Z^n_i$ in $t$ has the
 form $Z^n_i(a_1,\ldots,a_n)$, where $a_1,
 \ldots, a_n$ are pairwise distinct bound variables.
We denote by  $\Phi_t(Z^n_i)_k$ the $k$th element in the list $\Phi_t(Z^n_i)$.
\end{defi}


To provide a behaviour similar to that of CRSs, NRSs must maintain the relation among
argument lists occurring for a meta-variable $Z_i^n$ along a CRS meta-term.
This relation is one of position within the argument list such that all (possibly distinct)
variables at a position $k$ (where $1\leq k\leq n$), for each argument list adjacent to an occurrence of $Z_i^n$, are $\beta$-reduced by application of  binder $\underline{\lambda}x_k$ in a substitute for a valuation $\sigma(Z_i^n)$.
Furthermore,  consider a \emph{non left-linear} closed meta-term (i.e., more than one occurrence of a meta-variable exists) of the form $(\abs{a}Z_i^n(a),\abs{b}Z_i^n(b))$.
By the property of $\alpha$-equivalence in CRSs, $\abs{a}Z_i^n(a)$ and $\abs{b}Z_i^n(b)$ are considered syntactically equal.
A direct translation of this meta-term to the NRS term $(\abs{a}Z,\abs{b}Z)$ is not suitable: $\abs{a}Z$ and $\abs{b}Z$ are not $\alpha$-equivalent and, moreover, this is not a closed NRS term.
Both translation issues are approached by the same methodology: making use of the 
NRSs tools to check $\alpha$-equivalence of  terms, that is, swappings and the freshness relation.

To construct a list of swappings for an occurrence of a non-leftmost NRS variable $Z$ 
with respect to both $\Phi_t(Z_i^n)$ and a variable argument list $\tuple{b_1,\ldots,b_n}$ 
adjacent to that same CRS occurrence $Z_i^n$, we need to convert a permutation in two-line
notation into a series of $k$-cycles where $k\geq 2$ ($\Id$ permutations are discarded) followed
by a decomposition into swappings, as explained in Example~\ref{exa:cycleDecomp}.
Notice that the set of variables in $\Phi_t(Z_i^n)$ and $\tuple{b_1,\ldots,b_n}$ can be
equivalent, disjoint or sharing some variables, therefore it is not a bijection per se.
However,  the variables in each list are distinct among them and the length of the list is
the same in both cases, so there is a one-to-one correspondence
between each variable in $\Phi_t(Z_i^n)$ and $\tuple{b_1,\ldots,b_n}$.
This is enough to define a cycle notation (therefore forcing the construction of bijections).

\begin{exa}\label{ex:exoticCRSmetat}
Consider the closed CRS meta-term \[t\eq\abs{c}\tuple{\abs{a}\abs{b}Z\tuple{a,b,c},
\abs{x}\abs{y}Z\tuple{x,c,y}}\]
where  variable name $c$ is shared between meta-applications.
Application of $\Phi$ to $t$ generates $\Phi_t(Z)\eq [a,b,c]$ and, to convert the set of mappings
$\{a\ra x, b\ra c, c\ra y\}$, going from the leftmost argument list to the rightmost one, into a
list of swappings, we transform the implicit two-line notation (variables in the domain of the mapping on the top row and their respective image on the bottom row)
into the cycle notation $\tuple{a,x}\tuple{b,c,y}$ and finally into
the list of swappings $\pi_Z\eq\Tran{y}{b}\Tran{c}{b}\Tran{x}{a}$ as informally described in Section~\ref{sec:Symmetry}.
Then, $\pi_Z(a)\eq x,\pi_Z(b)\eq c,\pi_Z(c)\eq y$ as expected.
Also, $\pi_Z(x)\eq a$ and $\pi_Z(y)\eq c$, however both variables $x,y$ do not belong to the domain
 $\Phi_t(Z)$ of the mappings therefore their image is irrelevant for a correct translation of the permutation list.
Moreover, it will be shown in the translation (Definition~\ref{def:lCRS}) that both atoms
must be fresh for the nominal variable $Z$, i.e., $x\apart Z,y\apart Z\in\Delta$ to preserve closedness of the NRS translation.
Hence one could say that auxiliary function $\Psi$ upgrades \emph{directed paths} (i.e., finite sequence of edges connecting distinct vertices without repetition) to cycles.
\end{exa}
Next is the formal definition of $\Psi$.

When constructing cycles, the main difficulty originates when a variable chosen
by $\Psi$ as the initial element of the sequence is part of a directed path.
As a result, there may be
elements preceding in the sequence which are not reachable by
constructing the sequence rightwards from the initial element.
The solution: construct the sequence leftwards from the initial
element \textit{only} in the case $\Psi$ is not dealing with a real cycle (thus we
must know whether $\Psi$ is dealing with a cycle or a directed path at 
each recursive step).
Here is how.

\begin{defi}\label{def:Psi}
Let $s\eq [a_1,\ldots,a_n]$ and  $t\eq [b_1,\ldots,b_n]$ be any two pairs of lists of
length $n$ over the set $\mathcal{V}$ of variables, and
$f:[a_1,\ldots,a_n]\sm[b_1,\ldots,b_n]$, $f^{-1}:[b_1,\ldots,b_n]\sm[a_1,\ldots,a_n]$ a pair of mappings such that
$b_k\eq f(a_k)$ and $a_k\eq f^{-1}(b_k)$, for $1\leq k\leq n$.
Then, $\Psi(s,t)$ returns a list $\pi$  of swappings over the set $\mathcal{A}$ of atoms,
 recursively defined as follows:
\[
\begin{array}{lcl}
\Psi(nil,nil)&=&\Id\\
\Psi([a_1,\ldots,a_n],[b_1,\ldots,b_n])&=&\Tran{a_m}{b_k}\Tran{a_m}{b_j}\cdots\Tran{a_m}{b_1}\Tran{a_m}{a_1}\Tran{a_m}{a_i}\cdots\Tran{a_m}{a_l}\\&&\circ\ \Psi (s_1,t_1)\quad \mbox{where } 1\leq i,j,k,l,m\leq n \quad \mbox{and}
\end{array}
\]
\begin{itemize}
\item $\Tran{a_m}{b_k}\Tran{a_m}{b_j}\cdots\Tran{a_m}{b_1}\Tran{a_m}{a_1}\Tran{a_m}{a_i}\cdots\Tran{a_m}{a_l}$ is the $2$-cycle decomposition (see Property~\ref{def:toTrans}) of the
permutation in cycle form $C\eq(a_m,a_l,\ldots,a_i,a_1,b_1,\ldots,b_j,b_k)$;
\item $C$ is constructed by successive applications of functions $f$ and $f^{-1}$ 
over $a_1$  (as many times as possible) as follows:
\[
a_m\overset{f^{-1}(a_l)}\longmapsfrom a_l \cdots
\overset{f^{-1}(a_i)}\longmapsfrom a_i \overset{f^{-1}(a_1)}\longmapsfrom a_1
\overset{f(a_1)}\longmapsto b_1 \overset{f(b_1)}\longmapsto\cdots
 b_j \overset{f(b_j)}\longmapsto b_k
\]
where $f^{-1}(a_1)$ and $f(b_j)$ are only applicable when $b_k\not\eq a_1$.
Otherwise, if $b_k\eq a_1$ then the cycle form would be $(a_1,b_1,\ldots,b_j)$, generating then
a  list of swappings $\Tran{a_1}{b_j}\cdots\Tran{a_1}{b_1}$.

\item $s_1\eq s\setminus C$, 
\item $t_1\eq t\setminus C$.
\end{itemize}
\end{defi}

\begin{exa}
Using the definition on the ordered pair of variable lists $[a,b,c],[x,c,y]$ from the above example results in mappings:
\[
f(a)=x\quad f(b)=c\quad f(c)=y \qquad
f^{-1}(x)=a \quad f^{-1}(c)=b \quad f^{-1}(y)=c
\]
Then, function $\Psi$ begins by applying $f$ to variable $a$ such that
\[
a\overset{f(a)}\longmapsto x
\]
and since $x\not\in domain(f), a\not\in domain(f^{-1})$:
\[\Psi([a,b,c],[x,c,y])=\Tran{a}{x}\circ\Psi([b,c],[c,y])\]
where lists $[b,c],[c,y]$ are obtained
 by elimination of variables $a,x$ from both lists.
Finally, 
\[
\Psi([b,c],[c,y])=\Tran{b}{y}\Tran{b}{c}\circ\Id
\]
is created as follows.
Starting with variable $b$,
\[
b\overset{f(b)}\longmapsto c \overset{f(c)}\longmapsto y
\]
where $y\not\in domain(f)$ and $b\not\in domain(f^{-1})$ and thus the
cycle is $(b,c,y)$, converting to $\Tran{b}{y}\Tran{b}{c}$. Also, $\Psi(nil,nil)\eq\Id$.

Notice the positioning of swapping $\Tran{a}{x}$ is different from the previous example solution.
Previously we had $\pi_Z\eq\Tran{y}{b}\Tran{c}{b}\Tran{x}{a}$ whereas, by application
of Definition~\ref{def:Psi}, it resolves to $\Tran{x}{a}\Tran{y}{b}\Tran{c}{b}$ instead ($\Id$ is omitted).
However, Lemma~\ref{lem:disjointCycles} stated that disjoint cycles commute therefore both permutations perform equivalent actions.
\end{exa}

To translate a CRS rule $l \Rightarrow r$, two different functions,
called $\CLeft$ and $\CRight$, are applied to $l$ and $r$ respectively, both
parameterised by $\Phi_l$. We define them separately.
 
The  translation function $\CLeft$
 uses an auxiliary function working on pairs:
$(\Delta,l)_{\Phi_l}^{\Lambda}$, such that
$\CLeft(l)=\comp{(\emptyset,l)}{\emptyset}{l}$ where $l$
is a meta-term and $\Delta$, $\Lambda$ \emph{are recursively
  constructed}.  $\Delta$ has freshness constraints to avoid certain
names appearing, in order to keep the nominal term consistently named
throughout, since there are no naming conventions in NRSs.  $\Lambda$
is a set of variables such that the recursive call $(\comp{\Delta,
  Z^n_i(a_1,\ldots,a_n))}{\Lambda}{l}$ has in $\Lambda$ those
variables bound above $Z_{i}^{n}$.

For any occurrence of a meta-variable $Z^{n}_{i}$ in $l$ that is not 
the leftmost, $\Lambda\setminus\Phi_l(Z^n_i)$ contains the set of
atoms that must be fresh for $Z_i$ in the NRS translation  to translate consistently
from CRSs to NRSs and back again.
That is, $\Lambda\setminus\Phi_l(Z^n_i)$ is the set of all variables bound above $Z^{n}_{i}$ that cannot  occur in the translated NRS
term, otherwise the NRS term is no longer closed (see Definition~\ref{def:closed}).
We provide examples after the formal definition.

\begin{defi}[Left translation]\label{def:lCRS}
Let $s$ be a  closed CRS meta-term where all the meta-applications have the form $Z(a_1,\ldots,a_n)$ such that $a_1,\ldots,a_n$ are different bound variables. Let  $\Phi$ be the function given in Definition~\ref{def:Phi}
and $\Psi$ the function in Definition~\ref{def:Psi}.
Then $\CLeft(s)=\comp{(\emptyset,s)}{\emptyset}{s}$, where $\comp{(\Delta,s)}{\Lambda}{s}$ is inductively
defined as follows: 

\[
\begin{array}{lcl}
\comp{(\Delta,a)}{\Lambda}{s} & = & (\Delta,a)\\[1ex]
\comp{(\Delta,\abs{a}t)}{\Lambda}{s} & = & (\Delta',\abs{a}t'),  \\
 & & \mbox{where } (\Delta',t') = \comp{(\Delta,t)}{\Lambda\cup\{a\}}{s}\\[1ex]
 \comp{(\Delta,f t)}{\Lambda}{s} & = & (\Delta',f t'),  \\
 & & \mbox{where } (\Delta',t') = \comp{(\Delta,t)}{\Lambda}{s}\\[1ex]
\comp{(\Delta,(t_1,\ldots,t_n))}{\Lambda}{s} & = &
(\Delta',\tuple{t'_1,\ldots,t'_n}), \\
 & & \mbox{where } \comp{(\Delta,t_k)}{\Lambda}{s} = (\Delta_k,t'_k),\mbox{ for }1\leq k\leq n\\
 && \mbox{and } \Delta' = \bigcup_k \Delta_k\\[1ex] 
\comp{(\Delta,Z^n_i(a_1,\ldots,a_n))}{\Lambda}{s} & = & (\Delta\cup\Delta',\: {Z_i}) \mbox{ if leftmost occurrence of $Z^n_i$ in $s$}\\
[1ex] 
&&\lefteqn{\mbox{where }\Delta'=\{a\apart Z_i\mid a\in\Lambda\setminus\Phi_s(Z_i^n)\}}\\
\comp{(\Delta,Z^n_i(b_1,\ldots,b_n))}{\Lambda}{s} & =&
(\Delta\cup \Delta', \: \Psi(\Phi_s(Z_i^n),[b_1,\ldots,b_n])\act Z_i) \mbox{ otherwise,}\\
[1ex]
 &&\mbox{where }\Delta' = \{b\apart Z_i\mid b\in\Lambda\setminus\Phi_s(Z_i^n)\}
\end{array}
\]
\end{defi}\medskip

\noindent Therefore, the left-hand side translation function does the following:
for each meta-variable $Z^n_i$ in $s$, if $Z^n_i$ is the leftmost
subterm of the form $Z^n_i(a_1,\ldots,a_n)$, it is replaced by $Z_i$
and for each $a\in\Lambda$ but not in $Z^n_i(a_1,\ldots,a_n)$,
$a\apart Z_i$ is added to $\Delta$, whereas the rest of the subterms
with the form $Z^n_i(b_1,\ldots,b_n)$ are replaced by
$\Psi(\Phi_s(Z^n_i),[b_1,\ldots,b_n])\act Z_i$.  Additionally,
$b_j\apart Z_i$ is added to $\Delta$ for each
$b_j\in\Lambda\setminus\Phi_s(Z^n_i),\, 1\leq j\leq n$, that is, for
any binder with variables distinct to the binders in the leftmost
occurrence, a freshness constraint is added in the translated term to
preserve closedness.  No further freshness constraints are needed
since we are working with closed nominal rewriting.

For the examples in this and next sections, $X,Y,Z,\ldots$ range over the meta-variables instead of $Z^n_i$, without loss  of generality.
This notation is closer to nominal rewriting notation and we also find it more readable.
We write $Z^n_i$ when providing definitions to follow the standard CRS notation.

The following lemma is similar to Lemma~\ref{lem:preserVarsNRS2CRS}, but 
we are now translating from  CRS meta-terms to NRS terms.

\begin{lem}[Preservation of meta-variables as variables in the $\CLeft$ translation]
\label{lem:preserVarsCRS2NRS}
Suppose $t$ is a CRS meta-term and  $(\Delta,t')\eq \CLeft(t)$  its translation
by Definition~\ref{def:lCRS}.
Then,
$Z^n_i\tuple{a_1,\ldots,a_n}$ is an occurrence of a meta-application in $t$ with $Z^n_i$  at position $p$ if and only if
$\pi\act Z_i$ occurs in $t'$ at position $p$.
\end{lem}
\proof
By induction over the structure of $t$ and the fact that there is a one-to-one correspondence between the elements of $t$ and $t'$ as it is shown 
in the syntax-directed translation function in Definition~\ref{def:rCRS}.
Therefore, a meta-variable $Z^n_i$ occurs at position $p$ in $t$,
along with a list of variable arguments $\tuple{a_1,\ldots,a_n}$ (when $n>0$), if and only if there exists a variable $\pi\act Z_i$ at $p$ in its translation $t'$.

\begin{lem}[Preservation of free variables as unabstracted atoms in the $\CLeft$ translation]
\label{lem:preservation of atoms}
Let $l$ be a closed CRS meta-term, $\Phi_l$ as defined in
Definition~\ref{def:Phi} and $(\nabla,l')\eq \CLeft(l)$ as in
Definition~\ref{def:lCRS}.  Assume $\comp{(\Delta,s)}{\Lambda}{l}\eq
(\Delta',s')$ is computed in the translation of $l$, where $s$ is any
subterm of $l$ (e.g. $s\eq l$).  
Then,  $a$ is free in $s$ if and only if $a$
is an unabstracted atom in $s'$.  
Hence, there are no unabstracted
atom subterms in $l'$, since $l$ is closed.
\end{lem}
\proof
By induction on the definition of $\CLeft$.
\begin{description}
\item[(var)] If $s\eq a$ then $\comp{(\Delta,a)}{\Lambda}{l}\eq
  (\Delta, a)$ and the property holds.  Note that, since $l$ is closed,
  $s$ is a strict subterm of $l$ and there exists an abstraction for
  $a$ in $l$ and above $s$.  Therefore $\abs{a}$ occurs above $a$ in
  the translated term too, by the syntax-directed nature of the
  translation function.
\item[(Mvar)] If $s\eq Z^n_i t$,  then $\comp{(\Delta,Z^n_i t)}{\Lambda}{l}\eq(\Delta',\pi\act Z_i)$ and the property holds trivially.
\item[(abs)] If $s\eq\abs{a}t$ then $\comp{(\Delta,\abs{a}t)}{\Lambda}{l}\eq(\Delta',\abs{a}t')$ 
where $(\Delta',t')\eq\comp{(\Delta,t)}{\Lambda\cup\{a\}}{l}$ and the property holds for
$\comp{(\Delta,t)}{\Lambda\cup\{a\}}{l}$ by induction hypothesis.
\item[(fun)] If $s\eq ft$ then $\comp{(\Delta,ft)}{\Lambda}{l}\eq(\Delta',ft')$ 
where $(\Delta',t')\eq\comp{(\Delta,t)}{\Lambda\cup\{a\}}{l}$ and the property holds for
$\comp{(\Delta,t)}{\Lambda\cup\{a\}}{l}$ by induction hypothesis.
\item[(tuple)] If $s\eq\tuple{t_1,\ldots,t_n}$ then $\comp{(\Delta,\tuple{t_1,\ldots,t_n})}{\Lambda}{l}\eq(\Delta',\tuple{t'_1,\ldots,t'_n})$ 
where each $(\Delta'_i,t'_i)\eq\comp{(\Delta,t_i)}{\Lambda}{l}$, $\Delta'\eq\bigcup\Delta'_i$ for $1\leq i\leq n$ and the property holds for
each $\comp{(\Delta,t_i)}{\Lambda}{l}$ by induction hypothesis.
\end{description}
The other direction is similar.
\qed

\begin{exa}\label{ex:arity}
The CRS meta-term $f([a]X, [b]X)$ is translated using $\CLeft$ as
the closed nominal term
$$a\apart X, b\apart X\cent f([a]X, [b]X).$$

Thanks to the use of closed nominal rewriting, less freshness constraints are
needed in the translation than when using (standard) nominal rewriting.
However, in some cases freshness constraints are generated, even if a
translation without freshnesses might be possible. 
For example, the CRS meta-term $f([a]X(a), [b]X(b))$ produces the closed nominal translation
$$b\# X\cent f([a]X, [b]\Tran{a}{b}\act X)$$
where $b\# X$ ensures that the term is closed (see Definition~\ref{def:groundterm}); 
however, $\cent f([a]X, [a] X)$
would also be a correct translation (note that $b\# X\cent f([a]X, [b]\Tran{a}{b}\act X)\aleq f([a]X, [a] X)$).
\end{exa}

The  translation function $\CRight(\cdot)$ for the right-hand side of a CRS rule,
when applied to a closed  meta-term $r$, produces $(\Delta_r,\compil{r}{l})$, where subterms of
the form $Z^n_i(t_1,\ldots,t_n)$ in $r$ are translated as terms of the form
$(\pi\act Z^n_i)[\Phi_r(Z^n_i)_k\sm \compil{t_k}{l}]$ such that,
for the cases where $t_k\in (t_1,\ldots,t_n)$ is
not a variable in $r$, the translation to NRS introduces
 a notation for \emph{explicit atom substitution} by means of an extra set of rules (see Definition~\ref{def:explicitR}) added
 to the given system. These explicit substitution rules are handled separately from the specified rules in the NRS, namely they are
discarded if translating back  into CRSs by application of  the function defined in Section~\ref{sec:rules}.
 
On the other hand, a swapping $\Tran{\Phi_l(Z^n_i)_k}{t_k}$ is added to $\pi$
where $t_k$ is a variable occurring in $r$ 
(i.e. $t_k$ is a variable occurring bound above $Z^n_i$ by definition of a closed CRS meta-term).
Also $\Delta_r$ contains fresh atoms for each bound variable occurring in the term.

We remark that function $\Psi$ is not needed here.
CRSs are assumed to follow the usual naming conventions, that is, different bound variables are used in each abstraction.
Hence, there are no clashes among variable names both  in $\Phi_l(Z^n_i)$ and other occurrences of $Z^n_i$ in 
the right-hand side meta-term.

We use the notation for explicit atom substitution given in~\cite{Fernandez2004}, 
where $t[a\mapsto s]$
is an abbreviation for \texttt{sub}$([a]t, s)$.
We recall below the rules, then continue  by
formalising the  definition of the right-hand side rule translation.

\begin{defi}[Explicit substitution rules]
\label{def:explicitR} 
The following (closed) nominal rewrite rules define the behaviour of
the binary function symbol \texttt{sub}. The notation $t[a\mapsto s]$
is syntactic sugar for \texttt{sub}$([a]t, s)$
\[\begin{array}{lrlll}
(\sigma_{\vr})    &          \cent & a[a\sm X] & \to& X \\ 
(\sigma_\epsilon) & a\apart Y\cent & Y[a\sm X]& \to& Y \\ 
(\sigma_{f})     &           \cent & (f \: X )[a\sm Y]& \to & f X[a\sm Y]
\quad\mbox{for each $f$ in $\Sigma$} \\ 
(\sigma_{prod}) &           \cent &\tuple{X_1,\ldots,X_n}[a\sm Y] &\to& \tuple{X_1[a\sm
    Y],\ldots,X_n[a\sm Y]}\\ 
(\sigma_{abs}) & b\apart Y \cent &(\abs{b}X)[a\sm Y]& \to & \abs{b}(X[a\sm Y])
\end{array}
\]
\end{defi}

\begin{defi}[Right translation]\label{def:rCRS}
Let $l \Rightarrow r$ be a CRS rule. Let $\Phi_l$ be the function defined in 
Definition \ref{def:Phi} applied  to the CRS meta-term $l$.
Then $\CRight(r)=(\Delta_r,\compil{r}{l})$ where
$$\Delta_r = \{a_k \apart Z^n_i \mid a_k\mbox{ occurs bound above } Z^n_i \mbox{ in } r\}$$ and
$\compil{r}{l}$ is  defined by:
\[\begin{array}{lcl}
\compil{a}{l} & = & a\\
\compil{fs}{l} & = & f\compil{s}{l}\\
\compil{\abs{a}s}{l} & = & \abs{a}\compil{s}{l}\\
\compil{(t_1,\ldots,t_n)}{l} & = & \tuple{\compil{t_1}{l},\ldots,\compil{t_n}{l}}\\
\compil{Z^n_i t}{l} & = & \pi\act Z_i\mbox{ with } \pi=\Id\mbox{ if }n=0,\mbox{ otherwise}\\
 &&\lefteqn{n>0, t\eq\tuple{t_1,\ldots,t_n}\mbox{ such that }}\\
 &&\lefteqn{(\pi\act Z_i)[\Phi_l(Z^n_i)_{m1}\sm \compil{t_{m1}}{l}]\ldots[\Phi_l(Z^n_i)_{mk}\sm \compil{t_{mk}}{l}]\mbox{ where}}\\
&&\lefteqn{\pi =\Tran{\Phi_l(Z^n_i)_{j1}}{\compil{ t_{j1}}{l}}\act\ldots\act \Tran{\Phi_l(Z^n_i)_{jk}}{\compil{t_{jk}}{l}},\mbox{ and}}\\
&&\lefteqn{ j1\ldots jk, m1\ldots mk\in\{1,\ldots, n\},}\\
&&\lefteqn{\compil{ t_{j1}}{l},\ldots,\compil{ t_{jk}}{l}\in\mathcal{A}}
\end{array}\qquad\qquad\qquad
\]
\end{defi}\medskip

\noindent Similar to Definition~\ref{def:lCRS},
we take into account abstracted variables which do not appear in the meta-application.

Note that, compared with the translation given in~\cite{Fernandez2004}, we generate
less freshness constraints and  have also improved the flow when translating back and forth between CRS and NRS rules,
namely by converting  explicit substitution among atoms into  swappings added to $\pi$
 (see Example~\ref{exa:diff} and further examples in Section~\ref{sec:examples}).

The following pair of lemmas state that function $\CRight$ also preserves meta-variables as variables and free variables as unabstracted atoms during the translation of a CRS meta-term.

\begin{lem}[Preservation of meta-variables as variables in the $\CRight$ translation]
\label{lem:preserVarsCRS2NRS.right}
Suppose $t$ is a CRS meta-term and  $(\Delta,t')\eq \CRight(t)$  its translation
by Definition~\ref{def:rCRS}.
Then,
$Z^n_i\tuple{t_1,\ldots,t_n}$ is an occurrence of a meta-application in $t$ with $Z^n_i$  at position $p$ if and only if
$\pi\act Z_i$ occurs in $t'$ at position $p$.
\end{lem}
\proof
By induction over the structure of $t$.
The proof is similar to the proof in Lemma~\ref{lem:preserVarsCRS2NRS} and thus omitted.
\qed

\begin{lem}[Preservation of free variables as unabstracted atoms in the $\CRight$ translation]
\label{lem:preservation of atoms.right}
Let $l,r$ be a pair of closed CRS meta-terms, $\Phi_l$ as defined in
Definition~\ref{def:Phi} and $(\Delta,r')\eq \CRight(r)$ as in
Definition~\ref{def:rCRS}.
Assume $\compil{s}{l}\eq s'$ is computed in the translation of $r$, where $s$ is any
subterm of $r$ (e.g. $s\eq r$).  
Then, $a$ is free in $s$ if and only if $a$
is an unabstracted atom in $s'$.  
Hence, there are no unabstracted
atom subterms in $r'$, since $r$ is closed.
\end{lem}
\proof
By induction on the definition of $\CRight$.
The interesting case is that of a meta-variable since the rest of the cases are solved similarly to the proof given in Lemma~\ref{lem:preservation of atoms} (and thus omitted here).
\begin{description}
\item[(Mvar)] If $s\eq Z^n_i t$, then $\compil{Z^n_i t}{l}\eq\pi\act Z_i[b_k\sm t'_k]$
where $t'_k\eq\compil{t_k}{l}$,
for some $b_k\eq\phi_l(Z_i^n)_k$ and $t_k\in t|_p\wedge t_k\not\in\mathcal{V}$ where
$p$ is a position in $t$.
The property holds by induction hypothesis on $t'_k$.\qed
\end{description}\smallskip

\noindent Next, we prove that Definitions~\ref{def:lCRS} and \ref{def:rCRS} produce closed nominal
terms,  separately, and also as part of the translation of  a CRS rule.

\begin{lem}\label{lem:closedLeft}
Let $t$ be the left-hand side of a CRS rule following Barendregt's naming convention (i.e., variable names are pairwise distinct), and $\comp{(\emptyset, t) }{\emptyset}{t}=(\Delta, t')$ its translation as in Definition~\ref{def:lCRS}, where 
$\Phi$ is the function in Definition~\ref{def:Phi}.  
Then $\Nterm'$ is a closed nominal term-in-context.
\end{lem}
\proof
For $\Nterm'$ to be a closed term-in-context (see Definition~\ref{def:closed}) we must prove the following:
\begin{enumerate}
\item no unabstracted atoms occur in $t'$,
\item any pair $\piOf{a}{Z_i}$, $\piOf{b}{Z_i}$ occurring in  $t'$
satisfy conditions 2 \& 3 in Definition~\ref{def:closed}.
\end{enumerate}
Condition $(1)$ holds because, by definition, CRS rules are closed and the translation does
not introduce new atoms as proved in Lemma~\ref{lem:preservation of atoms}.

For condition $(2)$, 
assume $\metaVar{Z}\tuple{a_1,\ldots,a_n}$, $\metaVar{Z}\tuple{b_1,\ldots,b_n}$ are two occurrences of a meta-application for the same meta-variable in $t$ along with their respective list of distinct variables (possibly empty, thus $Z^0_i$), then their translation results in $\piOf{a}{Z_i}$, $\piOf{b}{Z_i}$
(Lemma~\ref{lem:preserVarsCRS2NRS}).\\
We must consider three possible cases: the case where the list of atoms is empty, $(a)$, the case where one of the occurrences is the leftmost, $(b)$, and finally the case where none of the occurrences is the leftmost, $(c)$. No other cases are possible.
\begin{enumerate}

\item[$(a)$] If the list of atoms is empty (thus $Z^0_i)$, both $support(\pi_a),\, support(\pi_b)\eq\emptyset$
in $\piOf{a}{Z_i}$, $\piOf{b}{Z_i}$ by definition of the translation function
where for each variable $a\in\Lambda$ in the recursive call
$\comp{(\Delta, Z^0_i)}{\Lambda}{t}$ (i.e., any abstraction occurring above the meta-variable),
$a\# Z_i$ is added to $\Delta$.
Hence the property of closedness holds for this case.

\item[$(b)$] If one of the occurrences is the leftmost in $t$, for instance $\metaVar{Z}\tuple{a_1,\ldots,a_n}$, then its translation is $\piOf{a}{Z_i}$ with $support(\pi_a)\eq\emptyset$ and $\Phi_t(Z^n_i)=[a_1, \ldots,a_n]$ where $(a_1, \ldots,a_n)\apart Z_i \not\in\Delta$.
On the other hand, $\metaVar{Z}\tuple{b_1,\ldots,b_n}$ is translated as $\piOf{b}{Z}$ with $\Psi(\Phi_t(Z^n_i),[b_1,\ldots,b_n])\eq\pi_b$ and $b_k\apart Z_i\in\Delta$ such that $b_k\in\Lambda\setminus\Phi_t(Z^n_i)$.\\
This means that $(a_1,\ldots,a_n)$ may appear unabstracted in  an instance of $Z_i$ but not any of $\{b_1,\ldots,b_n\}$ distinct from $\{a_1,\ldots,a_n\}$.
If so, $(a_1,\ldots,a_n)$ are all abstracted above $\piOf{a}{Z_i}$ since the meta-term is closed and $(\pi_b(a_1)=b_1,\ldots, \pi_b(a_n)=b_n)$ thus $(a_1,\ldots,a_n)$ are also abstracted above $\piOf{b}{Z_i}$.
For any other atom that may appear in an instance of $Z_i$, the atom is unabstracted above both occurrences.
Hence the property also holds for this case.

\item[$(c)$] If none of these occurrences is leftmost, let
  $\metaVar{Z}\tuple{c_1,\ldots,c_n}$ be the leftmost one.  Now
  $\Phi_t(Z_i^n)=[c_1,\ldots,c_n]$, and 
the translation of both occurrences in question is $\piOf{a}{Z_i}$ with
$\{c_1,\ldots,c_n,a_1,\ldots,a_n\}\!\in\! support(\pi_a)$ and
$\piOf{b}{Z_i}$ with $\{c_1,\ldots,c_n,b_1,\ldots,b_n\}\!\in\! support(\pi_b)$ where $a_k,b_k\apart Z_i\in\Delta$ for any $a_k\in\Lambda_a\setminus\Phi_t(Z^n_i)$ and $b_k\in\Lambda_b\setminus\Phi_t(Z^n_i)$ with 
$\Lambda_a,\Lambda_b$ the set of variables bound over meta-applications $Z^n_i\tuple{a_1,\ldots,a_n}$
and $Z^n_i\tuple{b_1,\ldots,b_n}$, respectively.\\
This means that any of $\{c_1,\ldots,c_n\}$ may appear unabstracted in an instance of  $Z_i$ but not any of
$\{a_1,\ldots,a_n,b_1,\ldots,b_n\}$ distinct from $\{c_1,\ldots,c_n\}$.
If so, $(\pi_a(c_1)=a_1,\ldots, \pi_a(c_n)=a_n)$ and $(\pi_b(c_1)=b_1,\ldots, \pi_b(c_n)=b_n)$ by Definition~\ref{def:Psi},
therefore $(c_1,\ldots,c_n)$ are abstracted above both $\piOf{a}{Z_i}$, $\piOf{b}{Z_i}$ respectively.
For any other atom that may appear in an instance of $Z_i$, the atom is unabstracted above both occurrences.

\end{enumerate}
Hence  $\Nterm'$ is a closed nominal term-in-context as expected.
\qed

\begin{lem}\label{lem:closedRight}
Let $t$ be the right-hand side meta-term of a CRS rule following Barendregt's naming convention,
$\Phi$ the function defined in Definition~\ref{def:Phi} and for each meta-variable $Z^n_i$ in $t$, $Z^n_i\in dom(\Phi_s)$
for the left-hand side meta-term $s$ such that $\Phi_s(Z^n_i)=[a_1,\ldots,a_n]$.

Let $\CRight(t)\eq (\Delta_t,\compil{t}{s})$ be the  translation as in Definition~\ref{def:rCRS},
then $\Delta_t\cent\compil{t}{s}$ is a closed term-in-context.
\end{lem}
\proof
Similarly to Lemma~\ref{lem:closedLeft}, for $\Delta_t\cent\compil{t}{s}$ to be a closed term-in-context we must
prove that no atom subterm  occurs unabstracted in $\compil{t}{s}$.
This is the case  due to CRS rules being closed by definition (see Definition~\ref{def:CRSrule}) and Lemma~\ref{lem:preservation of atoms.right}
stating that no unabstracted atoms are introduced in the term by the translation function $\compil{\cdot}{s}$.
Moreover, it must also be proved that any $\piOf{}{Z_i}$ occurring in  $t'$
satisfy conditions 2 \& 3 in Definition~\ref{def:closed}.
We have already proved this case in Lemma~\ref{lem:closedLeft} where none of the meta-variable occurrences is the leftmost (case $(c)$ in the proof).
Function $\compil{\cdot}{s}$ treats each meta-variable occurrence as a non-leftmost one
and  there exists a leftmost meta-application $Z_i^n\tuple{a_1,\ldots,a_n}$ in $s$.
Then, the result follows  using Lemma~\ref{lem:preserVarsCRS2NRS.right}
and the fact that $\Delta_t$ contains a freshness constraint, $b\apart X$, for
all atoms $b$ abstracted in $t'$ and all variables $X\in\Vars (t')$ (see Definition~\ref{def:rCRS}).

In the right-hand side of a rule, there is the possibility of explicit substitutions of form $r'[a_k\sm u']$ appearing in the translated nominal term where $a_k\in\Phi_s(Z^n_i)$.
Notice that
$r'[a_k\sm u']$ is syntactic sugar for $\mathtt{sub}(\abs{a_k}r',u')$ where $r,u$ are subterms of the closed meta-term $t$.
Then, any free variable in $r,u$ is in the scope of an abstraction in $t$
and, by Lemma~\ref{lem:preservation of atoms.right}, the same happens in their translations
$r'\eq\compil{r}{s}$ and $u'\eq\compil{u}{s}$
\qed

\begin{rem}[CRS term translation]\label{rem:groundCRS}
For any CRS term $t$, $t$ is
also a nominal ground term, trivially, since there are no meta-variables.
\end{rem}

\subsection{Transforming CRS Rules}\label{sec:fromCRSrules}
In this section we show how CRS rules are converted into closed NRS rules.

\begin{defi}\label{def:CRSrule}
We define the translation of the CRS rule $l \Rightarrow r$ as   
$\mathcal{C}^{\mathcal{R}}(l,r)= \Delta\cent l'\ra r'$, where
$\CLeft(l)= (\Delta_l,l')$, $\CRight(r)= (\Delta_r,r')$ and 
$\Delta = \Delta_l \cup \Delta_r$.
\end{defi}
We give some examples to illustrate the definition.

\begin{exa}\label{exa:diff}
The translation of the $\beta$-rule shown in Example~\ref{beta-CRS} according to Definition~\ref{def:CRSrule} is
$$ \cent \App(\lCRS(\abs{a}Z),Z') \ra Z[a\sm Z'].$$
Now, consider  the CRS \emph{differentiation operator} rule as taken from~\cite{Fernandez2004}:
$$\Diff(\abs{a}\Sin(Z(a))) \Rightarrow \abs{b}\Mult(\App(\Diff(\abs{c}Z(c)),b),\Cos(Z(b))).$$
The translation of this rule is
$$b\apart Z, c\apart Z\cent \mathtt{diff}(\abs{a}\mathtt{sin}(Z)) \ra \abs{b}\mathtt{mult}(\mathtt{app}(\mathtt{diff}(\abs{c}\Tran{a}{c}\act Z), b), \mathtt{cos}(\Tran{a}{b}\act Z))$$
where the freshness conditions are needed to preserve closedness, and mappings of atoms on the right-hand side are transformed into permutations.
\end{exa}

Further examples can be found in Section \ref{sec:examples}.

\begin{lem}
\label{lem:transNRS}
Let  $R\eq l\To r$ be a  CRS rule.
If
$\Delta \cent l' \ra r'$ is its translation according to Definition~\ref{def:CRSrule}, then $\Delta \cent l' \ra
r'$ is a closed nominal rewrite rule.
\end{lem}
\proof
Consequence of Lemmas~\ref{lem:closedLeft} and \ref{lem:closedRight}, and the fact that 
$\Phi_l$ 
is shared by the translation functions for both sides of the rule.
\qed

Next, we proceed by providing a mechanism to translate valuations to
nominal substitutions, that is, to convert each CRS substitute into a
ground nominal term.  Recall that we are assuming each binder binds a
different variable (Barendregt's convention), and that CRS terms
coincide with nominal ground terms (Remark~\ref{rem:groundCRS}).

The intuition behind the definition of the translation for a valuation, 
which is used to instantiate a CRS rule, is
 to rename bound variables from the list of binders, $\underline{\lambda}(\cdots)$,
 in a substitute for $Z^n_i$ to match those in the list of arguments for the leftmost meta-application of $Z_i^n$ in the left-hand side of the rule.
\begin{defi}[Valuation translation]
\label{def:valuation2NRS}
Let $t$ be a closed CRS meta-term, $\Phi_t$ as in Definition~\ref{def:Phi},
 and $\sigma$ a safe valuation such that
$\sigma\eq [Z_i^n\sm\underline{\lambda}\tuple{a_1,\ldots,a_n}.s_i]$
for $1\leq i\leq m$ where $dom(\sigma)\subseteq MV(t)$ and $\sigma(t)$ is
ground.
Then, $\langle\sigma\rangle_{\Phi_t}\triangleq[Z_i\sm\pi_i\act s_i]$ 
where $\pi_i\eq\Tran{a_n}{\Phi_t(Z^n_i)_n}\cdots\Tran{a_1}{\Phi_t(Z_i^n)_1}$.
\end{defi}

Below we denote by $(\nabla\cent t',\sigma')$ the result of 
$(\CLeft(t),\langle\sigma\rangle_{\Phi_t})$.

\begin{exa}
\label{ex:CRSvaluation}
We revisit the closed CRS meta-term $t = \abs{c}\tuple{\abs{a}\abs{b}Z\tuple{a,b,c},
\abs{x}\abs{y}Z\tuple{x,c,y}}$  given in Example~\ref{ex:exoticCRSmetat}, adding
a safe valuation $\sigma=[Z\sm\underline{\lambda}(d,e,f).g(d,e,f,z)]$.
where
\[\sigma(t)\eq\abs{c}\tuple{\abs{a}\abs{b}g\tuple{a,b,c,z},
\abs{x}\abs{y}g\tuple{x,c,y,z}}.\]
Then,  \[(\CLeft(t),\langle\sigma\rangle_{\Phi_t})= ( x\apart Z,y\apart Z\cent \abs{c}\tuple{\abs{a}\abs{b}Z,
\abs{x}\abs{y}\Tran{y}{b}\Tran{c}{b}\Tran{x}{a}\act Z},\sigma'\eq[Z\sm g(a,b,c,z)])\]
where $g(a,b,c,z)$ was obtained by applying $\Tran{f}{c}\Tran{e}{b}\Tran{d}{a}$ to $g(d,e,f,z)$.\\\
Since  $\sigma'$ satisfies the freshness constraints (i.e., $\emptyset\cent x\apart\sigma'(Z),y\apart\sigma'(Z)$),
the instantiation resolves to
\[t'\sigma'\eq \abs{c}\tuple{\abs{a}\abs{b}g\tuple{a,b,c,d},
\abs{x}\abs{y}g\tuple{x,c,y,d}}\]
which corresponds with the CRS term $\sigma(t)$.
\end{exa}

A proof of correctness for the translation of valuations follows.
Since there are two distinct functions to translate the left and right
hand side of rules, we must also provide a proof for each.

Lemma~\ref{lem:CLeftInstant} states the correctness property for the
left-hand side instantiation.  Suppose
$(\CLeft(t),\langle\sigma\rangle_{\Phi_t})\eq(\nabla\cent t',
\sigma')$.  The proof of correctness is more involved than the
instantiation proof in Lemma~\ref{lem:instantiation}, due to the
presence of the freshness context $\nabla$.  It is not sufficient to
prove that $\sigma'$ correctly instantiates $t'$, we must
also verify that $\sigma'$ satisfies the constraints in $\nabla$.
This is the case since the valuation $\sigma$ is safe with respect to
the CRS meta-term $t$ by Definition~\ref{def:safety}, hence free
variables in $\sigma$ cannot occur in $t$, and the function
$\langle\cdot\rangle_{\Phi}$ renames bound variables in
$\sigma(Z_i^n)$ to coincide with the list of variables
$\tuple{a_1,\ldots,a_n}$ in the leftmost occurrence of a
meta-application for each meta-variable $Z_i^n$ in $t$.  Recall that
$\CLeft$ creates freshness constraints for all abstractions in a
meta-term, except for the abstractions whose variables occur in the
list of arguments for the leftmost meta-application of each $Z^n_i$.

\begin{lem}[$\CLeft$ instantiation]
\label{lem:CLeftInstant}
Let $l$ be the left-hand side of a CRS rule (hence, a closed CRS meta-term);
$\nabla,\Delta,\Delta'$  freshness contexts and
$\Phi$ the function given in 
Definition~\ref{def:Phi}.
Assume $\CLeft(l) = \nabla\cent l'$ according to Definition~\ref{def:lCRS}, and
let $\sigma$ be a safe valuation such that $dom(\sigma)\subseteq MV(l)$ and
 $\sigma(l)$ is a CRS term.

Suppose $(\comp{(\Delta,s)}{\Lambda}{l},\langle\sigma\rangle_{\Phi_l})\eq((\Delta', s'),\sigma')$, for any subterm $s$ of $l$ (e.g., $s\eq l$), is a recursive call in the translation process.

Then, $\comp{(\emptyset,\sigma(s))}{\Lambda}{l}\eq (\emptyset, s'\sigma')$ where 
 $\sigma'$ satisfies $\Delta'$, that is,  $\cent\Delta'\sigma'$.
\end{lem}
\proof
By induction on the structure of $s$ and the fact that $\nabla\cent l'$ is a closed term-in-context by Lemma~\ref{lem:closedLeft}.
\begin{description}
\item[(var)] If $s\eq a$, the property holds trivially.
\item[(Mvar)] If $s\eq \metaVar{Z}t$, there are three cases to consider:
 $n\eq 0$,  $n>0$ and $\metaVar{Z} t$ is the leftmost occurrence of the meta-variable $Z^n_i$ in $l$,  and  $n>0$ and $\metaVar{Z} t$ is not the leftmost occurrence of $Z_i^n$ in $l$.
No other cases are possible.
We assume, without loss of generality,  $\sigma\eq [Z^n_i\sm \underline{\lambda}(a_1,\ldots,a_n).s_i]$ with $s_i$ a CRS term 
 for all $i$, where distinct variables $a_1,\ldots,a_n$ are bound in $s_i$.
Then,
\begin{enumerate}
\item In the case where $n\eq 0$, that is, the meta-variable has arity 0,
 it is a fact that $t\eq ()$  and also $\sigma\eq [Z^0_i\sm s_i]$.
Therefore,   $\comp{(\emptyset,\sigma(Z_i^0))}{\Lambda}{l}\eq (\emptyset, s_i)$ by Remark~\ref{rem:groundCRS}.
Now,  
$ (\comp{(\Delta,Z_i^0)}{\Lambda}{l},\langle\sigma\rangle_{\Phi_l})$ produces a pair,
 where the substitution is
$\sigma'=[Z_i\sm \pi_i\act s_i]$, and $\Delta'$ contains constraints of the form
 $a\apart Z$ (possibly empty) for each
$a\in\Lambda$ (i.e., for any abstraction occurring above $s$), and 
$\pi_i\eq\Id$ since $t\eq (a_1,\ldots,a_n)\eq ()$.
Then, $Z_i\sigma'\eq s_i$.

Finally, for each $a\apart Z\in\Delta'$, $\emptyset\cent a\apart s_i$ as a result of the safety conditions for rewriting in CRSs stated in Definition~\ref{def:safety},
which, in the nominal framework translates to:
$supp(s_i)\not\eq\Atoms(\Delta'\cent l')$.
Hence $\sigma'$ satisfies $\Delta'$, i.e., $\vdash \Delta'\sigma'$ as required.

\item In the case where $n>0$ and $\metaVar{Z} t$ is the leftmost occurrence of this metavariable in $l$, we proceed as follows: since the meta-variable has arity $n$, 
$t\eq\tuple{b_1,\ldots,b_n}$, where the variables $b_j,1\leq j\leq n,$ are all different, and also different from the variables in $(a_1,\ldots, a_n)$, since we are using Barendregt's convention.
Then,   \[\comp{(\emptyset,\sigma(Z^n_i t))}{\Lambda}{l}\eq 
(\emptyset, (\underline{\lambda}(a_1,\ldots,a_n).s_i)(b_1,\ldots,b_n)\Rightarrow_{\beta} s_i[a_1\sm b_1,\ldots, a_n\sm b_n])\,.\]
Note that \[(\comp{(\Delta,Z_i^n\tuple{b_1,\ldots,b_n})}{\Lambda}{l},\langle\sigma\rangle_{\Phi_l})= ((\Delta', Z_i),\sigma'=[Z_i \sm \pi_i\act s_i])\,,\]
where $\Delta'$ contains constraints of the form  $a\apart Z$ (possibly empty) for each
$a\in\Lambda\setminus\{b_1,\ldots,b_n\}$, and $\pi_i\eq\Tran{a_n}{b_n}\cdots\Tran{a_1}{b_1}$.
Then, $Z_i\sigma'\eq \pi_i\act s_i$ where $b_1,\ldots,b_n\not\in\Atoms(s_i)$ following the convention.

Finally, for each $a\apart Z\in\Delta'$, $\cent a\apart \pi_i\act s_i$
as a result of the safety conditions for rewriting in CRSs (see~Definition~\ref{def:safety});
note that  $a\not\in \{b_1,\ldots,b_n\}$ since this tuple is the leftmost argument list in $l$ for $Z^n_i$.
Hence $\sigma'$ satisfies $\Delta'$, i.e., $\cent \Delta'\sigma'$ as required.

\item In the case where $n>0$ and $\metaVar{Z} t$ is not the leftmost occurrence, the proof 
is similar to the previous case:
$t\eq\tuple{b_1,\ldots,b_n}$, where the  variables $b_j,1\leq j\leq n$ are all different and also different from the variables in $(a_1,\ldots, a_n)$.
Then,   
\[\comp{(\Delta,\sigma(Z^n_i t))}{\Lambda}{l}\eq 
(\emptyset,(\underline{\lambda}(a_1,\ldots,a_n).s_i)(b_1,\ldots,b_n)\Rightarrow_{\beta} s_i[a_1\sm b_1,\ldots, a_n\sm b_n])\,.\]
Also,  we have 
\[(\comp{(\Delta,Z_i^n\tuple{b_1,\ldots,b_n})}{\Lambda}{l},\langle\sigma\rangle_{\Phi_l})=((\Delta'\sigma', \pi'\act Z_i),\sigma'=[Z_i\sm \pi_i\act s_i])\,,\] 
where $\Delta'$ contains constraints of the form $a\apart Z$  (possibly empty) for all
$a\in\Lambda\setminus\Phi_l(Z^n_i)$;
$\pi_i\eq\Tran{a_n}{\Phi_l(Z_i^n)_n}\cdots\Tran{a_1}{\Phi_l(Z_i^n)_1}$ and $\pi'$ maps $\pi'(\Phi_l(Z_i^n)_1)\eq b_1,\ldots,\pi'(\Phi_l(Z_i^n)_n)\eq b_n$ by
Definition~\ref{def:Psi} ($\Psi$).
Then, $(\pi'\act Z_i)\sigma'\eq \pi'\act(\pi_i\act s_i)\eq (\pi'\circ\pi_i)\act s_i$, by application of Definition~\ref{def:permAct},
such that $(\pi'\circ\pi_i)(a_1)\eq b_1,\ldots,(\pi'\circ\pi_i)(a_n)\eq b_n$ where
$\Phi_l(Z_i^n)_1,\ldots \Phi_l(Z_i^n)_n,b_1,\ldots, b_n\not\in\Atoms(s_i)$ by convention.

To conclude, for each $b_i\in\{b_1,\ldots,b_n\}\setminus\Lambda$, $b_i\apart Z\in\Delta'$,
and $\emptyset\cent b_i\apart \pi_i\act s_i$
as a result of the safety conditions for rewriting in CRSs stated in Definition~\ref{def:safety},
thus it is a  fact that $b_i$  does not range over $\{a_1,\ldots,a_n\}\cup\Lambda$.
Hence $\sigma'$ satisfies $\Delta'$, as required.
\end{enumerate}
Then, the result follows.

\item[(abs)] 
If $s\eq \abs{a}t$, then 
$\tuple{\Delta,\abs{a}t}^{\Lambda}_{\Phi_l}= \tuple{\Delta',[a]t'}$, where
$ \tuple{\Delta,t}^{\Lambda\cup\{a\}}_{\Phi_l} = \tuple{\Delta',t'}$. 

By the induction hypothesis, $\tuple{\emptyset,\sigma(t)}^{\Lambda\cup\{a\}}_{\Phi_l} =
\tuple{\emptyset, t'\sigma'}$. Also by induction, 
$\cent\Delta'\sigma'$.
The result follows, since $(\abs{a}t')\sigma'\eq\abs{a}t'\sigma'$.

\item[(fun)] 
If $s\eq ft$, then 
$\tuple{\Delta,ft}^{\Lambda}_{\Phi_l} = (\Delta', ft')$, where $\tuple{\Delta,t}^{\Lambda}_{\Phi_l} = (\Delta', t')$. By induction hypothesis, 
$(\emptyset,t'\sigma')\eq \tuple{\emptyset,\sigma(t)}^{\Lambda}_{\Phi_l}$, and $\cent\Delta'\sigma'$. The result follows, since $\sigma(ft) = f\sigma(t)$.

\item[(tuple)] If $s\eq \tuple{s_1,\ldots ,s_n}$, then 
$\tuple{\Delta,\tuple{s_1,\ldots ,s_n}}^{\Lambda}_{\Phi_l} = (\Delta',\tuple{s'_1,\ldots, s'_n})$, where $\tuple{\Delta,s_i}^{\Lambda}_{\Phi_l}= (\Delta'_i,s'_i)$ and $\Delta'\eq\Delta'_1\cup\cdots\cup\Delta'_n$. By induction, $\tuple{\emptyset,\sigma(s_i)}^{\Lambda}_{\Phi_l}= s_i\sigma'$, and $\cent\Delta'_i\sigma'$ for $1\leq i\leq n$. Therefore
 $\cent\Delta'\sigma'$, and the result follows since $\tuple{s'_1,\ldots, s'_n}\sigma' = \tuple{s'_1\sigma',\ldots, s'_n\sigma'}$.\qed
\end{description}\medskip

\noindent The proof for function $\CRight$ is similar to $\CLeft$ because of the syntax-directed nature of the translation. However, there is a difference in the case of a meta-application $Z_i^n t$, because in the right-hand side of a CRS rule the list of arguments $t$ may contain any kind of CRS meta-term, not just variables. The translation function $\CRight$ deals with non-variable arguments by introducing explicit substitutions for atoms. Unlike Lemma~\ref{lem:CLeftInstant}, the instantiation lemma for $\CRight$ relies on the use of the explicit substitution rules (see Definition~\ref{def:explicitR}).  We first state the correctness property for the rules in Definition~\ref{def:explicitR}, namely that the rules indeed specify the non-capturing atom-substitution operation.

\begin{defi}[$\sigma$-normal form of a term-in-context]
\label{def:nfTerm}
We denote by $\nf(\Nterm)$ the normal form of the term-in-context $\Nterm$
with respect to the rules in Definition~\ref{def:explicitR}. It is
uniquely defined modulo $\alpha$-equivalence~\cite{Fernandez2004}.
\end{defi}

\begin{lem}[Correctness of explicit substitution rules]
\label{lem:corrsub}
Let $t$,  $s$ be  CRS terms (and therefore also nominal terms).
Then $\nf(t[a\sm s]) \aleq t\{a\sm s\}$, where in the right-hand side 
$t\{a\sm s\}$ denotes the term obtained by substituting (using the capture-avoiding
 substitution of the CRS) $a$ by $s$ in $t$.
\end{lem}
\proof
The proof by induction is omitted since it is standard in explicit substitution calculi.
\qed

We have already proved in Lemma~\ref{lem:closedRight} that $\CRight$
translates closed CRS meta-terms to closed NRS terms, and thanks to
the safety conditions (see Definition~\ref{def:safety}) and
Barendregt's convention (see Remark~\ref{rem:Barendregt}), bound
variable names are all different, and also different from free
variables, both in meta-terms and valuations.  This ensures that, in
the nominal translation, the  explicit substitutions for atoms
preserve the semantics of terms.

\begin{lem}[$\CRight$ instantiation]\label{lem:CRightInst}
Let $l \Rightarrow r$ be a CRS rule and
$\Phi_l$ the function given in 
Definition~\ref{def:Phi} applied to $l$.
Assume $\CRight(r)\eq\Delta\cent r'$, 
where according to  Definition~\ref{def:rCRS}
$\Delta\eq\{a_k\apart Z^n_i\mid a_k \mbox{ occurs bound above } Z^n_i \mbox{ in } r\}$.
Let $\sigma$ be a safe valuation such that $dom(\sigma) \subseteq MV(l)$ (hence,
$\sigma(r)$ is a CRS term).

Suppose $s'\eq\compil{s}{l}$ is a recursive call in the translation of $\CRight(r)$, for any
subterm $s$ of $r$ (e.g., $s\eq r$), and $\sigma'\eq\langle\sigma\rangle_{\phi_l}$ by Definition~\ref{def:valuation2NRS}.

Then, $\compil{\sigma(s)}{\emptyset}\aleq \nf(s'\sigma')$ and 
$\sigma'$ satisfies $\Delta$, i.e.,
$\cent\Delta\sigma'$.
\end{lem}
\proof
By induction on the structure of $s$, using the fact that $\Delta\cent r'$ is  closed  by Lemma~\ref{lem:closedRight},
the syntactic equivalence between CRS terms and ground nominal terms (Remark~\ref{rem:groundCRS})
and the safety conditions given in Definition~\ref{def:safety} and Remark~\ref{rem:Barendregt}.

Interesting cases to consider are $s\eq\tuple{s_1,\ldots,s_n}$ and also $s \eq Z^n_i t$.
\begin{itemize}
\item The case $s\eq\tuple{s_1,\ldots,s_n}$.\
By Definition~\ref{def:rCRS}
 we have $(\compil{\tuple{s_1,\ldots,s_n}}{l}\eq\tuple{s'_1,\ldots,s'_n}$
where
$\Vars \tuple{s'_1,\ldots,s'_n}\subseteq dom(\sigma')$ and,
by Remark~\ref{rem:Barendregt}, $\Atoms (img(\sigma'))\cap\Atoms \tuple{s'_1,\ldots,s'_n}\eq\emptyset$
such that $\cent \Delta\sigma'|_{\Vars \tuple{s'_1,\ldots,s'_n}}$ follows by induction hypothesis.

Now, \[\tuple{s'_1,\ldots,s'_n}\sigma'\eq\tuple{s'_1\sigma',\ldots,s'_n\sigma'}\quad\mbox{and}\quad
\compil{\sigma\tuple{s_1,\ldots,s_n}}{\emptyset}\eq
\tuple{\compil{\sigma(s_1)}{\emptyset},\ldots,\compil{\sigma(s_n)}{\emptyset}}\] such that
\[\compil{\sigma(s_1)}{\emptyset}\aleq\nf(s'_1\sigma'),\ldots,
\compil{\sigma(s_n)}{\emptyset}\aleq\nf(s'_n\sigma')\]
by induction hypothesis.\smallskip

\item The case ($s \eq Z^n_i t$).
For the case where $t\eq ()$, the result follows similarly to case (1) of (Mvar) in Lemma~\ref{lem:CLeftInstant}

For the case where $t\eq\tuple{t_1,\ldots,t_n}$,
there is  $\Phi_l(Z^n_i)=[a_1,\ldots, a_n]$ by Definition~\ref{def:Phi}
where $[a_1,\ldots,a_n]$ is a list of distinct atoms by Remark~\ref{rem:groundCRS} and then
\[\compil{\metaVar{Z} \tuple{t_1,\ldots,t_n}}{l}= (\Tran{a_{j1}}{\compil{t_{j1}}{l}}\cdots\Tran{a_{jk}}{\compil{t_{jk}}{l}}\act{Z_i})[a_{m1}\sm \compil{t_{m1}}{l}]\ldots[a_{mk}\sm \compil{t_{mk}}{l}]\,,\]
where $j1,\ldots,jk, m1,\ldots,mk\in\{1...n\}$ and $\compil{t_{j1}}{l},\ldots,\compil{t_{jk}}{l}\in\mathcal{A}$ by Definition~\ref{def:rCRS}.

Assume, without loss of generality, $\sigma(Z^{n}_i)\eq \underline{\lambda}(b_{1} \ldots b_{n}).s_i$.
Following Definition~\ref{def:valuation2NRS} for $\langle\sigma\rangle_{\Phi_l}$, $\sigma'(Z_i)\eq \Tran{a_1}{b_{1}}\cdots\Tran{a_n}{b_{n}}\act s_i$
where $s_i$ is a NRS ground term by Remark~\ref{rem:groundCRS} where $\{a_1,\ldots,a_n\}\not\in\Atoms (s_i)$ by Remark~\ref{rem:Barendregt}.
Moreover, following the same remark  we have
$V(\sigma)\cap V(r)\eq\emptyset$, therefore it is the case that $\sigma'|_{Z_i}$ satisfies $\Delta$, $\cent \Delta\sigma'|_{Z_i}$.

Hence, by application of explicit rules in Definition~\ref{def:explicitR},
$\nf(s'\sigma')\aleq$\\
 $s_i\{b_1\sm \compil{t_{1}}{l}\sigma'\}\cdots\{b_n\sm \compil{t_{n}}{l}\sigma'\}$
since  $\nf(\cdot)$ correctly  computes the substitution as stated in Lemma~\ref{lem:corrsub}.

Now, $\compil{\sigma(Z_i^{n}\tuple{t_1,\ldots,t_n})}{\emptyset}\eq s_i\{b_1\sm \compil{\sigma(t_1)}{l}\}\cdots\{b_n\sm \compil{\sigma(t_n)}{l}\}$, and the result follows by induction hypothesis.
\end{itemize}
The rest of the cases are omitted.
\qed

Let us denote by $\mathcal{R}$ the nominal rewriting system obtained
by translating all the rules of the CRS $R$ according to
Definition~\ref{def:CRSrule} and adding 
the  rules for explicit substitution in Definition~\ref{def:explicitR}.

We are now ready  to prove the correctness of the translated reduction relation.

\begin{thm}[Translation of CRS Rewrite Steps]\label{theo:CRSRewRel}
Let $R\eq l\To r$ be a CRS rule.
Let $u$ be a CRS term.\\
If $u\To_R v$ then $~\cent u\xrightarrow[\mathcal{R\cup\sigma}]{+}^c v$ using  $\mathcal{R}=\mathcal{C}^{\mathcal{R}}(l,r)$
and the explicit substitution rules.
\end{thm}

\proof
By 
Definition~\ref{def:CRSrule}, 
$\mathcal{R}=\mathcal{C}^{\mathcal{R}}(l,r)=\nabla_l\cup\nabla_r\cent l'\ra r'$. It is 
a closed NRS rule by Lemma~\ref{lem:transNRS}.
Since $u,v$ are terms in CRS $R$, they are also ground terms in 
NRS $\mathcal{R}$ by Remark~\ref{rem:groundCRS}, and without loss of 
generality we can assume that $\nabla_l\cup\nabla_r\cent l'\ra r'$ does not
 mention any atom in $u$ (i.e., it is already freshened for $u$).
If $u\To_R v$ then there exists $p,\sigma$ such that 
$u|_p=\sigma(l)$ where $p$ is a position in $u$, $\sigma$ a 
CRS valuation where   $MV(l)= dom(\sigma)$ and $v=u[\sigma(r)]_p$.
Let $\sigma'\eq\langle\sigma\rangle_{\Phi_l}$  according to
 Definition~\ref{def:valuation2NRS}. 
By Lemma~\ref{lem:CLeftInstant}, if $\CLeft(l) = (\nabla_l,l')$ then 
$\CLeft (l\sigma)= (\emptyset, l'\sigma')$ where $\sigma'$
satisfies $\nabla_l$.
Then,
 it is also the case that $\cent u|_p\aleq l'\sigma'$ by 
Remark~\ref{rem:groundCRS}. 
Moreover, $\sigma'$ also satisfies $\nabla_r$ since we are using Barendregt convention.

It remains to prove that $u[r'\sigma']_p\to^* v$.
But this follows from Lemma~\ref{lem:CRightInst}, since 
$\CRight (\sigma(r))=(\emptyset,  \nf{(r'\sigma')})$ 
where  $\CRight (r)=(\nabla_r, r')$. Hence, 
$u\To_R v$ with $v=u[\sigma(r)]_p$ implies 
$~\cent u\xrightarrow[\mathcal{R\cup\sigma}]{+}^c v$ where $v=u[nf_{\sigma}(r'\sigma')]_p$.
\qed

 We have designed an algorithm that correctly transforms CRS rules inot
 NRS closed rules. It improves over the function defined
 in~\cite{Fernandez2004} in two ways: we have fixed a bug in the
 translation of  meta-variables in the scope of abstractions, and by
 using closed rewriting (see Definition~\ref{def:closedR}) we are able
 to simulate the variable naming convention without adding extra
 freshness constraints, as shown in Example~\ref{exa:diff}.

Next, we provide examples where both transformations are applied
(NRSs to CRSs and back).

\section{Examples}\label{sec:examples}

After describing the tools required to translate NRSs to CRSs (section \ref{sec:rules}) and back to NRSs (section \ref{sec:toNRS}), in this section we give two examples, which have been translated using the implementation available from~\cite{Dominguez2014d}.

\subsection{Prenex Normal Form}

We present here a translation back to NRSs by application of Definition~\ref{def:CRSrule} to
the CRS rules obtained in Example~\ref{ie:prenex}.
Beforehand  we have applied the usual naming  convention in rules (renaming bound variable $a$ to $b$ on the right-hand side). The resulting NRS is:
\[\begin{array}{lcll}
b\apart P, b\apart Q  &\cent & \mbox{\sf and}(P, \mbox{\sf forall}([a]Q)) &\ra \mbox{\sf forall}( [b] \mbox{\sf and}(P, (ab)\act Q))\\
b\apart P, b\apart Q &\cent & \mbox{\sf and}(\mbox{\sf forall}( [a]Q) , P) &\ra \mbox{\sf forall}( [b] \mbox{\sf and}((ab)\act Q, P))
\\
b\apart P, b\apart Q &\cent & \mbox{\sf or}(P , \mbox{\sf forall}( [a]Q)) &\ra \mbox{\sf forall}( [b] \mbox{\sf or}(P, (ab)\act Q))\\
b\apart P, b\apart Q &\cent & \mbox{\sf or}(\mbox{\sf forall}( [a]Q) , P) &\ra \mbox{\sf forall}( [b] \mbox{\sf or}((ab)\act Q, P))\\

b\apart P, b\apart Q &\cent & \mbox{\sf and}(P, \mbox{\sf exist
s}( [a]Q)) &\ra \mbox{\sf exists}( [b] \mbox{\sf and}(P, (ab)\act Q))\\
b\apart P, b\apart Q &\cent & \mbox{\sf and}(\mbox{\sf exists}([a]Q),  P) &\ra \mbox{\sf exists}([b] \mbox{\sf and}((ab)\act Q, P))
\\
b\apart P, b\apart Q &\cent & \mbox{\sf or}(P,  \mbox{\sf exists}( [a]Q) &\ra \mbox{\sf exists} ([b] \mbox{\sf or}(P, (ab)\act Q))\\
b\apart P, b\apart Q &\cent & \mbox{\sf or}(\mbox{\sf exists}( [a]Q), P) &\ra \mbox{\sf exists} [b] \mbox{\sf or}((ab)\act Q, P)
\\
~~~~	~~~b\apart Q &\cent & \mbox{\sf not}(\mbox{\sf exists}( [a]Q)) &\ra \mbox{\sf forall}( [b] \mbox{\sf not}((ab)\act Q))\\
~~~~	~~~b\apart Q &\cent & \mbox{\sf not}(\mbox{\sf forall}( [a]Q)) &\ra \mbox{\sf exists}( [b] \mbox{\sf not}((ab)\act Q)).
\end{array}
\]\smallskip

\noindent Notice that a matching $\sigma=[Q\mapsto a]$ with the left-hand side of a rule leads to the expected result when applied to the right-hand side of the rule thanks to the swapping $(a\ b)$.
Also, the freshness condition $a\apart P$ in the initial set of NRS rules (see Example~\ref{ex:fol}) is shown here as $b\apart P$
because of the variable convention applied beforehand,
$b\apart Q$ is added to the set of freshness conditions (so, the rules remain closed).
This does not alter the semantics, as we can see by translating back. When we translate
 back to CRS the first NRS rule \[b\apart P, b\apart Q  \cent  \mbox{\sf and}(P, \mbox{\sf forall}([a]Q)) \ra \mbox{\sf forall}( [b] \mbox{\sf and}(P, (ab)\act Q))\] using Definition~\ref{def:rules}, we obtain the CRS rule \[ \mbox{\sf and}(P, \mbox{\sf forall}([a]Q(a))) \ra \mbox{\sf forall}( [b] \mbox{\sf and}(P, Q(b)))\] as expected.

\subsection{Simulating $\beta$-reduction}

Consider the $(\beta_{\sf lam})$ rule given in Example~\ref{ie:beta}.
First, we apply Barendregt's convention to the CRS rule so that each bound variable is distinct, obtaining:
$$\App(\lCRS([a] \lCRS ([b]X(a, b))), Y)  \Rightarrow \lCRS([d] \App(\lCRS([c]X(c, d)), Y))$$
Its translation to NRSs is the following:
$$   d\apart Y, d\apart X, c\apart X\cent \App(\lCRS[a] \lCRS [b]X, Y)  \ra
\lCRS[d] \App(\lCRS[c]\Tran{a}{c}\Tran{b}{d}\act X, Y)$$
And when translated back to CRSs by means of Definition~\ref{def:rules},
we obtain the same CRS rule as expected.

\section{Conclusions and Future Work}\label{sec:conclusions}
We have shown two extensions of first-order rewriting, CRSs and NRSs,
to be closely related.  The main differences are in the meta-language
used, NRSs do not rely on the $\lambda$-calculus, employing instead
permutations of atoms and a freshness relation to axiomatise
$\alpha$-equivalence. Also NRS rules are more general than CRS rules
in that unabstracted atoms may occur in rules.  On the other hand,
CRSs rely on the $\lambda$-calculus to generate its rewrite relation
and CRS rules are closed by definition.

We have shown that despite these differences it is possible to
translate between these formalisms, under certain conditions. We have
given a translation function which transforms the class of closed
nominal rewriting systems into CRS systems.  We have shown some
non-trivial examples to support our work;
see~\cite{Dominguez2014d,Dominguez2014c} for a Haskell implementation.

Moreover, existing translation algorithms between CRSs and  HRSs~\cite{Oostrom1994}, $\rho$-calculus~\cite{Bertolissi2006} and ERSs~\cite{Glauert2005} allow transformations from NRSs to any of these systems and vice versa.

Although previous translations between nominal and higher-order syntax exist~\cite{Levy2012,Cheney2005}, our work differs from these by focusing on a syntax-directed translation of terms and rewrite rules that preserves the rewriting relation, which is key to the translation of properties such as confluence and termination.
We have also corrected and  improved a previous algorithm translating CRSs
to NRSs originally found in \cite{Fernandez2004}.

Since nominal terms have good algorithmic properties,  
we could translate CRSs to NRSs in order to take advantage of 
existing nominal procedures (i.e., orderings, completion) and 
then transfer back the results.
The extension to typed systems (adapting the type systems developed for 
NRSs~\cite{Fairweather2011,Fernandez2007a} to  CRSs) is also left for future work.

\paragraph*{Acknowledgements}
We thank Elliot Fairweather and Christian Urban for many helpful
discussions, and Jamie Gabbay for providing the macro for $\new$.  We
also thank the anonymous referees for their valuable comments.

\bibliographystyle{abbrv}
\bibliography{long,BibJournal}

\begin{thebibliography}{10}

\bibitem{Baader1988}
F.~Baader and T.~Nipkow.
\newblock {\em Term rewriting and all that}.
\newblock Cambridge University Press, 1988.

\bibitem{Barendregt1984}
H.~P. Barendregt.
\newblock {\em The lambda calculus, its syntax and semantics}, volume 103 of
  {\em Studies in Logic and the Foundation of Mathematics}.
\newblock North-Holland, revised edition, 1984.

\bibitem{Bertolissi2006}
C.~Bertolissi, H.~Cirstea, and C.~Kirchner.
\newblock Expressing combinatory reduction systems derivations in the rewriting
  calculus.
\newblock {\em Higher-Order and Symbolic Computation}, 19:00110869, 2006.

\bibitem{Dershowitz2003}
M.~Bezem, J.~W. Klop, and R.~de~Vrijer, editors.
\newblock {\em Term rewrite systems by {TeReSe}}, volume~55 of {\em Cambridge
  Tracts in Theoretical Computer Science}.
\newblock Cambridge University Press, 2003.

\bibitem{Calves2009}
C.~Calv\`es and M.~Fern\'andez.
\newblock Matching and alpha-equivalence check for nominal terms.
\newblock {\em Journal of Computer and System Sciences}, 2009.
\newblock Special issue: Selected papers from WOLLIC 2008.

\bibitem{Cheney2005}
J.~Cheney.
\newblock Relating nominal and higher-order pattern unification.
\newblock In {\em Proceedings of International Workshop in Unification}, pages
  104--119, 2005.

\bibitem{Cirstea2001}
H.~Cirstea and C.~Kirchner.
\newblock The rewriting calculus --- {Part~I}.
\newblock {\em Logic Journal of the Interest Group in Pure and Applied Logics},
  9(3):427--463, May 2001.

\bibitem{Cirstea2001a}
H.~Cirstea and C.~Kirchner.
\newblock The rewriting calculus - {Part~II}.
\newblock {\em IGPL}, 9(3):377--410, 2001.

\bibitem{Dominguez2014c}
J.~Dom\'{\i}nguez.
\newblock A tool to apply nominal recursive path ordering to nominal rules.
\newblock 2014.
\newblock Available from http://www.inf.kcl.ac.uk/pg/domijesu/nrpo.tgz.

\bibitem{Dominguez2014d}
J.~Dom\'{\i}nguez.
\newblock A tool to translate between closed {NRSs} and {CRSs}.
\newblock 2014.
\newblock Available from http://www.inf.kcl.ac.uk/pg/domijesu/NRS2CRS.tar.gz.

\bibitem{Dominguez2014b}
J.~Dom{\'{\i}}nguez and M.~Fern{\'{a}}ndez.
\newblock Relating nominal and higher-order rewriting.
\newblock In {\em Mathematical Foundations of Computer Science 2014 - 39th
  International Symposium, Budapest, Hungary, August 25-29, 2014. Proceedings,
  Part {I}}, pages 244--255, 2014.

\bibitem{Fairweather2011}
E.~Fairweather, M.~Fern\'andez, and M.~J. Gabbay.
\newblock Principal types for nominal theories.
\newblock {\em Lecture Notes in Computer Science}, 6914 LNCS:160--172, 2011.

\bibitem{Fernandez2007a}
M.~Fern\'{a}ndez and M.~J. Gabbay.
\newblock Curry-style types for nominal terms.
\newblock In T.~Altenkirch and C.~McBride, editors, {\em Types for Proofs and
  Programs}, volume 4502 of {\em Lecture Notes in Computer Science}, pages
  125--139. Springer Berlin Heidelberg, 2007.

\bibitem{Fernandez2007}
M.~Fern\'{a}ndez and M.~J. Gabbay.
\newblock Nominal rewriting.
\newblock {\em Information and Computation}, 205(6):917--965, 2007.

\bibitem{Fernandez2010}
M.~Fern{\'a}ndez and M.~J. Gabbay.
\newblock Closed nominal rewriting and efficiently computable nominal algebra
  equality.
\newblock In {\em Proceedings of International Workshop on Logical Frameworks
  and Meta-Languages: Theory and Practice}, pages 37--51, 2010.

\bibitem{Fernandez2004}
M.~Fern\'{a}ndez, M.~J. Gabbay, and I.~Mackie.
\newblock Nominal rewriting systems.
\newblock In {\em Proceedings of the 6th ACM SIGPLAN international conference
  on Principles and practice of declarative programming}, PPDP '04, pages
  108--119, New York, NY, USA, 2004. ACM.

\bibitem{Fernandez2012}
M.~Fern\'{a}ndez and A.~Rubio.
\newblock Nominal completion for rewrite systems with binders.
\newblock In A.~Czumaj, K.~Mehlhorn, A.~Pitts, and R.~Wattenhofer, editors,
  {\em Automata, Languages, and Programming}, volume 7392 of {\em Lecture Notes
  in Computer Science}, pages 201--213. Springer Berlin Heidelberg, 2012.

\bibitem{Gabbay2008}
M.~J. Gabbay and A.~Mathijssen.
\newblock Capture-avoiding substitution as a nominal algebra.
\newblock {\em Formal Aspects of Computation}, 20(4-5):451--479, 2008.

\bibitem{Gabbay2002}
M.~J. Gabbay and A.~M. Pitts.
\newblock A new approach to abstract syntax with variable binding.
\newblock {\em Formal Aspects of Computing}, 13(3-5):341--363, 2002.

\bibitem{Glauert2005}
J.~Glauert, D.~Kesner, and Z.~Khasidashvili.
\newblock Expression reduction systems and extensions: An overview.
\newblock In A.~Middeldorp, V.~van Oostrom, F.~van Raamsdonk, and R.~Vrijer,
  editors, {\em Processes, Terms and Cycles: Steps on the Road to Infinity},
  volume 3838 of {\em Lecture Notes in Computer Science}, pages 496--553.
  Springer Berlin Heidelberg, 2005.

\bibitem{Hamana2010}
M.~Hamana.
\newblock Semantic labelling for proving termination of combinatory reduction
  systems.
\newblock In S.~Escobar, editor, {\em Functional and Constraint Logic
  Programming}, volume 5979 of {\em Lecture Notes in Computer Science}, pages
  62--78. Springer Berlin Heidelberg, 2010.

\bibitem{Jouannaud2005}
J.-P. Jouannaud.
\newblock Higher-order rewriting: Framework, confluence and termination.
\newblock In A.~Middeldorp, V.~van Oostrom, F.~van Raamsdonk, and R.~Vrijer,
  editors, {\em Processes, Terms and Cycles: Steps on the Road to Infinity},
  volume 3838 of {\em Lecture Notes in Computer Science}, pages 224--250.
  Springer Berlin Heidelberg, 2005.

\bibitem{Khasidashvili1990}
Z.~Khasidashvili.
\newblock Expression reduction systems.
\newblock In {\em Proc. of I. Vekua Institute of Applied Mathematics},
  volume~36, pages 200--220, 1990.

\bibitem{Klop1980}
J.~W. Klop.
\newblock {\em Combinatory reduction systems}.
\newblock PhD thesis, Utrecht University, 1980.

\bibitem{Klop1993}
J.~W. Klop, V.~{van Oostrom}, and F.~{van Raamsdonk}.
\newblock Combinatory reduction systems: Introduction and survey.
\newblock {\em Theoretical Computer Science}, 121:279--308, 1993.

\bibitem{Kop2011}
C.~Kop.
\newblock Simplifying algebraic functional systems.
\newblock In F.~Winkler, editor, {\em Algebraic Informatics}, volume 6742 of
  {\em Lecture Notes in Computer Science}, pages 201--215. Springer Berlin
  Heidelberg, 2011.

\bibitem{Levy2012}
J.~Levy and M.~Villaret.
\newblock Nominal unification from a higher-order perspective.
\newblock {\em ACM Transactions in Computer Logic}, 13(2):10:1--10:31, 2012.

\bibitem{Mayr1998}
R.~Mayr and T.~Nipkow.
\newblock Higher-order rewrite systems and their confluence.
\newblock {\em Theoretical Computer Science}, 192(1):3 -- 29, 1998.

\bibitem{Nipkow1991}
T.~Nipkow.
\newblock Higher-order critical pairs.
\newblock In {\em Proceedings of IEEE Symposium on Logic in Computer Science},
  pages 342--349, 1991.

\bibitem{Nipkow1998}
T.~Nipkow and C.~Prehofer.
\newblock Higher-order rewriting and equational reasoning.
\newblock In W.~Bibel and P.~Schmitt, editors, {\em Automated Deduction --- A
  Basis for Applications. Volume I: Foundations}, volume~8 of {\em Journal of
  Applied Logic}, pages 399--430. Kluwer, 1998.

\bibitem{Pitts2003}
A.~M. Pitts.
\newblock Nominal logic, a first order theory of names and binding.
\newblock {\em Information and Computation}, 186:165--193, 2003.

\bibitem{PittsA:nomsetbook}
A.~M. Pitts.
\newblock {\em Nominal sets: names and symmetry in computer science}, volume~57
  of {\em Cambridge Tracts in Theoretical Computer Science}.
\newblock Cambridge University Press, 2013.

\bibitem{sagan2001}
B.~Sagan.
\newblock {\em The symmetric group}, volume 203.
\newblock Springer-Verlag New York, 2 edition, 2001.

\bibitem{Urban2003}
C.~Urban, A.~M. Pitts, and M.~Gabbay.
\newblock Nominal unification.
\newblock In M.~Baaz and J.~Makowsky, editors, {\em Computer Science Logic},
  volume 2803 of {\em Lecture Notes in Computer Science}, pages 513--527.
  Springer Berlin Heidelberg, 2003.

\bibitem{Urban2004}
C.~Urban, A.~M. Pitts, and M.~J. Gabbay.
\newblock Nominal unification.
\newblock {\em Theoretical Computer Science}, 323(1–3):473 -- 497, 2004.

\bibitem{Oostrom1994}
V.~van Oostrom and F.~van Raamsdonk.
\newblock Comparing combinatory reduction systems and higher-order rewrite
  systems.
\newblock In J.~Heering, K.~Meinke, B.~Möller, and T.~Nipkow, editors, {\em
  Higher-Order Algebra, Logic, and Term Rewriting}, volume 816 of {\em Lecture
  Notes in Computer Science}, pages 276--304. Springer Berlin Heidelberg, 1994.

\bibitem{Raamsdonk1999}
F.~{van Raamsdonk}.
\newblock Higher-order rewriting.
\newblock In P.~Narendran and M.~Rusinowitch, editors, {\em Rewriting
  Techniques and Applications}, volume 1631 of {\em Lecture Notes in Computer
  Science}, pages 220--239. Springer Berlin Heidelberg, 1999.

\end{thebibliography}
\end{document}